\documentclass[aps,prd,amssymb,amsmath]{revtex4}
\usepackage{latexsym}
\usepackage{graphics}
\usepackage{graphicx}
\usepackage{feynmf}
\usepackage{color}
\usepackage{MnSymbol}
\newcommand{\be}{\begin{equation}}
\newcommand{\ee}{\end{equation}}
\newcommand{\ba}{\begin{eqnarray}}
\newcommand{\ea}{\end{eqnarray}}

\newcommand{\pa}{\partial}

\newcommand{\bw}{\begin{widetext}}
\newcommand{\ew}{\end{widetext}}

\newcommand{\ts}{\tilde{\sigma}}
\newcommand{\mb}{\mathbf}

\begin{document}
\title{Bulk viscosity and the phase transition of the linear sigma model}
\author{Antonio Dobado}
\email{dobado@fis.ucm.es}
\affiliation{Departamento de F\'isica
Te\'orica I, Universidad Complutense de Madrid, 28040 Madrid, Spain}
\author{Juan M. Torres-Rincon}
\email{jtorres@physics.umn.edu, j.torres@fis.ucm.es}
\affiliation{School of Physics and Astronomy, University of Minnesota, Minneapolis, Minnesota 55455, USA\footnote{On leave of absence from Departamento de F\'isica Te\'orica I, Universidad Complutense de Madrid, 28040 Madrid, Spain.}}

\begin{abstract}
In this work we deal with the critical behavior of the bulk
viscosity in the linear sigma model (L$\sigma$M) as an example of a
system which can be treated by using different techniques. Starting
from the Boltzmann-Uehling-Uhlenbeck equation we compute  the bulk
viscosity over entropy density  of the L$\sigma$M in the large-$N$
limit. We search for a possible  maximum of $\zeta/s$  at the
critical temperature of the chiral phase transition.  The
information about this critical temperature, as well as the
effective masses, is obtained from the effective potential. We find
that the expected maximum  (as a measure of the conformality loss)
  is absent in the large-$N$ in agreement with
other models in the same limit. However, this maximum appears when,
instead of the large-$N$ limit, the Hartree approximation within the
Cornwall-Jackiw-Tomboulis formalism is used. Nevertheless, this
last approach to the L$\sigma$M  does not give rise to the Goldstone
theorem and also predicts a first-order phase transition instead of
the expected second-order one. Therefore both, the large-$N$ limit
and the Hartree approximations, should
be considered relevant and informative for the study of the critical behavior of the bulk viscosity in the L$\sigma$M.

\end{abstract}
\maketitle
\section{Introduction}

Transport properties are essential to understand the equilibration of the hadronic
matter created at heavy-ion colliders. In this context the most studied transport
 coefficient has been the shear viscosity $\eta$.
When two relativistic nuclei collide, the initial spatial anisotropy
of the interacting region is converted into momentum anisotropy due
to the hydrodynamical evolution. The momentum-anisotropy
equilibration is mainly controlled by the shear viscosity and it
influences the value of the flow coefficients $v_n$. Comparing the
measured value of the elliptic flow
$v_2$~\cite{Adare:2006ti,Aamodt:2010pa} with the numerical
hydrodynamic simulations, a value of $\eta/s \simeq 0.1$ has been
estimated~\cite{Luzum:2008cw,Song:2008hj,Bozek:2011ph} (where $s$ is
the entropy density). This small value, very close to the Kovtun,
Son and Starinets (KSS) minimum conjectured in~\cite{Kovtun:2004de},
has characterized the matter created in relativistic heavy-ion
collisions as a nearly perfect fluid. It has also been shown that
the KSS coefficient $\eta/s$  has a minimum at the liquid-gas phase
transition of common fluids~\cite{Csernai:2006zz,Dobado:2008vt}. A
minimum near the phase transition of the linear sigma model
(L$\sigma$M) in the large-$N$ limit has also been
found~\cite{Dobado:2009ek} and this minimum is also expected in the
deconfinement phase transition of
QCD~\cite{Dobado:2008vt,Dobado:2008ri}.

Another transport coefficient that relates momentum flux with a velocity
gradient is the bulk viscosity $\zeta$. It is
sensitive to uniform expansions of the system in such a way that it
is closely related to the scale invariance of the
fluid~\cite{Weinberg:1971mx,Canuto:1978jj}. For conformal systems
the bulk viscosity identically vanishes. This coefficient has been
assumed to be much smaller than the shear viscosity as it is the
case for common fluids. Moreover, in perturbative QCD the ratio
between the bulk viscosity and the shear viscosity has been
estimated around $10^{-3}-10^{-8}$~\cite{Arnold:2006fz}. In the low
temperature phase, the hadronic medium, this coefficient has been
calculated using the Green-Kubo formalism and kinetic theory
in~\cite{FernandezFraile:2008vu} and~\cite{Dobado:2011qu},
respectively. In this phase, the ratio $\zeta/\eta$ has been found
to be around $10^{-3}-3 \cdot 10^{-3}$~\cite{TorresRincon:2011sa}.

However, near the critical point $\zeta$ can be larger than the
shear viscosity. For those systems belonging to the dynamical
universality class of model H (like the liquid-gas transition) the
bulk viscosity diverges~\cite{Onuki} with the correlation length
as $\zeta \simeq \xi^{2.8}$ (or as a function of the reduced
temperature [$t=T/T_c-1$] as $\zeta \simeq t^{-1.8}$). In
lattice QCD calculations of the bulk viscosity over entropy density
above the critical temperature~\cite{Meyer:2007dy,Karsch:2007jc}
this coefficient seems to diverge at $T_c$. Other authors have
pointed out that the bulk viscosity presents a maximum in the
critical temperature for some different models like the $O(N)$
model, even at mean-field level~\cite{Paech:2006st,Kharzeev:2007wb,Li:2009by,Chakraborty:2010fr}.
However, this maximum at $T_c$ is not seen in other systems as in
the Gross-Neveu model in the limit of a large number of fermion
fields~\cite{FernandezFraile:2010gu}, where the bulk viscosity is a
monotonically decreasing function when temperature increases. In the
context of the $O(N)$ model it has been recently shown~\cite{Nakano:2011re} that it belongs to model C if $N=1$ and to
model G if $N>1$ (particularly the large-$N$ limit) with a
divergence with the correlation length if $N=1$ going like $\zeta \sim \xi^2$ and being finite for
$N>1$ as $\zeta \sim \xi^0$.

In the hydrodynamic calculations of relativistic heavy-ion
collisions, the bulk viscosity over entropy density can also be
extracted. For example in~\cite{Bozek:2011ph} it is found that a
value $\zeta/s=0.04$ is compatible with the integrated elliptic flow
measured by RHIC and ALICE collaborations. In~\cite{Dusling:2011fd}
a similar value is found ($\zeta/s \lesssim 0.05$) near the
freeze-out temperature.

In principle our goal is to study the behavior of the bulk viscosity
over entropy density in the L$\sigma$M in the large-$N$ limit following
the same lines of our previous article~\cite{Dobado:2009ek}. We will
compare the phase diagram of the L$\sigma$M in the large-$N$ limit
with the coefficient $\zeta/s$ in order to ascertain if a maximum of
$\zeta/s$ is found near $T_c$. In Sec.~\ref{sec:LSM} we present the
L$\sigma$M and its effective potential at finite temperature in the
large-$N$ limit. In Sec.~\ref{sec:shearvisco} we review the
calculation of the shear viscosity over entropy density in this
model. In Sec.~\ref{sec:bulkvisco} we will perform the detailed
computation of the bulk viscosity over entropy density and discuss
the results in terms of the conformality loss of the system. In
Sec.~\ref{sec:bulkcjt} we repeat the calculation for $\zeta/s$
 in the context of the Cornwall-Jackiw-Tomboulis (CJT)
formalism in the Hartree approximation. Finally, in
Sec.~\ref{sec:conclusions} we present our main conclusions.

\section{L$\sigma$M Lagrangian in the large-$N$ limit \label{sec:LSM}}

In this section we will review the dynamics of the L$\sigma$M at
finite temperature, deriving the effective potential in the large-$N$ limit.
As we have already detailed this procedure
in~\cite{Dobado:2009ek} we will only sketch the key steps of the
calculation.

The bare Euclidean Lagrangian of the L$\sigma$M reads~\footnote{With
respect to the Lagrangian in~\cite{Dobado:2009ek} we explicitly show
the $N$ dependence of the coupling constant. The Lagrangian
shown there was in Minkowski space.} \be \label{eq:lsmlagrangian}
\mathcal{L}  = \frac{1}{2} \pa_{\mu} \Phi^T \pa^{\mu} \Phi -
\overline{\mu}^2 \Phi^T \Phi + \frac{\lambda}{N} \left(\Phi^T \Phi
 \right)^2 -\epsilon \Phi_{N+1} \ , \ee
where the multiplet $\Phi=(\pi_{i},\sigma)$ ($i=1,N$) contains $N+1$
scalar fields. $\lambda$ is positive in order to have a potential
bounded from below and we consider $\overline{\mu}^2$ to be positive
in order to provide a spontaneous symmetry breaking (SSB). When
$\epsilon=0$, the SSB pattern is $SO(N+1) \rightarrow SO(N)$. In the
following we will refer for historical reasons to the $\pi$ fields
as pions (even if it is well established that the L$\sigma$M is not
an accurate model for chiral dynamics) and to the $\sigma$ related
degree of freedom as the Higgs.

The factor $\epsilon=m^2_{\pi} f_{\pi}$ is responsible for the
physical pion mass and, as it is well known,  it gives rise to an
explicit breaking of the $SO(N+1)$ symmetry. In the limit  $T=0$,
the potential has a nonzero vacuum expectation value (VEV) and one
expects SSB. Choosing the VEV, which will be denoted by  $f_{\pi}$,
pointing in the $N+1$ direction, we get the equation: \be
\label{eq:min} - 2 \overline{\mu}^2 f_{\pi} + \frac{4\lambda}{N}
f^3_{\pi} - \epsilon =0 \ . \ee For small $\epsilon$ the solution to
this equation is: \be \label{eq:fpihelp} f_{\pi} = \sqrt{\frac{N \overline{\mu}^2}{2
\lambda}} + \frac{\epsilon}{4 \overline{\mu}^2} = f_{\pi}
(\epsilon=0) + \frac{\epsilon}{4 \overline{\mu}^2}
 \ . \ee

The VEV can be written in terms of  the $N$-independent $F$
parameter defined as: \be \langle \Phi^T \Phi \rangle = \langle
\sigma^2 (T=0) \rangle = f^2_{\pi} = NF^2 \ . \ee

In our notation we will call $f_{\pi} (\epsilon=0)$ and
$f_{\pi}$  the VEV at $T=0$ for the
cases without and with explicit symmetry breaking term, respectively.
In the next sections, the VEV at arbitrary temperature will be
denoted by $v(T)$, in such a way that $f_{\pi} = v(T=0)$.

At  $T=0$ the low-energy dynamics is controlled by the broken phase.
In this case  $\langle \pi^a \rangle =0$. Then the relevant degrees
of freedom are the pions which
 correspond to the (pseudo) Goldstone bosons when ($\epsilon \neq 0$)  $\epsilon=0$. Fluctuations
  along the $\sigma$ direction will be denoted by $\ts$ and they correspond to the Higgs,
the massive mode which is relevant at higher energies (or temperatures):
\be \label{eq:sigma} \sigma = f_{\pi} + \ts \rightarrow \langle  \sigma  \rangle = f_{\pi} \ . \ee

The Lagrangian (\ref{eq:lsmlagrangian}) written in terms of $\pi^a$, $f_{\pi}$ and $\ts$ reads:

\begin{eqnarray}{}
\nonumber  \mathcal{L} & = &  \frac{1}{2} \pa_{\mu} \pi^a \pa^{\mu} \pi^a+\frac{1}{2} \pa_{\mu} \ts \pa^{\mu} \ts - \overline{\mu}^2 \pi^a \pi^a -
\overline{\mu}^2 \ts^2
 +   \frac{\lambda}{N} \left[ 2 f^2_{\pi} \left( \pi^a \pi^a + 3 \ts^2 \right) + (\pi^a \pi^a)^2 + 2 \pi^a  \pi^a \ts^2 + \ts^4 + 4 f_{\pi} \ts^3 +4 \pi^a \pi^a f_{\pi} \ts
 \right]  \\
\label{eq:lagtzero} & + &  \left(-\epsilon + \frac{4 \lambda}{N} f^3_{\pi} - 2 \overline{\mu}^2 f_{\pi} \right) \ts - \overline{\mu}^2 f^2_{\pi} + \frac{\lambda}{N} f^4_{\pi} - \epsilon f_{\pi}
\end{eqnarray}

Note that the tadpole term vanishes because of Eq.~(\ref{eq:min}).
The tree-level Higgs mass can be read from the Lagrangian
in~(\ref{eq:lagtzero}) \be M^2_{\ts} = - 2\overline{\mu}^2 +
\frac{12 \lambda}{N} f^2_{\pi} \ , \ee and it can also be written
as: \be \label{eq:effmasssigma} M^2_{\ts} = 4
\overline{\mu}^2+ 3 \frac{\epsilon}{f_{\pi}} = \frac{8 \lambda
f^2_{\pi}}{N} + \frac{\epsilon}{f_{\pi}} \simeq \frac{8 \lambda
f^2_{\pi}(\epsilon=0)}{N} + 3\frac{\epsilon}{f_{\pi}} \ , \ee
using Eq.~(\ref{eq:fpihelp}) for the last identity.

The pion mass reads: \be m^2_{\pi} = - 2 \overline{\mu}^2 + \frac{4
\lambda}{N} f_{\pi}^2 \ , \ee which, as expected, depends only on
the explicit symmetry breaking term \be m^2_{\pi} =
\frac{\epsilon}{f_{\pi}} \ ,\ee and vanishes when $\epsilon=0$
(obviously all pions are degenerate). The pion and Higgs masses are
related through \be \label{eq:relmasses} M^2_{\ts}= m^2_{\pi} +
\frac{8 \lambda f^2_{\pi}}{N} \ . \ee

For small enough explicit symmetry breaking we have:
\begin{eqnarray}
\overline{\mu}^2 & = & \frac{M^2_{\ts} - 3 m^2_{\pi}}{4} \ , \\
\label{eq:lambda} \lambda & = & \frac{N}{8 f^2_{\pi}} (M^2_{\ts}-m^2_{\pi}) \ .
\end{eqnarray}

Equation~(\ref{eq:lambda}) can be expressed in terms of the VEV at $\epsilon=0$:
\be \lambda = \frac{N}{8 f^2_{\pi} (\epsilon=0)} \frac{M^2_{\ts}-m^2_{\pi}}{\alpha^2} \ , \ee
where $\alpha$ is a multiplicative constant relating $f_{\pi}$ and $f_{\pi} (\epsilon=0)$
\be f_{\pi} = \alpha f_{\pi} (\epsilon=0) \ , \ee
which can be written as:
\be \alpha= \frac{M^2_{\ts}-3 m^2_{\pi}}{M^2_{\ts}-4 m^2_{\pi}} \ , \ee
where we have used Eq.~(\ref{eq:fpihelp}) and consider small $\epsilon$.

In order to obtain the effective potential we will integrate out the
fluctuations from the partition function $\mathcal{Z}$. To do that it
is convenient to introduce
 an auxiliary scalar field. This field allows for a systematic counting of $N$ factors and provides  some simplifications in the large-$N$ limit.
Then, the integration of the pions and the Higgs is performed by standard Gaussian integration.

The integration of the fluctuations is done regardless of their wavelenghts. All the frequency modes of the scalar fields are treated at the same
footing and this ``unorganized'' integration produces two undesirable features in the effective potential. First, an imaginary part of the effective potential appears. This imaginary
part has been given the interpretation of a decay rate per unit volume of the unstable vacuum state by Weinberg and Wu in~\cite{Weinberg:1987vp}.

The second characteristic is the nonconvexity of the quantum
effective potential. However, the effective potential, defined
through a Legendre transformation of the generating functional of
the connected diagrams~\cite{Iliopoulos:1974ur}, should always be
convex. This nonconvexity problem and the imaginary part appear as
long as a perturbative method is used to calculate the effective
potential~\cite{vanKessel:2008ht}.

A possible solution for these problems can be given by using a
nonperturbative method to generate the effective potential. For
example, the functional renormalization group generates an
effective potential in such a way that an organized integration of
the fluctuations is performed. Following the ideas of the
renormalization group, only the low-wavelength components of the
quantum fluctuations are integrated out at each step. Thus, the UV
components are infinitesimally integrated step by step and the final
effective potential (defined in the infrared scale) does not acquire
an imaginary part and it remains convex at every scale (at the IR
point, the Maxwell construction is dynamically generated by the
renormalization flow)~\cite{Alexandre:1998ts}.

In spite of the previous discussion, we understand  that it is not
necessary to perform a more sophisticated method to obtain the
effective potential. The only relevant information for us is the
location of the effective potential minimum, which eventually gives
the position of the critical temperature. This minimum appears
always outside of the nonconvex region. On the other hand, the
possible presence of an imaginary part (whose domain in fact
coincides with the domain of the nonconvex part of the effective potential)
is not relevant for us in this work.

\subsection{Auxiliary field method and effective potential at $T\neq 0$}

To compute the effective potential in the large-$N$ limit we start by
considering the partition function: \be \mathcal{Z} = \int
\mathcal{D} \pi^a \mathcal{D} \sigma \ \exp \left( -\int d^4 x \
\mathcal{L} \right) \ , \ee with the Lagrangian of
Eq.~(\ref{eq:lsmlagrangian}). Then, we introduce an auxiliary field
$\chi$ in order to deal with the quartic coupling by using  the
Gaussian integral: \be \label{eq:auxfield} \exp \left( - \int d^4x \frac{\lambda}{N}
\Phi^4 \right) \propto \int \mathcal{D} \chi \exp \left[ 
\int d^4x \left( \frac{N}{8 \lambda} \chi^2 - \frac{\sqrt{2}}{2} \chi \Phi^2
\right) \right] \ . \ee
To prove the equivalence between the two Lagrangians note that this auxiliary field has
introduced a mass term and a coupling with $\Phi^2$ in the
Lagrangian. However, it has no kinetic term which means that $\chi$
has not a true dynamics. The Euler-Lagrange equation for $\chi$ simply gives $\chi= 2 \sqrt{2} \lambda \Phi^2/N$. Introducing this solution into the right-hand side of Eq.~(\ref{eq:auxfield})
one obtains the Lagrangian
\be \mathcal{L} = -\frac{N}{8 \lambda} \chi^2 + \frac{\sqrt{2}}{2} \chi \Phi^2 = - \frac{\lambda}{N} \Phi^4 + 2 \frac{\lambda}{N} \Phi^4 = \frac{\lambda}{N} \Phi^4 \ , \ee
which is the original interaction Lagrangian.

The partition function can then be written as:
\be
  \nonumber \mathcal{Z}  =  \int \mathcal{D} \pi^a \mathcal{D} \sigma \mathcal{D} \chi \  \exp \left[ -\int d^4 x \
\left( \frac{1}{2} \pa_{\mu} \pi^a \pa^{\mu} \pi^a + \frac{1}{2} \pa_{\mu} \sigma \pa^{\mu} \sigma - \overline{\mu}^2 \pi^a \pi^a
- \overline{\mu}^2 \sigma^2 -  \frac{N}{8\lambda} \chi^2 + \frac{\sqrt{2}}{2}  \chi \pi^a \pi^a + \frac{\sqrt{2}}{2} \chi \sigma^2 - \epsilon \sigma \
\right) \right] \ .
\ee

Thus the action in terms of the $\pi^a$, $\sigma$ and $\chi$ fields reads:
\be \label{eq:actionnofinal} S = \int d^4 x \ \left[
\frac{1}{2} \pi^a \left( - \square_E - 2\overline{\mu}^2 + \sqrt{2} \chi \right) \pi^a
+ \frac{1}{2} \sigma \left( - \square_E - 2\overline{\mu}^2 + \sqrt{2} \chi \right) \sigma
-  \frac{N}{8 \lambda} \chi^2 - \epsilon \sigma \right] \ . \ee

Before identifying properly the pion propagator one must get rid of
the unphysical $\sigma$ tadpole. We have already seen that this term
vanishes at $T=0$. Now, we perform a shift
of the $\sigma$ field $\sigma = v + \ts$ which also produces a
change in the auxiliary field $\chi = \tilde{\chi} + 2 \sqrt{2}
\frac{\lambda}{N} v^2 + 4 \sqrt{2} \frac{\lambda}{N} v \ts$
allowing to cancel the tadpole term for $\ts$ and the unphysical mass mixing term between $\ts$ and $\tilde{\chi}$. After some manipulations we obtain the action:
\be S=  \int d^4 x \ \left[
\frac{1}{2} \pi^a \left( - \square_E + G_{\pi}^{-1} [0,\chi] \right) \pi^a + \frac{1}{2} \ts \left( - \square_E + G_{\pi}^{-1} [0,\chi] +8 \frac{\lambda}{N} v^2 \right) \ts
+ \frac{1}{2} v^2 G_{\pi}^{-1} [0,\chi] - \frac{N}{8 \lambda} \tilde{\chi}^2 - \frac{\sqrt{2}}{2} \chi v^2 + \frac{\lambda}{N} v^4 - \epsilon v \right] \ , \ee 
where we have introduced the function:
\be \label{eq:relgandchi} G^{-1}_{\pi} [q,\chi] \equiv q^2 - 2\overline{\mu}^2 + \sqrt{2} \chi \ .\ee
Now comes our approximation for the auxiliary field. $\chi$ is not going to be integrated out, but treat it at mean-field level, so that it contains no fluctuations. In particular, we apply this
simplification for the quadratic term in the action. At mean field (note that we will keep the same notation $\chi$, instead of using $\langle \chi \rangle$ as they coincide in this approximation):
\be \tilde{\chi} = \chi - 2 \sqrt{2} \frac{\lambda}{N} v^2 \ . \ee
So the quadratic term reads
\be - \frac{N}{8 \lambda} \tilde{\chi}^2 = - \frac{N}{8 \lambda} \chi^2 + \frac{\sqrt{2}}{2} v^2 \chi - \frac{\lambda}{N} v^4 \ . \ee
The last two factors are cancelled in the action (\ref{eq:actionnofinal}). We are going to trade $\chi$ by $G^{-1}_{\pi} [0,\chi]$. Using (\ref{eq:relgandchi}), the quadratic terms is:
\be - \frac{N}{8 \lambda} \chi^2 = - \frac{N}{16 \lambda} (G^{-1}_{\pi} [0,\chi])^2 - \frac{N}{4 \lambda} \overline{\mu}^2 G_{\pi}^{-1} - \frac{N}{4 \lambda} \overline{\mu}^4 \ . \ee
We will neglect the last term in the action because it is a constant. Finally we use the relations
\be \frac{N \overline{\mu}^2}{4 \lambda} = \frac{f^2_{\pi} (\epsilon=0)}{2} = \frac{f_{\pi}^2}{2\alpha^2} = \frac{NF^2}{2\alpha} \ee
to introduce $F=v(T=0)$, i.e. the VEV at zero temperature.

The final action becomes:

\begin{eqnarray}  S[\pi^a, v , \ts, G^{-1}_{\pi}[0,\chi]] & = &  \int d^4 x \ \left[
\frac{1}{2} \pi^a \ \left(-\square_E + G^{-1}_{\pi}[0,\chi] \right) \pi^a + \frac{1}{2} \ts \ \left(-\square_E + G^{-1}_{\pi}[0,\chi] + 8\frac{\lambda}{N} v^2 \right) \ \ts \nonumber \right. \\
 & & \left. + \frac{1}{2} v^2 G^{-1}_{\pi} [0,\chi]  -  \frac{NF^2}{2\alpha^2} G^{-1}_{\pi} [0,\chi] - \frac{N}{16 \lambda} (G^{-1}_{\pi} [0,\chi])^2- \epsilon v \right] \ , \end{eqnarray}

Notice that  $G^{-1}_{\pi} [q,\chi]$ is nothing but the inverse of the pion propagator in Fourier space. The inverse propagator of the $\ts$ field is:
\be G^{-1}_{\ts} [q,\chi] = G^{-1}_{\pi} [q,\chi] + 8 \frac{\lambda}{N} v^2 \ .  \ee

In order to generate the effective potential for $v$ we now integrate out the fluctuations.
By performing  a standard Gaussian integration of the pions we get:
\be
\int \mathcal{D} \pi^a \exp \left(- \int d^4x \ \frac{1}{2} \pi^a \left[ - \square_E + G_{\pi}^{-1} [0,\chi] \right] \pi^a \right)
= \int d^4x \exp \left(  \frac{N}{2} \int T \sum_{n \in \mathcal{Z}}\frac{d^3 q}{(2\pi)^3} \log G_{\pi}^{-1} [q,\chi] \right)  \ ,
\ee
where  $q_0=2 \pi n T$ is the well-known Matsubara frequency appearing in finite temperature computations. The same procedure can be applied to integrate out $\ts$. To be able to perform the integrations
 we have assumed that the auxiliary field (or $G_{\pi}^{-1} [0,\chi]$) is homogeneous, i.e. it does not depend on $x$. This assumption is also taken for the VEV of the $\sigma$ field and allows to obtain a simple representation
in Fourier space, where different modes do not mix between them and the integration is straightforward.

The effective potential (density) reads
\be V_{eff} (v,G^{-1}_{\pi} [0,\chi]) = \frac{1}{2} \left( v^2 - \frac{NF^2}{\alpha^2} \right) G_{\pi}^{-1} [0,\chi] - \frac{N}{16 \lambda } (G_{\pi}^{-1} [0,\chi]) ^2 -\epsilon v + \frac{N}{2} \sumint_{\beta} \log G^{-1}_{\pi} [q,\chi]+\frac{1}{2} \sumint_{\beta} \log G^{-1}_{\ts} [q,\chi] \ , \ee
with
\be  \sumint_{\beta} = T \sum_{n \in \mathcal{Z}} \int \frac{d^3q}{(2\pi)^3} \ .\ee

Looking at the $N$ power counting of the different terms in the
effective potential one finds that all of them behave as
$\mathcal{O}(N)$ except the last one. Therefore the
contribution of the Higgs to the effective potential is suppressed
in the large-$N$ limit by one power of $N$ and it will be neglected
in the following. Then the effective potential becomes:
 \be V_{eff} (v,G^{-1}_{\pi} [0,\chi])=
\frac{1}{2} \left( v^2 - \frac{NF^2}{\alpha^2} \right) G_{\pi}^{-1}
[0,\chi] - \frac{N}{16 \lambda } (G_{\pi}^{-1} [0,\chi]) ^2
-\epsilon v + \frac{N}{2} \sumint_{\beta} \log G^{-1}_{\pi} [q,\chi]
\ .\ee

The last term needs to be regulated because it contains a
divergence. This divergence can be absorbed by a proper
renormalization of the  quartic coupling~\cite{Dobado:2009ek}. Thus
the renormalized ($\mu$-independent) effective potential finally
reads:

\be \label{eq:finaleffpot}
V_{eff} (v,G^{-1} [0,\chi])  =  \frac{1}{2} \left(v^2 - \frac{NF^2}{\alpha^2} \right) G^{-1}_{\pi} [0,\chi] - \epsilon v - \frac{N}{2} g_0 (T, G^{-1}_{\pi} [0,\chi])  - \frac{N}{16} (G^{-1}_{\pi} [0,\chi])^2 \left[
\frac{1}{\lambda_R (\mu)} - \frac{1}{4 \pi^2} \log \left( \frac{\sqrt{e} G^{-1}_{\pi} [0,\chi]}{\mu^2} \right)
\right] \ .
\ee
where the function $g_0(T,G_{\pi}^{-1}[0,\chi])$ is defined in App.~\ref{app:moments1}.

The stationary conditions are given by:
\be \label{eq:gapeq} \left. \frac{d V_{eff}}{d G^{-1}_{\pi} [0,\chi]} \right|_{G^{-1}_{\pi} [0,\chi]=G^{-1}_{\pi,0}[0,\chi]} =0, \quad  \left. \frac{d V_{eff}}{dv} \right|_{v=v_0}=0 \ , \ee
and they provide  $G^{-1}_{\pi,0} [0,\chi]$ and the order parameter $v_0$ in this approximation. The effective mass of the pion is obtained as $m^2_{\pi}= G^{-1}_{\pi, 0} [0,\chi]$ and the effective Higgs mass as
$M^2_{R} = G^{-1}_{\pi,0} [0,\chi] +8 \frac{\lambda}{N} v_0^2 $, which has the same form as Eq.~(\ref{eq:relmasses}).

\begin{figure}[t]
\begin{center}
\includegraphics[scale=0.3]{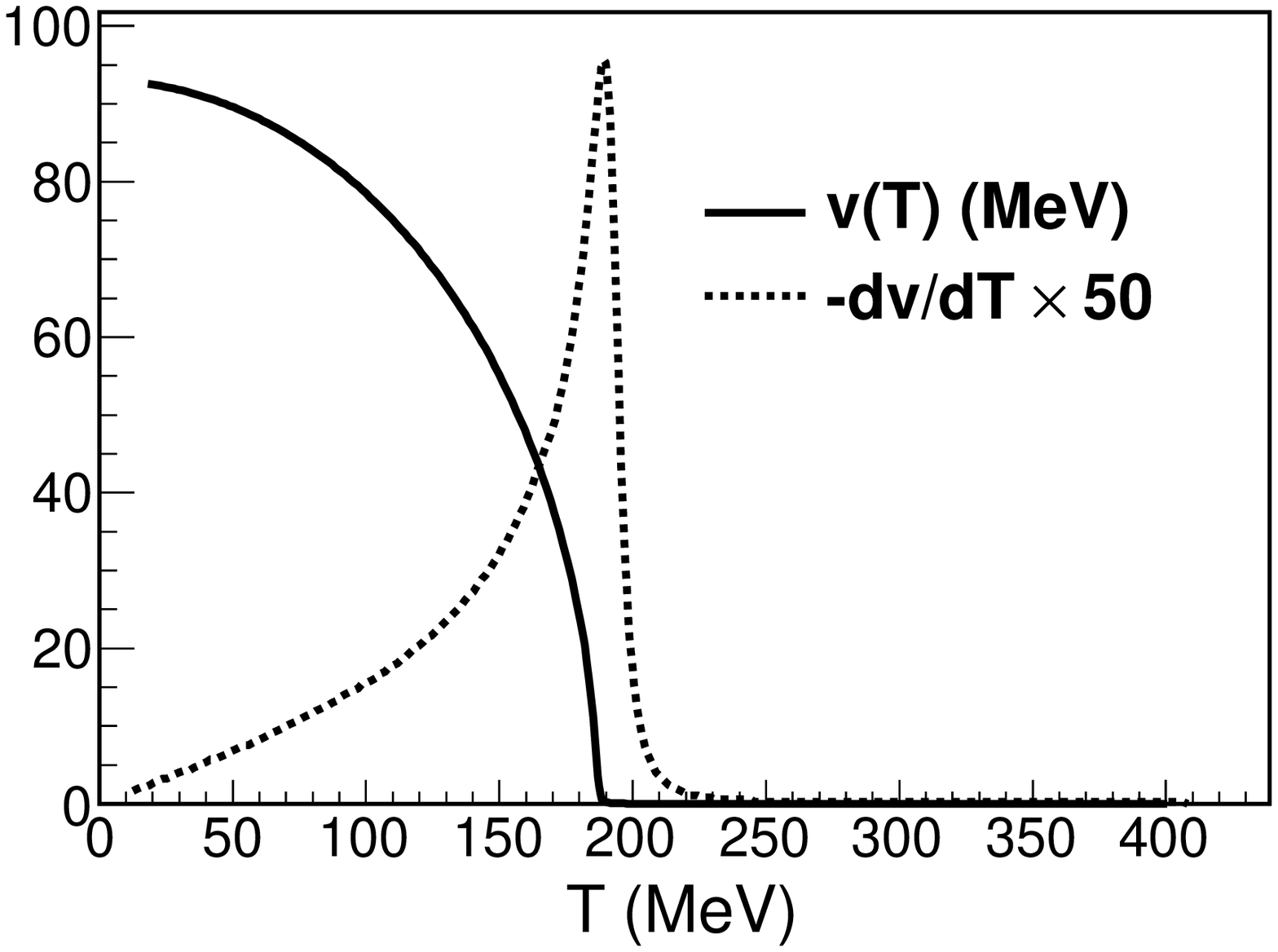}
\includegraphics[scale=0.3]{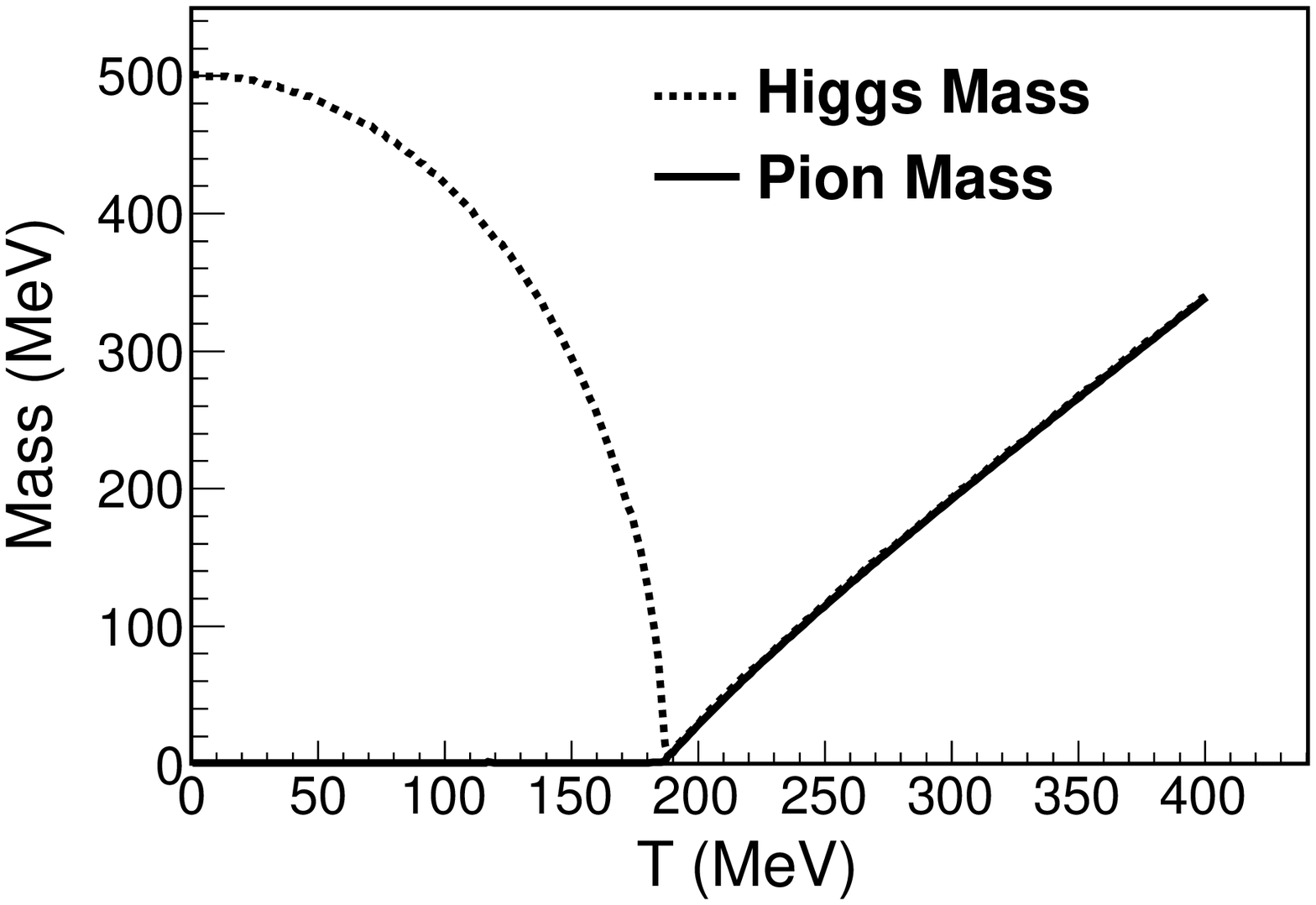}
\caption{\label{fig:sec_ord} Second-order phase transition. Left panel: Order
 parameter or VEV for $\ts$ and its susceptibility. Right panel: Higgs and pion effective masses.}
\end{center}
\end{figure}

For the $\epsilon=0$ case there is no explicit symmetry-breaking
 term and we expect to have a second-order phase transition defined
  by the critical temperature $T_c=\sqrt{12} F = \sqrt{12/N} f_{\pi}$. To obtain
   the numerical results appearing in Fig.~\ref{fig:sec_ord}  we have used
 a vanishing pion mass at $T=0$ (with $N=3$), a Higgs mass of $M_R=500$
  MeV and  $v(T=0)=93$ MeV. In the left panel we show the behavior
of $v(T)$ that follows the analytical  solution $v(T) = f_{\pi}
\sqrt{1-T^2/T_c^2}$. The numerical critical temperature coincides
with the theoretical value of $T_c = 2 f_{\pi} = 186$ MeV. We
also show the susceptibility defined as minus the $T-$derivative of
the order parameter. Its peak shows the position of the critical
temperature. In the right panel we show the effective masses as a
function of the temperature. The effective mass of the pions at
$T<T_c$ must be always zero according to the Goldstone theorem
(numerically it is fixed at $0.5$ MeV at $T=0$ in order to avoid
computational problems). At $T_c$ it starts growing with temperature
in the symmetric phase. The effective mass of the Higgs follows the
same pattern as the order parameter becoming zero at $T_c$. For
higher temperatures it increases until being degenerate
with the effective pion mass as expected because of the restoration of the $SO(N+1)$ symmetry.

\begin{figure}[t]
\begin{center}
\includegraphics[scale=0.3]{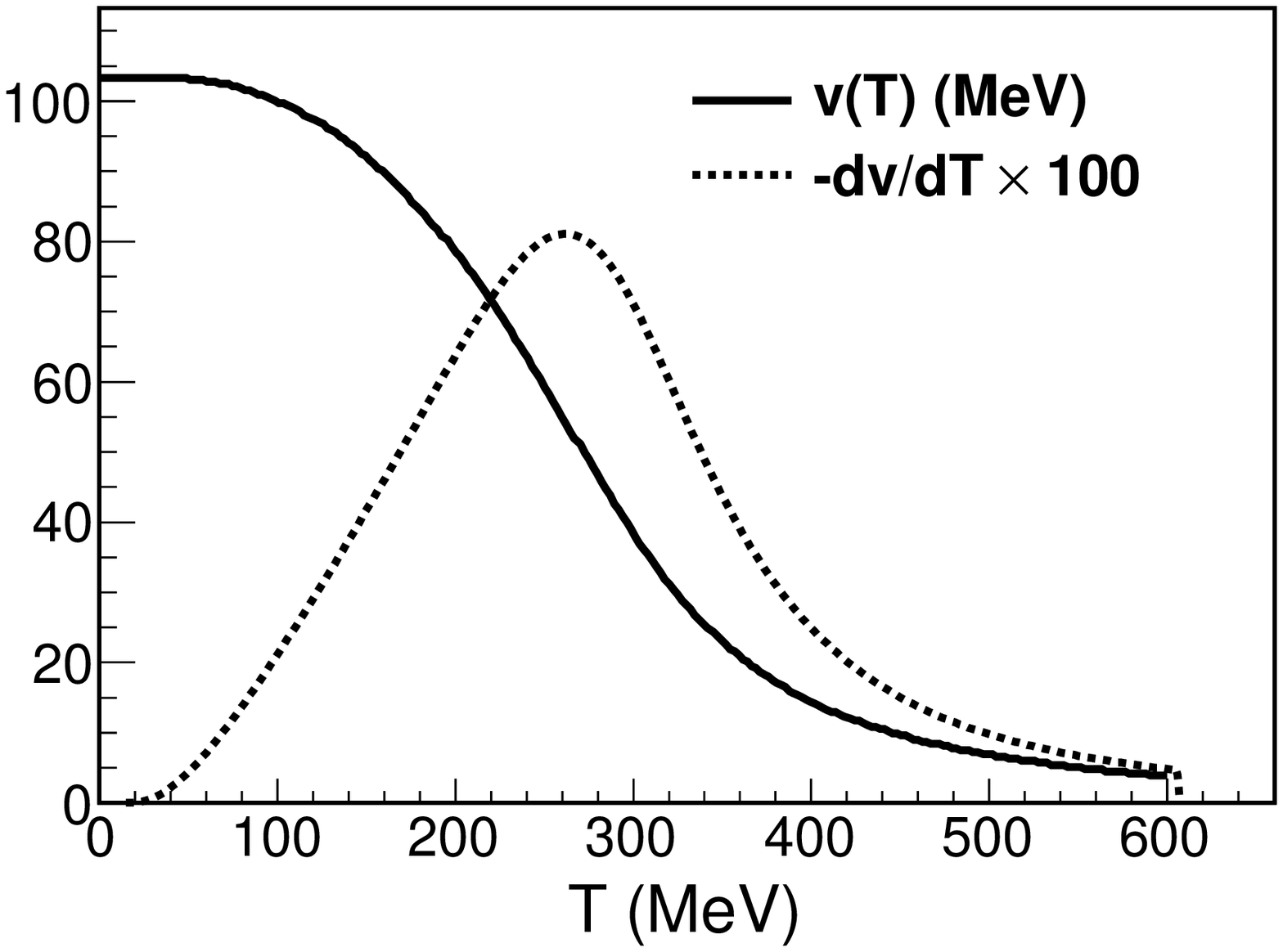}
\includegraphics[scale=0.3]{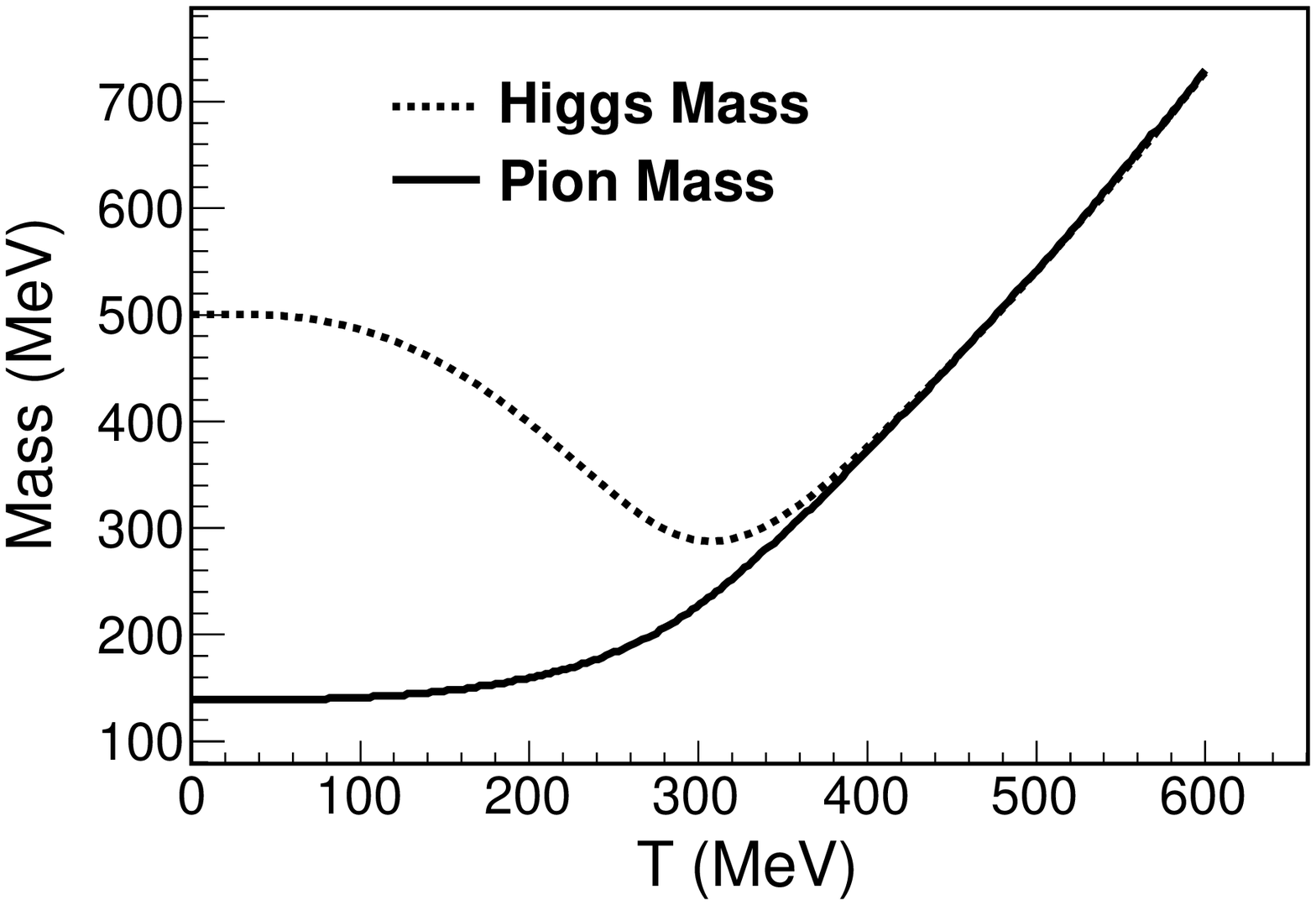}
\caption{\label{fig:crossover} Same as Fig.~\ref{fig:sec_ord} but in the crossover case, when the pion mass is fixed at $T=0$ to $m_{\pi}=138$ MeV.}
\end{center}
\end{figure}

For the $\epsilon \neq 0$ case a crossover instead of a real phase
transition is expected (this is the same situation as adding an
external magnetic field to a ferromagnet). We fix $m_{\pi}=138$ MeV
at zero temperature and the same values for $M_R$ and $f_{\pi}$
taken for the $\epsilon = 0$ case. The results are shown in
Fig.~\ref{fig:crossover}. The left panel shows the order parameter
$v(T)$ that decreases with temperature and never becomes exactly
zero. The crossover temperature can be defined as the position of
the peak in the susceptibility that we plot in dotted line. In the
right panel we show the pion thermal mass (solid line) and the Higgs
mass (dotted line). They are degenerate for high temperatures.

\section{Shear viscosity over entropy density \label{sec:shearvisco}}

The shear viscosity for the L$\sigma$M was obtained
in~\cite{Dobado:2009ek}. In the following we will briefly review the
method used there with some minimal changes. The transport equation
for the one-particle distribution function $f_p(t,\mb{x}) $ is: \be
\frac{df_p(t,\mb{x})}{dt} = C [f_p ] \ . \ee
Taking into account
only elastic collisions (we will discuss later the influence of
inelastic terms) this equation reads \be \frac{ \pa f_p (t,
\mb{x})}{ \pa t} + \frac{\mb{p}}{E_p} \cdot \nabla f_p (t,\mb{x}) =
\frac{N}{2} \int d\Gamma_{12,3p} \ [f_1 f_2 (1+f_3) (1+f_p) - f_3
f_p (1+f_1) (1+f_2)] \ , \ee
where
\be d\Gamma_{12,3p} = \frac{1}{2E_p}  \prod_{i=1}^3 \frac{d \mathbf{k}_i}{(2 \pi)^3 2 E_i} \ \overline{|T|^2} \ (2\pi)^4 \delta^{(4)} (k_1 +k_2 -k_3 -p)\  . \ee
At first order in the Chapman-Enskog expansion the
distribution function is expressed as the
local equilibrium distribution function plus a small correction: \be
f_p = n_p + f^{(1)}_p \ . \ee
where $n_p$ is the local Bose-Einstein distribution function 
\be n_p(t,\mb{x}) = \frac{1}{e^{\frac{
p^{\alpha} u_{\alpha} (t,\mb{x}) - \mu(t,\mb{x}) }{T(t,\mb{x})}}-1} \ . \ee

The Boltzmann-Uehling-Uhlenbeck (BUU) equation is then linearized in $f^{(1)}_p$:
\be E_p \frac{ \pa n_p (t, \mb{x})}{ \pa t} + \mb{p} \cdot \nabla n_p (t,\mb{x}) = -\frac{N E_p}{2} \int d\Gamma_{12,3p} (1+n_1) (1+n_2) n_3 n_p \left[ \frac{f_3^{(1)}}{n_3 (1+n_3)} + \frac{f_p^{(1)}}{n_p (1+n_p)} -
\frac{f_1^{(1)}}{n_1 (1+n_1)} - \frac{f_2^{(1)}}{n_2 (1+n_2)} \right] \ . \ee
The left-hand side depends only on the space-time derivatives of $n_p$, which can be explicitly obtained using the Euler and continuity equations.

The shear viscosity can be also expressed in terms of $f^{(1)}_p$:
\be \label{eq:shearvisco} 2 \eta  \tilde{V}_{ij} = N  \int \frac{d^3p}{(2\pi^3) E_p }f_p^{(1)} p_i p_j \ , \ee
where $\tilde{V}_{ij} = \pa_i V_j + \pa_j V_i - \frac{1}{3} \delta_{ij} \nabla \cdot \mb{V}$ is the shear gradient of the velocity field.

The left-hand side of the BUU equation also carries the same gradient (neglecting the influence of other transport coefficients)
\be E_p \frac{ \pa n_p (t, \mb{x})}{ \pa t} + \mb{p} \cdot \nabla n_p (t,\mb{x}) = \beta n_p (1+n_p) p^i p^j \tilde{V}_{ij} \ .\ee

In order to cancel out this factor from the BUU equation and also from (\ref{eq:shearvisco})  the function $f^{(1)}_p$ is taken as
\be \label{eq:f1out} f_p^{(1)} = - n_p (1+n_p) \beta^3 \ B(p) \ p^i p^j \tilde{V}_{ij} \ ,  \ee
with $B(p)$ being an unknown function of $p$.
Inserting Eq.~(\ref{eq:f1out}) into (\ref{eq:shearvisco}) one gets:
\be \eta = \frac{N}{30 \pi^2 T^3} \int dp \frac{p^6}{E_p} \ n_p (1+n_p) \ B(p) \ . \ee

It is convenient to write this formula in terms of the adimensional variables
\be \label{eq:adivar} x =\frac{E_p}{m_{\pi}} \ , \quad y = \frac{m_{\pi}}{T} \ .  \ee

Introducing the integration measure $d\mu_{\eta}(x;y)$ defined in App.~\ref{app:moments2}, we find:
\be \eta = \frac{Nm^6_{\pi} }{30 \pi^2 T^3} \int_{\Omega} d \mu_{\eta}(x;y) \ B(x) \ . \ee

This $B(x)$ function can be expanded in terms of the polynomial basis $P_n(x)$ defined in App.~\ref{app:moments2},

\be \label{eq:expshear} B(x)= \sum_{n=0}^{\infty} b_n P_n(x) \ . \ee

By projecting the BUU equation into the space generated by $P_n$ one gets
\be K^0 \delta_{l0} = \sum_{n=0}^N b_n \ \mathcal{C}^{\eta}_{nl} \ , \ee
where the functions $K^i$ are also defined in App.~\ref{app:moments2}. The collision integrals read
\begin{eqnarray}
\nonumber \mathcal{C}^{\eta}_{nl} & = & \frac{N \pi^2}{4 m_{\pi}^2 T^2 } \int \prod_{m=1}^4 \frac{d^3k_m}{(2 \pi)^3 2 E_m} \overline{|T|^2} (2\pi)^4 \delta^{(4)} (k_1+k_2-k_3-p) (1+n_1) (1+n_2) n_3 n_p \\
& & \times \nonumber \left[ \frac{p_i p_j}{m_{\pi}^2} P_l (p)+ \frac{k_{3i} k_{3j}}{m_{\pi}^2} P_l (k_3) - \frac{k_{1i} k_{1j}}{m_{\pi}^2}  P_l(K_1) - \frac{k_{2i} k_{2j}}{m_{\pi}^2} P_l (k_2)\right] \\
 & &\times  \left[ \frac{p^i p^j}{m_{\pi}^2} P_n (p)+ \frac{k^{3i} k^{3j}}{m_{\pi}^2} P_n (k_3) - \frac{k^{1i} k^{1j}}{m_{\pi}^2} P_n(k_1) - \frac{k^{2i} k^{2j}}{m_{\pi}^2} P_n(k_2) \right]  \ . \end{eqnarray}

At the lowest order in the expansion (\ref{eq:expshear}) the
shear viscosity can be written as \be \eta = \frac{Nm^6_{\pi}}{30 \pi^2 T^3}
\frac{K_0^2}{\mathcal{C}^{\eta}_{00}} \ , \ee where the pion effective mass
$m_{\pi}$ is now taken from the gap equations of the effective
potential and

\begin{eqnarray}
\nonumber \mathcal{C}^{\eta}_{00} & = & \frac{N \pi^2}{4 m_{\pi}^2 T^2 } \int \prod_{m=1}^4 \frac{d^3k_m}{(2 \pi)^3 2 E_m} \overline{|T|^2} (2\pi)^4 \delta^{(4)} (k_1+k_2-k_3-p) (1+n_1) (1+n_2) n_3 n_p \\
& & \times \left[ \frac{p_i p_j}{m_{\pi}^2} + \frac{k_{3i} k_{3j}}{m_{\pi}^2} - \frac{k_{1i} k_{1j}}{m_{\pi}^2}   - \frac{k_{2i} k_{2j}}{m_{\pi}^2} \right] \left[ \frac{p^i p^j}{m_{\pi}^2}
+ \frac{k^{3i} k^{3j}}{m_{\pi}^2}  - \frac{k^{1i} k^{1j}}{m_{\pi}^2}  - \frac{k^{2i} k^{2j}}{m_{\pi}^2}  \right]  \ . \end{eqnarray}

The scattering amplitude in the large-$N$ limit has been extensively
described in~\cite{Dobado:2009ek} which includes the tree-level
$\pi-\pi$ amplitude and its resummation in pion loops (as they
are of the same order in $N$). The  finite pion mass gives rise to
new couplings to the pion scattering~\cite{Dobado:1994fd} and the
corresponding amplitude should be added to the chiral one. In the
large-$N$ limit, the $s$-channel scattering is the leading one. As
the amplitude is $\mathcal{O} (1/N)$, the total cross section is
$\mathcal{O} (1/N)$, but the average cross section (which is included
in the collision integral) is $\mathcal{O} (1/N^2)$.

Note that, as $\mathcal{C}^{\eta}_{nl} \sim \mathcal{O} (1/N)$,  the shear viscosity is $\mathcal{O} (N^2)$. This  is expected~\cite{Aarts:2004sd}
since the shear viscosity is proportional to the inverse  coupling constant squared, and this is suppressed by one power of $N$.

\begin{figure}[t]
\begin{center}
\includegraphics[scale=0.35]{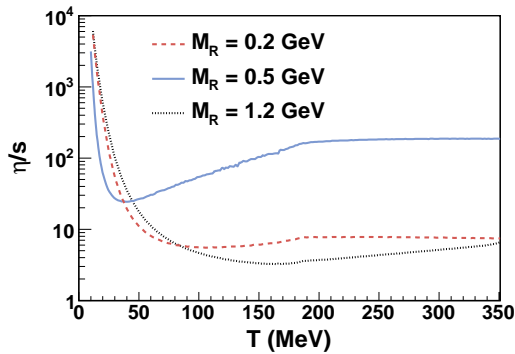}
\caption{\label{fig:kssdiffmr} Viscosity over entropy density in the L$\sigma$M at large $N$ for different values of $M_R$.}
\end{center}
\end{figure}

The obtained numerical results for $N=3$, $m_{\pi} (T=0)=0$ and Higgs masses
$M_R=0.2,0.5$ and $1.2$ GeV are shown in Fig.~\ref{fig:kssdiffmr}. They are
similar to that appearing in our Ref.~\cite{Dobado:2009ek} where we have used a slightly different
parametrization for $f^{(1)}_p$. We have found a minimum of $\eta/s$
for the three cases, always greater than the KSS bound $1/(4\pi)$.
However the exact position of the minimum depends on the value of $M_R$.

\begin{figure}[t]
\begin{center}
\includegraphics[scale=0.35]{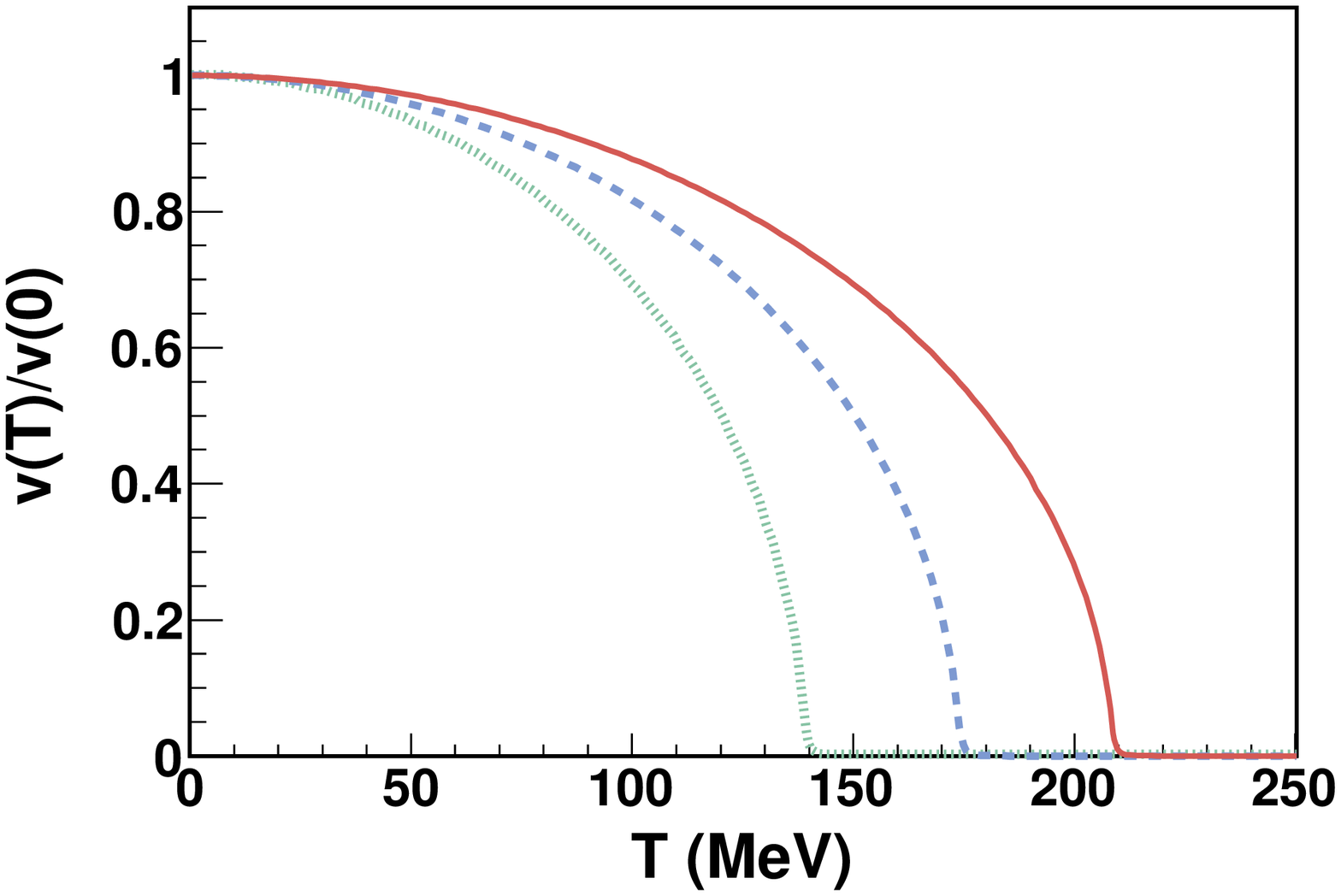}
\includegraphics[scale=0.35]{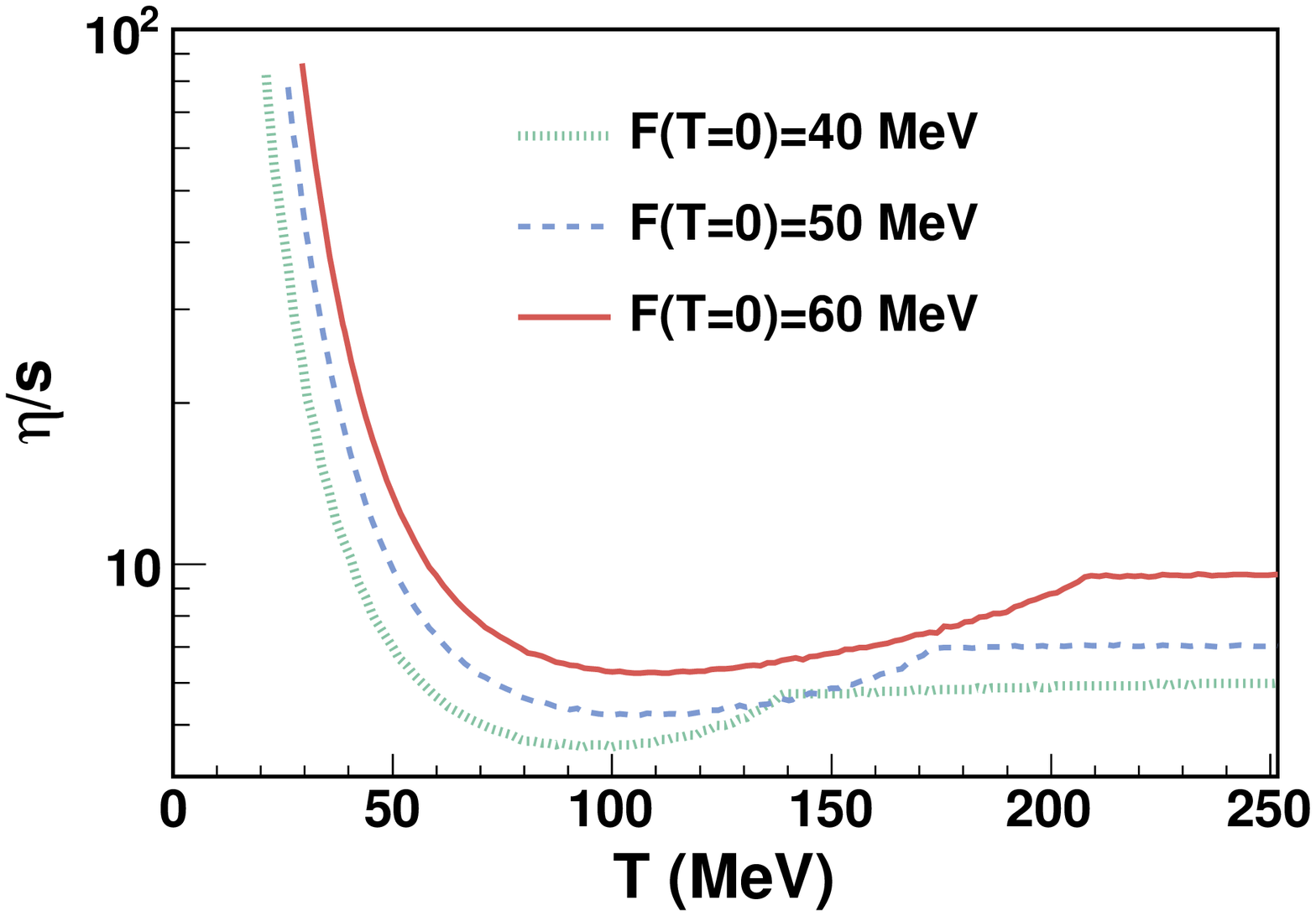}
\caption{\label{fig:ksskey} Comparison between the minimum of the viscosity over entropy density and the position of the critical temperature. The minimum is located close to it but slightly below.}
\end{center}
\end{figure}

To check whether the minimum of $\eta/s$ corresponds to the location
of the critical temperature one must compare the previous plot with
the order parameter. We show this in Fig.~\ref{fig:ksskey} at
different values of $F(T=0)$. In the left panel we show the
normalized value of $v(T)$. The position of the critical temperature
(where the order parameter first vanishes) depends linearly on $F$. The
same behavior is followed by the minimum viscosity over entropy
density shown in the right panel. The position of this minimum is
close to $T_c$, but not exactly there (as shown in
Fig.~\ref{fig:ksskey}) but slightly below. Notice that $T_c$ is
$M_R$-independent in the large-$N$ approximation considered here.

\section{Bulk viscosity over entropy density \label{sec:bulkvisco}}

In this section we will perform the calculation of the bulk viscosity in the L$\sigma$M in the large-$N$ approximation.
We start by considering both elastic and inelastic scatterings. Therefore, we do not introduce a pion chemical potential
in the calculation, as the pion number is not conserved in principle.
This fact simplifies the thermodynamics with respect to our previous work~\cite{Dobado:2011qu} since, instead of using the isochoric speed of
sound $v_n$ and the compressibility $\kappa^{-1}_{\epsilon}$, one only needs to consider the adiabatic speed of sound
\be v_S^2 = \left( \frac{\pa P}{\pa \epsilon} \right)_{s/n} \ , \ee
where $P$ is the pressure and $\epsilon$ the energy density.

Additionally, as we will use a quasiparticle description of the scalar fields, we must introduce a nonvanishing term $dm_{\pi}/dT$ that enters in the left-hand side of the BUU equation:
 \be p_{\mu} \pa^{\mu} n_p (x)|_{\zeta}= \beta n_p (1+n_p) \left( \frac{p^2}{3} - E_p^2 v_S^2 + T m_{\pi} \frac{dm_{\pi}}{dT} v_S^2 \right) \mb{\nabla} \cdot \mb{V} \ . \ee

Here it is useful to define a new $T$-dependent parameter
$\tilde{m}$ that includes the derivative of the thermal mass:
 \be \tilde{m}^2 \equiv m_{\pi}^2 - T^2 \frac{dm_{\pi}^2}{dT^2} \ . \ee
Then the left-hand side  reads
 \be \label{eq:lhsbuu} p_{\mu} \pa^{\mu} n_p (x)|_{\zeta}= \beta n_p (1+n_p) \left[ \frac{p^2}{3} - v_S^2 (p^2 + \tilde{m}^2) \right] \mb{\nabla} \cdot \mb{V} \ .
 \ee

The first-order Chapman-Enskog correction to the distribution function $f_p = n_p + f_p^{(1)}$ is
 \be f_p^{(1)} = - n_p (1+n_p) \beta \ A(p) \ \mb{\nabla} \cdot \mb{V}  \ . \ee
 The linearized BUU equation reads
 \begin{eqnarray} n_p(1+n_p) \left[ \frac{p^2}{3} - v_S^2 (p^2 + \tilde{m}^2) \right]  = \mathcal{C}_{el} + \mathcal{C}_{in} \nonumber \\
 \label{eq:buuforlsm} = \frac{N E_p}{2} \int d\Gamma_{12,3p} (1+n_1) (1+n_2) n_3 n_p [ A(p) + A(k_3) - A(k_1) - A(k_2) ] + \mathcal{C}_{in}\ ,
  \end{eqnarray}
 where we represent by $\mathcal{C}_{el}$ the elastic collision operator and by $\mathcal{C}_{in}$ the part of the collision operator including inelastic scattering (that we do not explicitly detail here).
As we have not fixed the pion chemical potential, this term should be
present in order to allow particle number changing processes.

 The stress-energy tensor  $\tau^{\mu \nu}$ in  presence of a thermal mass is written as~\cite{Jeon:1994if,Chakraborty:2010fr,Dusling:2011fd}:
 \be \tau^{\mu \nu} = N \int \frac{d^3p}{(2\pi)^3 E_p} \left( p^{\mu} p^{\nu} - u^{\mu} u^{\nu} T^2 \frac{dm^2}{dT^2} \right) f^{(1)}_p \ , \ee
 that reduces to the usual stress-energy tensor when the mass is $T$ independent.

This new term does not modify the expression for the bulk viscosity in terms of $f^{(1)}_p$ because the former only depends on the spatial components
 of the stress-energy tensor and in the local rest frame one has $u^i=0$. Therefore we have:
 \be \zeta = \frac{N}{T} \int \frac{d^3p}{(2\pi)^3 E_p} n_p (1+n_p) A(p) \frac{p^2}{3} . \ee
 However, this new term in the stress-energy tensor does change the form of the Landau-Lifschitz condition~\cite{lifshitz1981physical} $\tau^{00}=0$ (notice that this is the only condition,
  since the one fixing the particle density number out of equilibrium does not apply here).
 \be \tau^{00}=N \int \frac{d^3p}{(2\pi)^3 E_p} \left( p^2 + \tilde{m}^2 \right) f^{(1)}_p =0 \ , \ee
 As usual, this condition can be used to add a vanishing contribution to the bulk viscosity, making  much easier the comparison with the BUU equation left-hand side~(\ref{eq:lhsbuu}),
 \be \zeta = \frac{N}{T} \int \frac{d^3p}{(2 \pi)^3 E_p} n_p (1+n_p) A(p) \left[ \frac{p^2}{3} - v_S^2 \left( p^2 + \tilde{m}^2\right) \right]  \ . \ee

Next, we consider again  the same adimensional variables we have used for the shear viscosity ($x=E_p/m_{\pi}$, $y=m_{\pi}/T$) and also
an integration measure $d\mu_{\zeta}$. This
integration measure is described in detail  in App.~\ref{app:moments1} including its corresponding  scalar product, the norm, the moments, the functions $I^i$
and the polynomial basis. The bulk viscosity is then expressed as
the following scalar product:
 \be \zeta = \frac{Nm_{\pi}^4}{2 \pi^2 T} \langle A(x)| P_2 (x) \rangle_{\zeta} . \ee

The projected BUU equation is obtained by multiplying both sides of Eq.~(\ref{eq:buuforlsm}) by  $\frac{1}{4\pi m_{\pi}^4 E_p} P_l(x) d^3p$ and integrating over the three-momentum:
\be
\langle P_l(x) | P_2(x) \rangle  = \frac{N}{8\pi m_{\pi}^4} \int d^3p \int d\Gamma_{12,3p} (1+n_1)(1+n_2) n_3 n_p \ P_l(x) [ A(p) + A(p_3) - A(p_1 ) - A(p_2) ] + \langle \mathcal{C}_{in} \rangle\ .
\ee

There is one important remark concerning the solution of the linearized BUU. In this linearized equation only one zero mode is present.
 When $A(p)$ is proportional to $x$ the right-hand side is zero due to the energy-conservation law.
However, due to the presence of the inelastic collision operator, the zero mode associated to the particle conservation ($A(p) \propto 1$) is absent.

In the left-hand side it is easy to check that whereas $\langle P_1 | P_2 \rangle = 0$ we have $\langle P_0 | P_2 \rangle \neq 0$, and therefore consistent with the previous remark. The BUU equation
is solvable in the entire Hilbert space of solutions~\footnote{However, the component of $A(p)$ parallel to $x$ is not fixed by the BUU equation. If needed, it can be fixed by the Landau-Lifschitz condition.}.

Due to arguments of final phase-space and suppression in the large-$N$ limit (that will be detailed later), we will not consider the inelastic processes in the collision integral
 and retain only the $2\to2$ processes. This simplification causes an inconsistency in the BUU equation, as the projection of the BUU onto $P_0$ gives different results on both sides
  of the equation. This inconsistency is avoided by solving the equation in the  subspace perpendicular to the ``accidental'' zero mode $P_0=1$.

Therefore we expand the $A(x)$ solution:
 \be A(x) = \sum_{n=1}^\infty a_n P_n(x) \ . \ee
After symmetrization of the collision integral the BUU equation finally reads
\be
\delta_{l2} ||P_2 (x)||^2   = \sum_{n=1}^{\infty} a_n \mathcal{C}_{nl} \ ,
\ee
 with
\be \mathcal{C}_{nl} = \frac{N \pi^2}{4m_{\pi}^4} \int \prod_{i=1}^4 \frac{d^3k_i}{(2\pi)^3 2 E_i} \overline{|T|^2} (2\pi)^4 \delta^{(4)} (k_1+k_2-k_3-p)
 (1+n_1) (1+n_2) n_3 n_p  \ \Delta[P_n(x)] \Delta[P_l(x)] \ ,  \ee
where $\Delta[P_n(x)] \equiv P_n (x) + P_n(x_3) - P_n(x_1) - P_n(x_2)$.

For $l=1$ one gets the identity $0=0$ which does not determine the coefficient $a_1$.
The first nontrivial case corresponds to $l=2$, for which the solution of the BUU equation reads:
\be a_2 = \frac{||P_2(x)||^2}{\mathcal{C}_{22}} \ , \ee
By introducing it in the  formula for the bulk viscosity one finds:
\be \label{eq:finalbulk} \zeta = \frac{Nm_{\pi}^4}{2 \pi^2 T} a_2 ||P_2(x)||^2 = \frac{Nm_{\pi}^4}{2 \pi^2 T } \left( \frac{1}{3} - v_S^2 \right) \frac{1}{\mathcal{C}_{22}} \left( I_4 - \frac{I_2 I_3}{I_1} \right) \ . \ee

In the following we  proceed to the discussion of the results for the bulk viscosity over entropy density. They will depend on the value of the pion mass at zero temperature, i.e. on having  a second-order phase transition
or a crossover. We will begin with the latter.

\subsection{Crossover}

We start with the explicit symmetry-breaking case, for which we set
a nonzero pion mass at $T=0$. In order to get numerical results a
possible choice could be the physical pion mass $m_{\pi} (T=0)=138$
MeV. The effective mass depends on temperature in such a way that
$m_{\pi}(T) > m_{\pi}(T=0)=138$ MeV. For this reason, and
analogously to the physical pion gas in~\cite{Dobado:2011qu}, the
inelastic terms in the collision operator ($1 \to 3$ or $2 \to 4$) are
suppressed by the Boltzmann exponential factor in the final phase space $e^{-2
m_{\pi}/T}$ (the inverses being too improbable to occur if the gas
is dilute).

However, the crossover case does not have a precise definition of the ``critical'' temperature. Here we will define it as the point
where the minus derivative of the order parameter (the susceptibility) peaks. From the left panel of Fig.~\ref{fig:crossover}
we obtain a value of $T_{cr}=261$ MeV.

\begin{figure}[t]
\begin{center}
\includegraphics[scale=0.35]{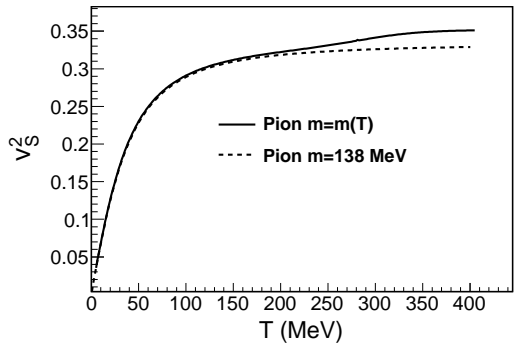}
\caption{\label{fig:speedofsoundcross} $v_S^2$ in the L$\sigma$M at large $N$ for the crossover case (solid line) compared with the speed of sound of a pion mass with a constant mass of 138 MeV.}
\end{center}
\end{figure}

In Fig.~\ref{fig:speedofsoundcross} we show the squared speed of sound for the pions together with the result from a pion gas with a constant (temperature-independent) mass of $m_{\pi}=138$ MeV. The difference between the two curves
is attributed to the effective mass of the quasiparticles. It is important to remark here  that both results correspond to the noninteracting gas, i.e. ideal gas. The introduction of interactions in the thermodynamic functions
(for example, through the free energy obtained from the effective potential) would be inconsistent with the conception of the Chapman-Enskog expansion at first order.

\begin{figure}[t]
\begin{center}
\includegraphics[scale=0.35]{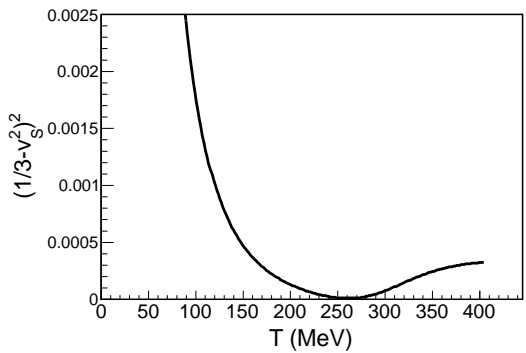}
\caption{\label{fig:conformality} $(1/3-v_S^2)^2$ as a measure of
the violation of the conformality of the system. The bulk viscosity
qualitatively follows this factor.}
\end{center}
\end{figure}

The speed of sound turns out to be a monotonic function of the
temperature. It takes the conformal value $v_S^2=1/3$ at $T=259$
MeV, very close to the crossover temperature. By
Eq.~(\ref{eq:speedsound}) one can check that at this precise point:

 \be\label{eq:peculiar} \frac{dm_{\pi}}{dT} = \frac{m_{\pi}}{T} \ . \ee
When $v_S^2=1/3$ the squared norm of the source $P_2$ identically
vanishes. Therefore, the bulk viscosity is zero because of
Eq.~(\ref{eq:finalbulk}). More generally, the bulk viscosity features
the factor $(\frac{1}{3} - v_S^2)^2$ which measures the violation of
conformality in the system~\cite{Chakraborty:2010fr}. Thus, when the
speed of sound takes values far from the conformal one,
the bulk viscosity will be non-negligible, whereas if $v_S^2$ is
near the conformal value~\footnote{The conformal value of the
squared speed of sound depends on the space-time dimensions $D$ and reads
\[ \left. v_S^2 \right|_{CFT} = \frac{1}{D-1} \ . \]}, the bulk viscosity will be close to
zero. We plot the factor $(\frac{1}{3} - v_S^2)^2$ in
Fig.~\ref{fig:conformality}.

\begin{figure}[t]
\begin{center}
\includegraphics[scale=0.35]{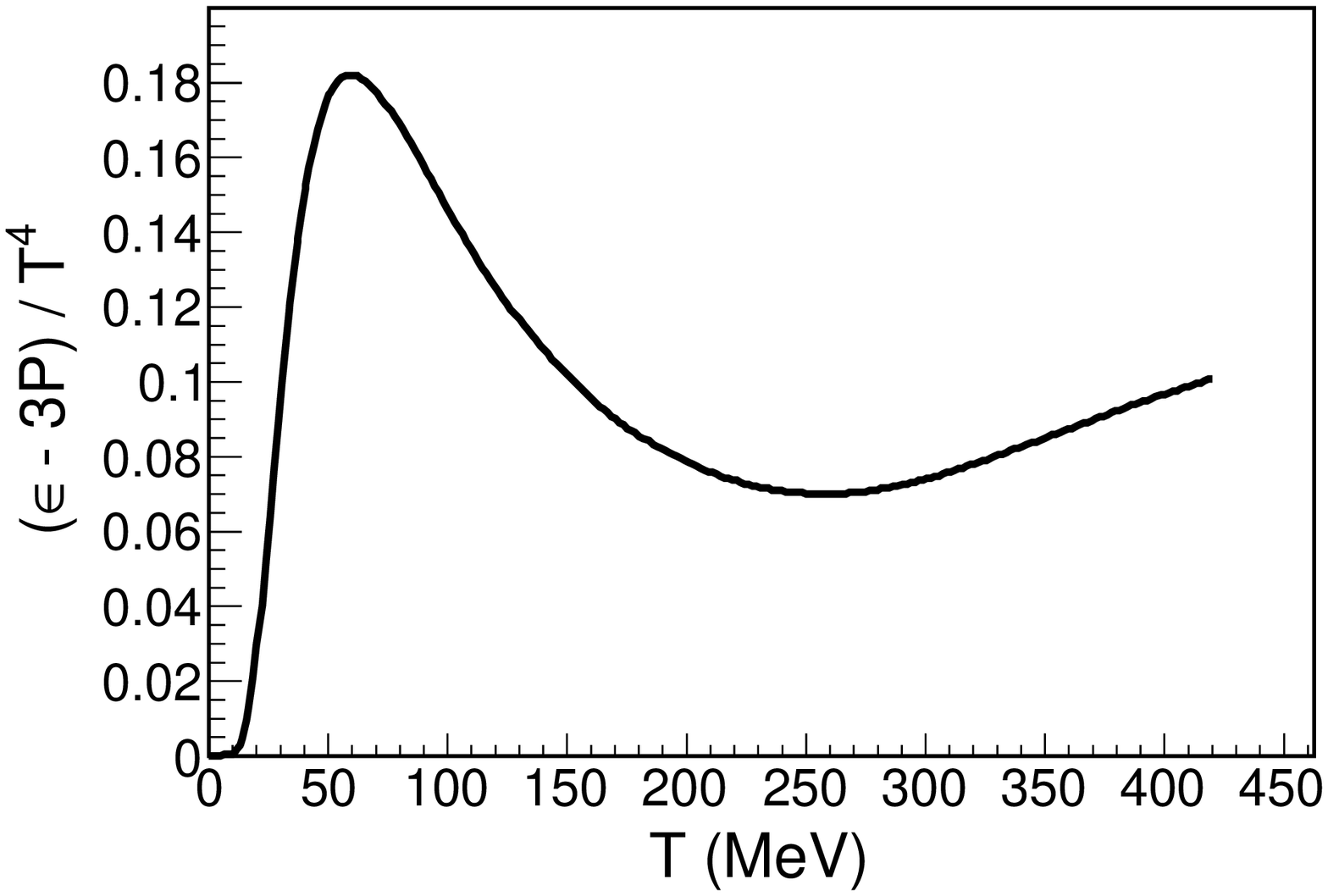}
\includegraphics[scale=0.35]{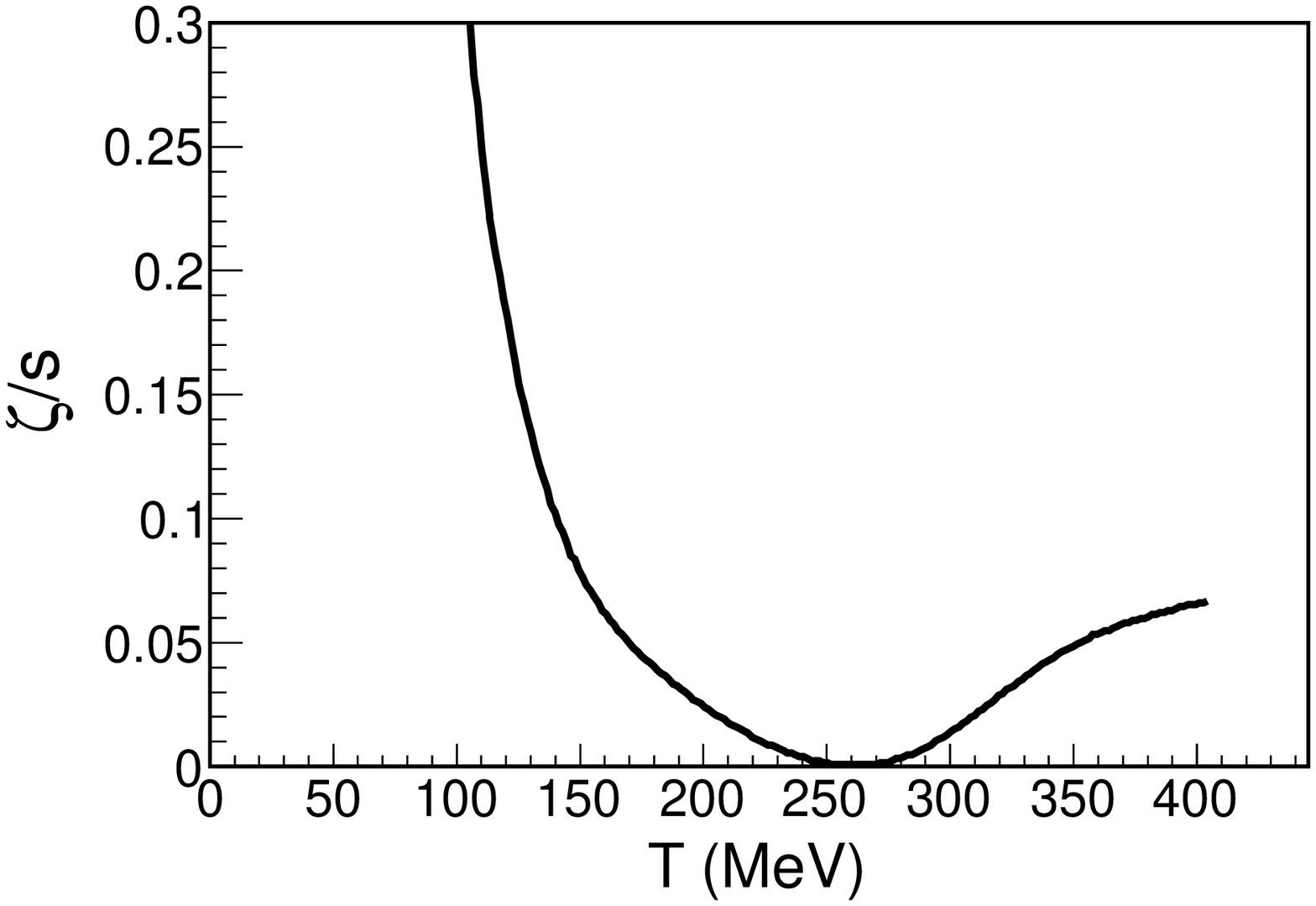}
\caption{\label{fig:lsmzetaoverscross} Left panel: Interaction measure for the crossover case as a signal of loss of conformality in the system. Right panel: $\zeta/s$ in the L$\sigma$M at large $N$ for the crossover case.}
\end{center}
\end{figure}

Another measure of the loss of conformality is the so-called interaction measure. In $D=3+1$ dimensions it is defined as $\langle \theta \rangle_T = \epsilon-3P$, where $\epsilon$ is the energy density of the gas and $P$ its pressure.
Using the relativistic equation of state $P+ \epsilon = Ts$ it can be written as $\langle \theta \rangle_T=Ts-4P$ as a function of the pressure and the entropy density. The interaction measure is shown in the left panel
of Fig.~\ref{fig:lsmzetaoverscross}. The result for the bulk viscosity over entropy density is plotted in the right panel of the same figure. As stated before, the qualitative shape of the $\zeta/s$ follows the behavior of $(1/3-v_S^2)^2$ shown in
Fig.~\ref{fig:conformality}.

A maximum in the bulk viscosity over entropy density is not seen at the crossover temperature. However, a minimum
 is found for  $v_S^2=1/3$. Later, we will discuss the apparent contradiction with the claim of a maximum of $\zeta/s$  appearing at the phase transition.

To end this section, we want to stress that this result is perfectly
consistent (with due respect to the inevitable differences) with the
result in~\cite{FernandezFraile:2010gu} for the Gross-Neveu model in
the large-$N$ limit. In that case, no peak is seen in $\zeta/s$.
More specifically, in the crossover case (which is the equivalent to
this one) the speed of sound is monotonic and structureless like in
our case. However, the speed of sound does not cross the conformal
value because if $D=1+1$ then $v_S|_{CFT}=1$ and this is the value
to which the speed of sound approaches asymptotically at large
temperatures~\footnote{If the speed of sound is a monotonically
increasing function with temperature, as the relativistic limit is
$v^2_S=1$, this value can only be reached at $T \rightarrow
\infty$.}. Following our considerations, the bulk viscosity in that
model should be a monotonic decreasing function on the temperature,
going to zero at asymptotically higher temperatures (where the
conformal value of the speed of sound is reached at $T \rightarrow
\infty$).

In order to give a consistency check of the previous result we will
repeat the calculation in the chiral limit (at zero temperature) for pions.

\subsection{Second-order phase transition}

If the pion mass vanishes at $T=0$, then a second-order phase transition is expected. In this case we can neglect the inelastic processes
by a different argument. In the broken phase, as the pion mass vanishes, there is no Boltzmann suppression in the final phase space. However, in the large-$N$ limit the tree-level amplitude for inelastic processes ($2 \rightarrow 4$) carries
one extra $1/N$ factor. The average cross section (or the average scattering amplitude inside the collision operator) is order $1/N$ for elastic scattering, but it is order $1/N^3$ for the inelastic case and therefore it is suppressed in the large-$N$ limit.

\begin{figure}[t]
\begin{center}
\includegraphics[scale=0.35]{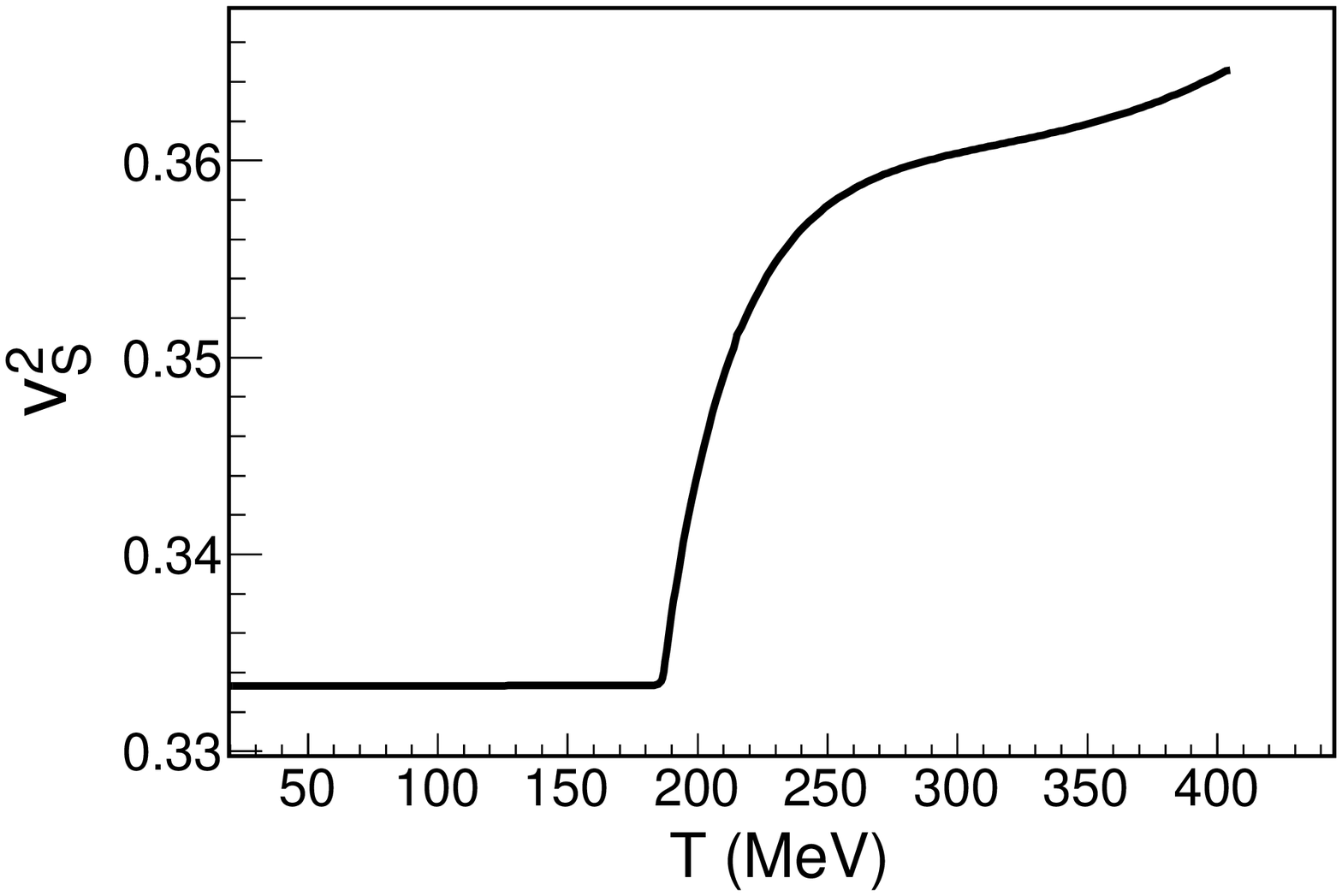}
\includegraphics[scale=0.35]{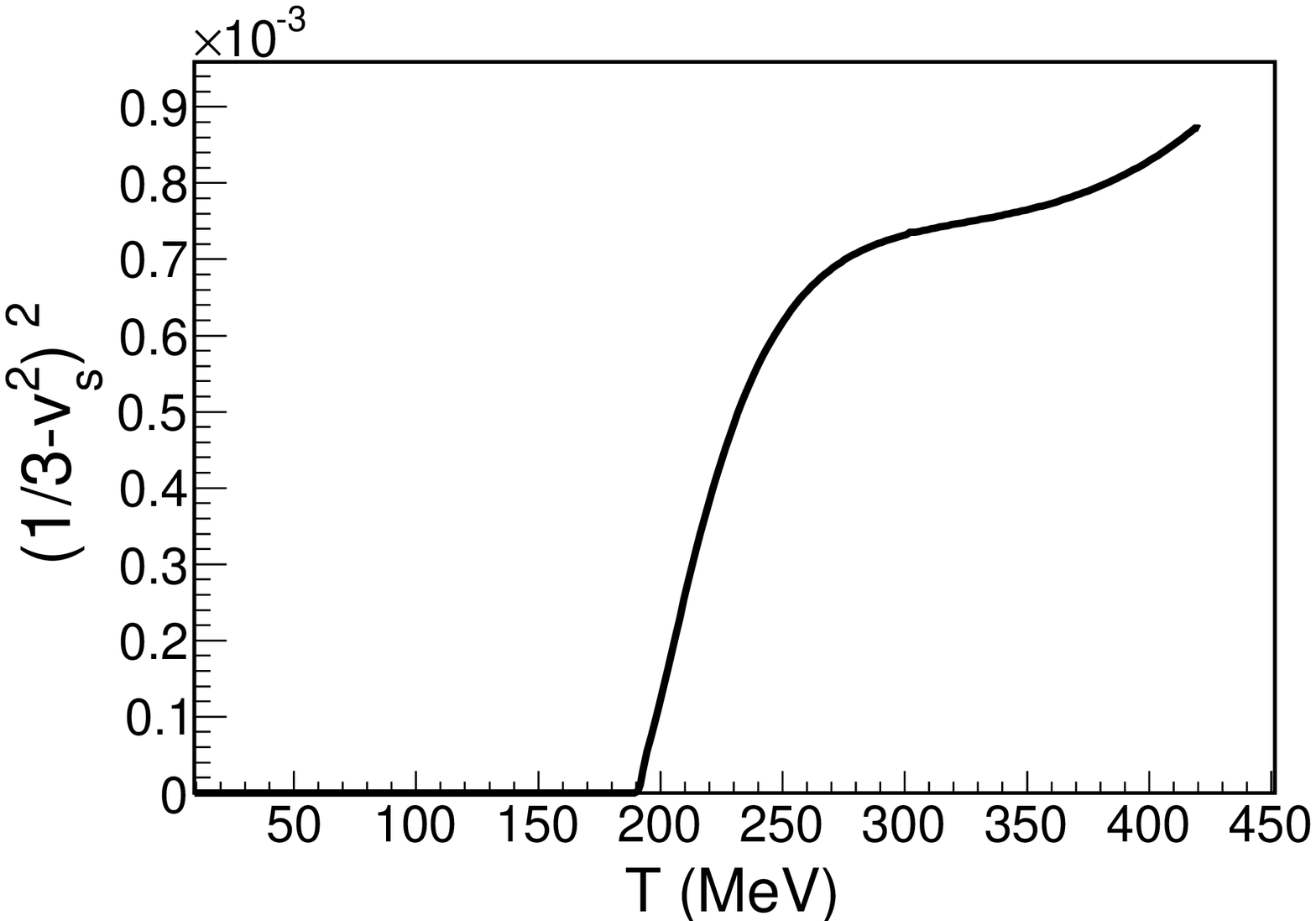}
\caption{\label{fig:speedofsoundsec} Left panel: $v_S^2$ in the L$\sigma$M at large $N$ for the second-order phase transition case. Right panel: Factor $(1/3- v_S^2)^2$ that measures the violation of conformality.}
\end{center}
\end{figure}

Before looking at $\zeta/s$ we show the speed of sound in Fig.~\ref{fig:speedofsoundsec}. In the broken phase the pions are massless and the speed of sound takes the value of a conformal gas $v_S^2=1/3$
(note that the Higgs does not contribute because its effect is suppressed in the large-$N$ limit). When the critical temperature $T_c=186$ MeV is reached, the mass of pions starts to grow and the speed of sound separates
from the conformal value. In the right panel we show the factor of loss of conformality $(1/3 - v_S^2)^2$ which gives an idea of the behavior of the bulk viscosity. Of course, we expect the bulk viscosity to vanish in the
broken phase and to grow for $T>T_c$.

\begin{figure}[t]
\begin{center}
\includegraphics[scale=0.35]{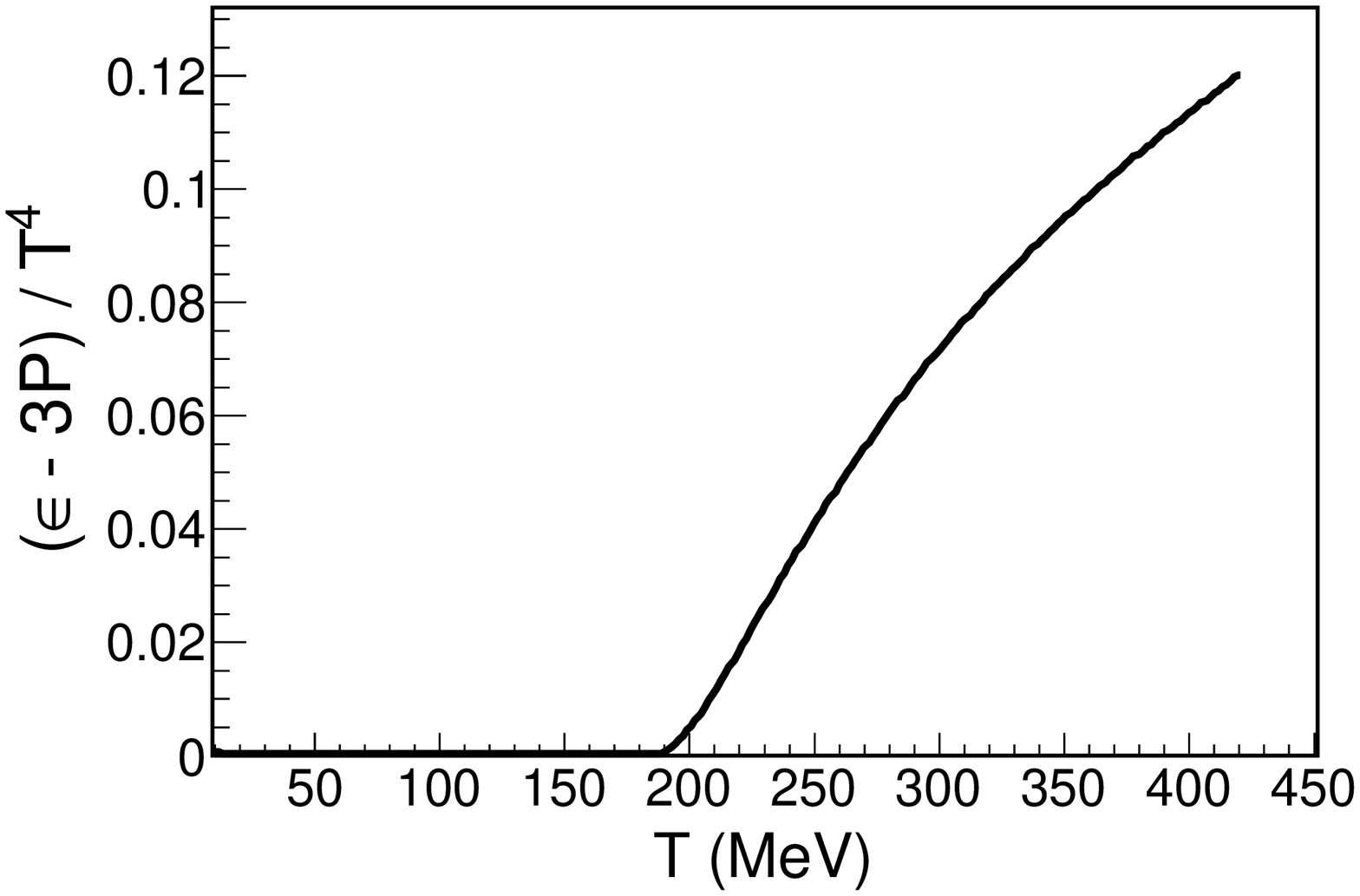}
\includegraphics[scale=0.35]{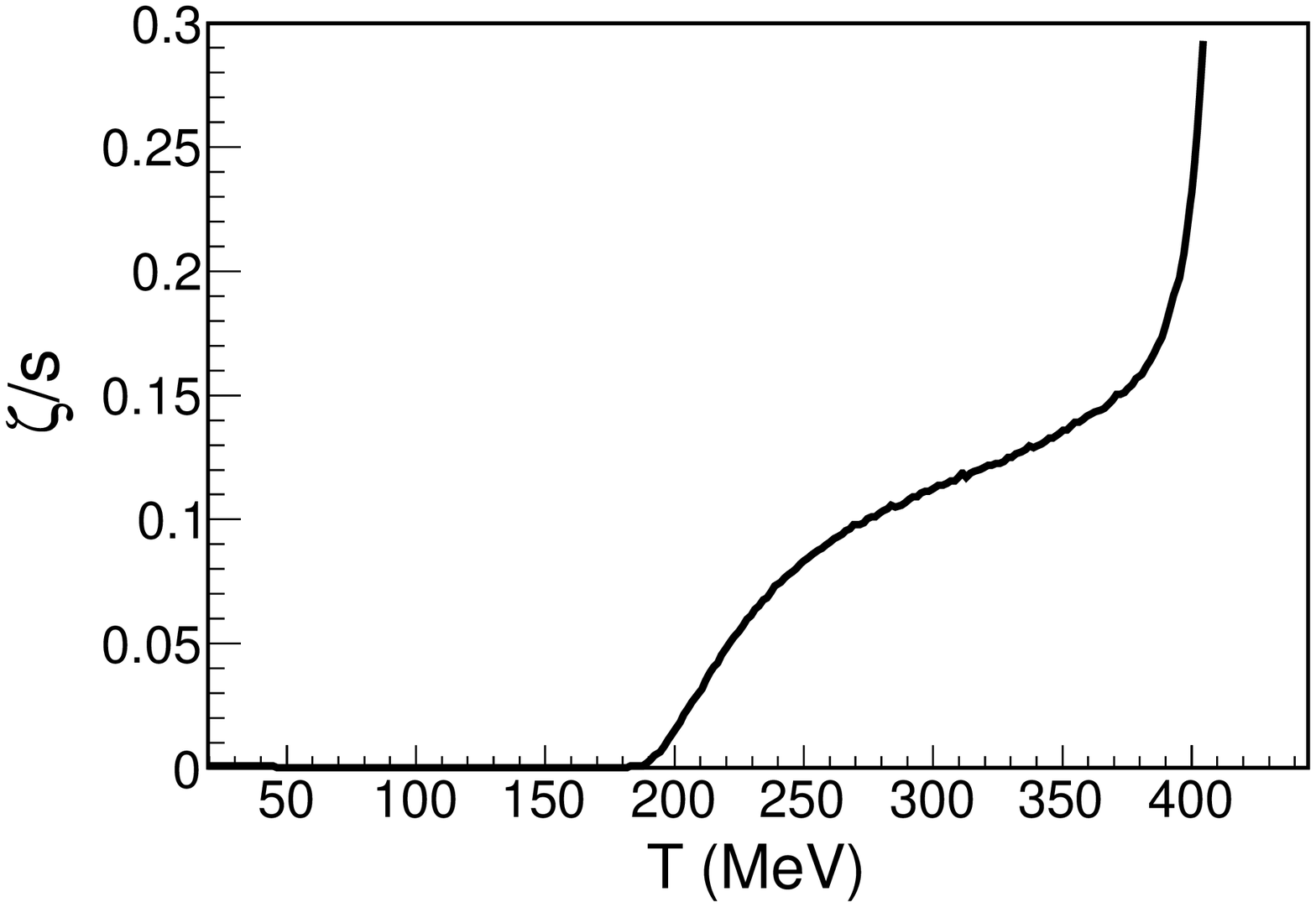}
\caption{\label{fig:lsmzetaoverssec} Left panel: Interaction measure as a signature of loss of conformality of the system. Right panel: $\zeta/s$ in the L$\sigma$M at large $N$ for the second-order phase transition.}
\end{center}
\end{figure}

In the left panel of Fig.~\ref{fig:lsmzetaoverssec} the
interaction measure is plotted. In the broken phase it also vanishes as the gas
is conformal. The $\zeta/s$ coefficient is also shown in
Fig.~\ref{fig:lsmzetaoverssec}. The bulk viscosity is
zero in the broken phase but it starts to grow and takes a monotonic
behavior when temperature is increased.

Note that again this result is consistent with the Gross-Neveu model at large $N$ in the chiral limit of fermion mass~\cite{FernandezFraile:2010gu}. In the high temperature phase
the mass of the fermion field is exactly zero, and the speed of sound
takes the conformal value of $v_S^2=1$. Therefore, the bulk viscosity turns out to be zero. No maximum is seen at the phase transition temperature. When decreasing the temperature (the fermion
thermal mass increases), the bulk viscosity has a finite discontinuity and increases as $T \rightarrow 0$. This is exactly what happens in our case, but with the high and low-temperature limits reversed.

\begin{figure}[t]
\begin{center}
\includegraphics[scale=0.35]{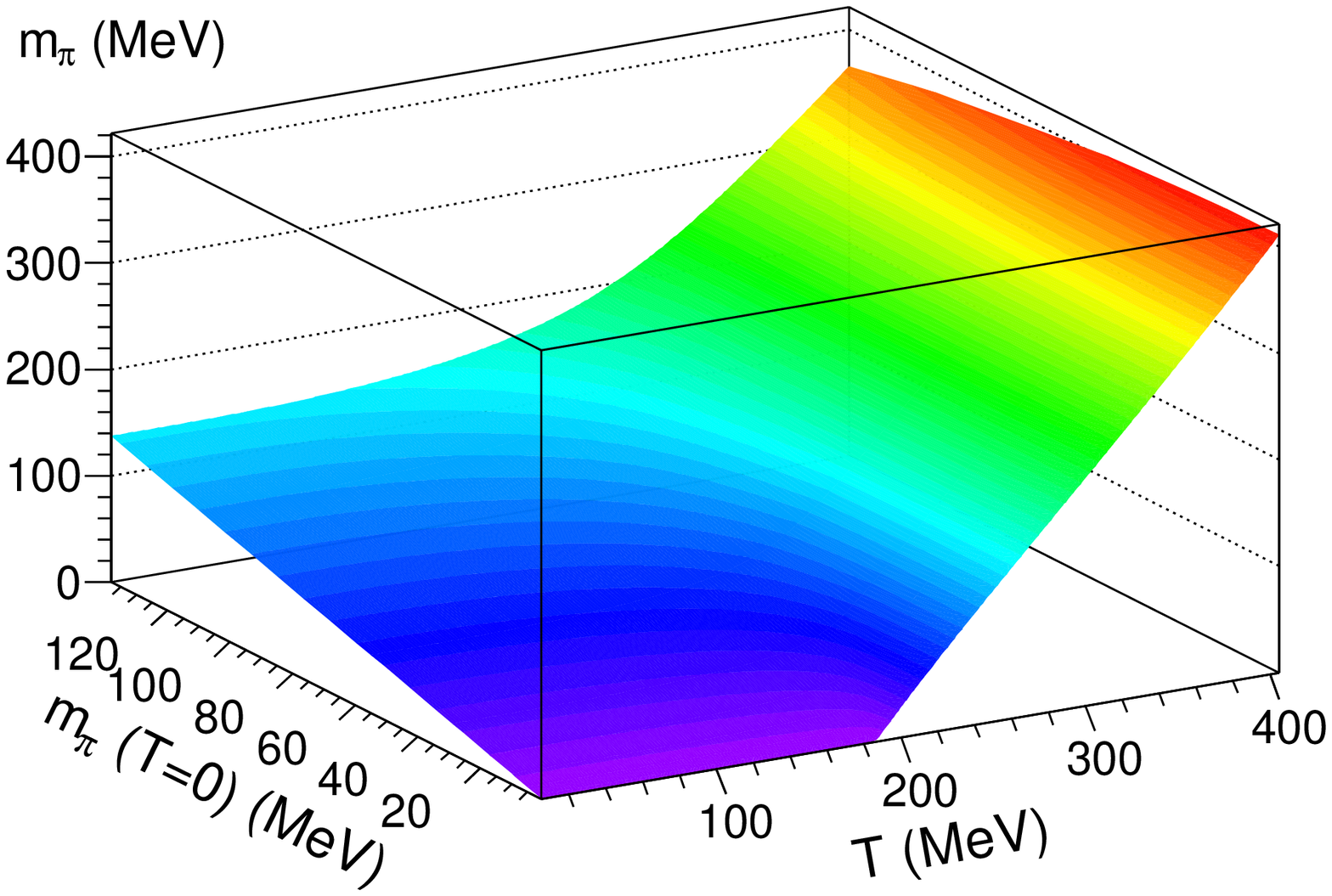}
\includegraphics[scale=0.35]{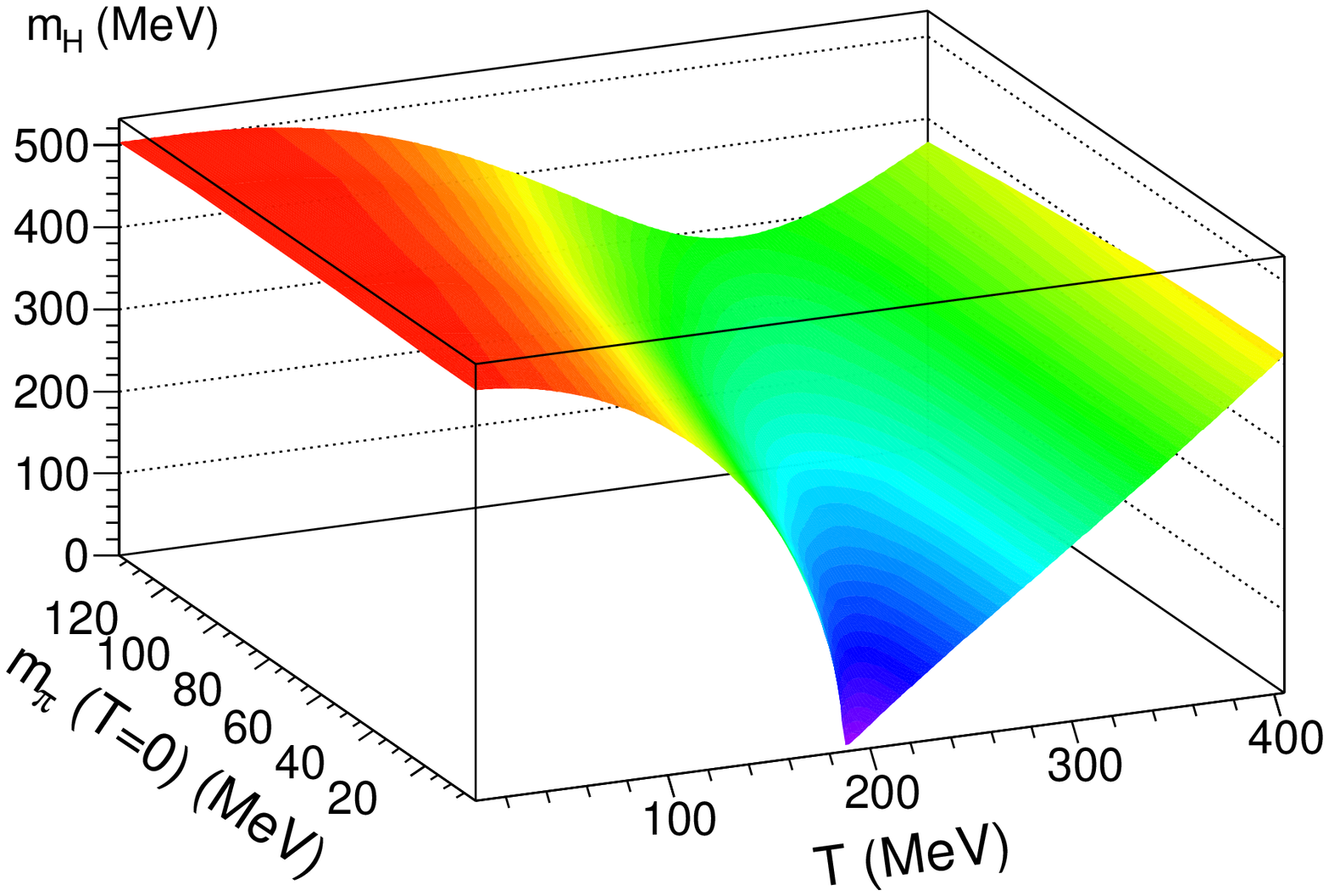}
\includegraphics[scale=0.35]{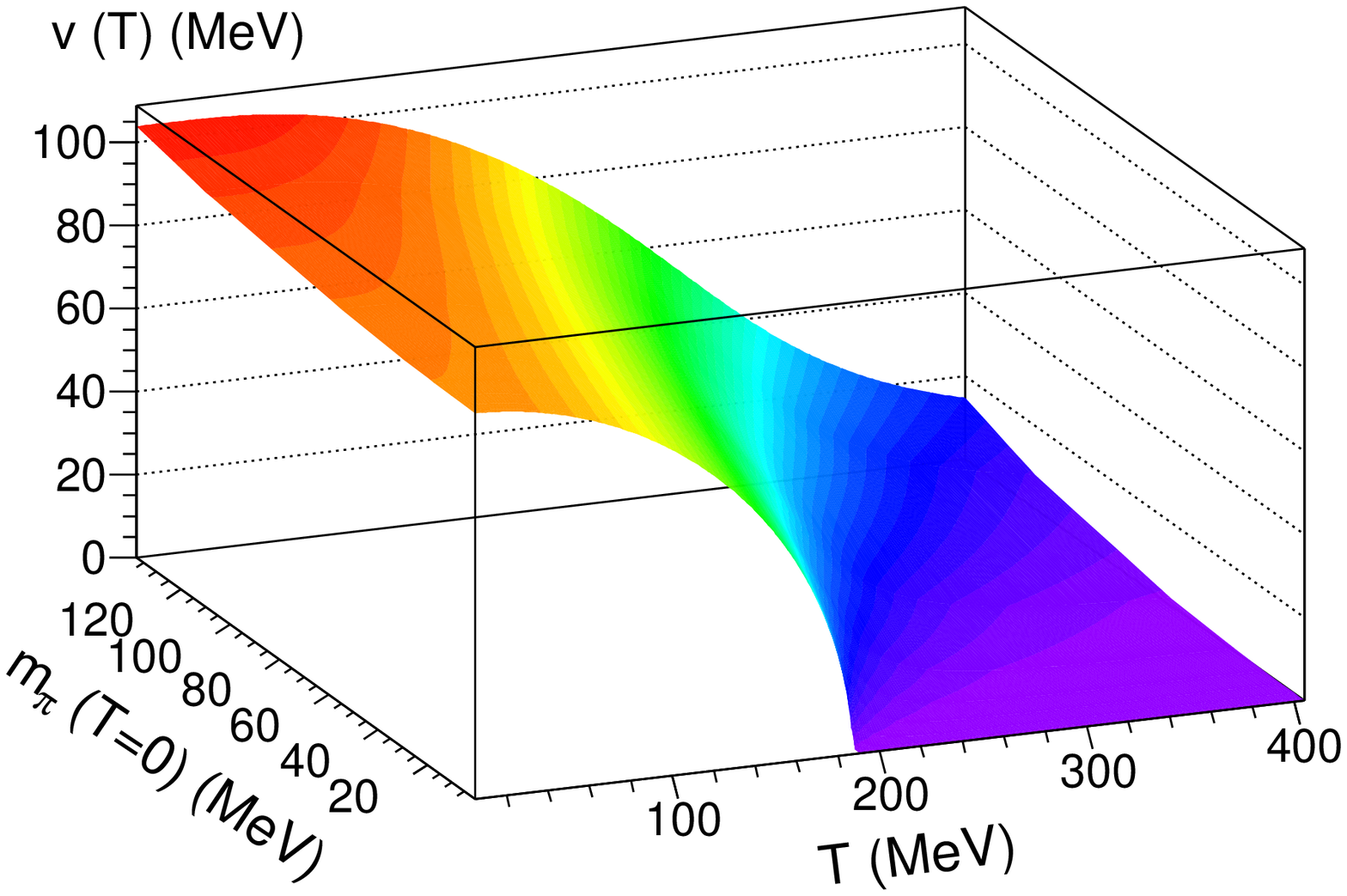}
\caption{\label{fig:3dmasses} Top panels: Pion and Higgs thermal masses. Bottom panel: Order parameter. All of them are plotted as a function of the temperature and the pion mass at $T=0$.}
\end{center}
\end{figure}

\begin{figure}[t]
\begin{center}
\includegraphics[scale=0.35]{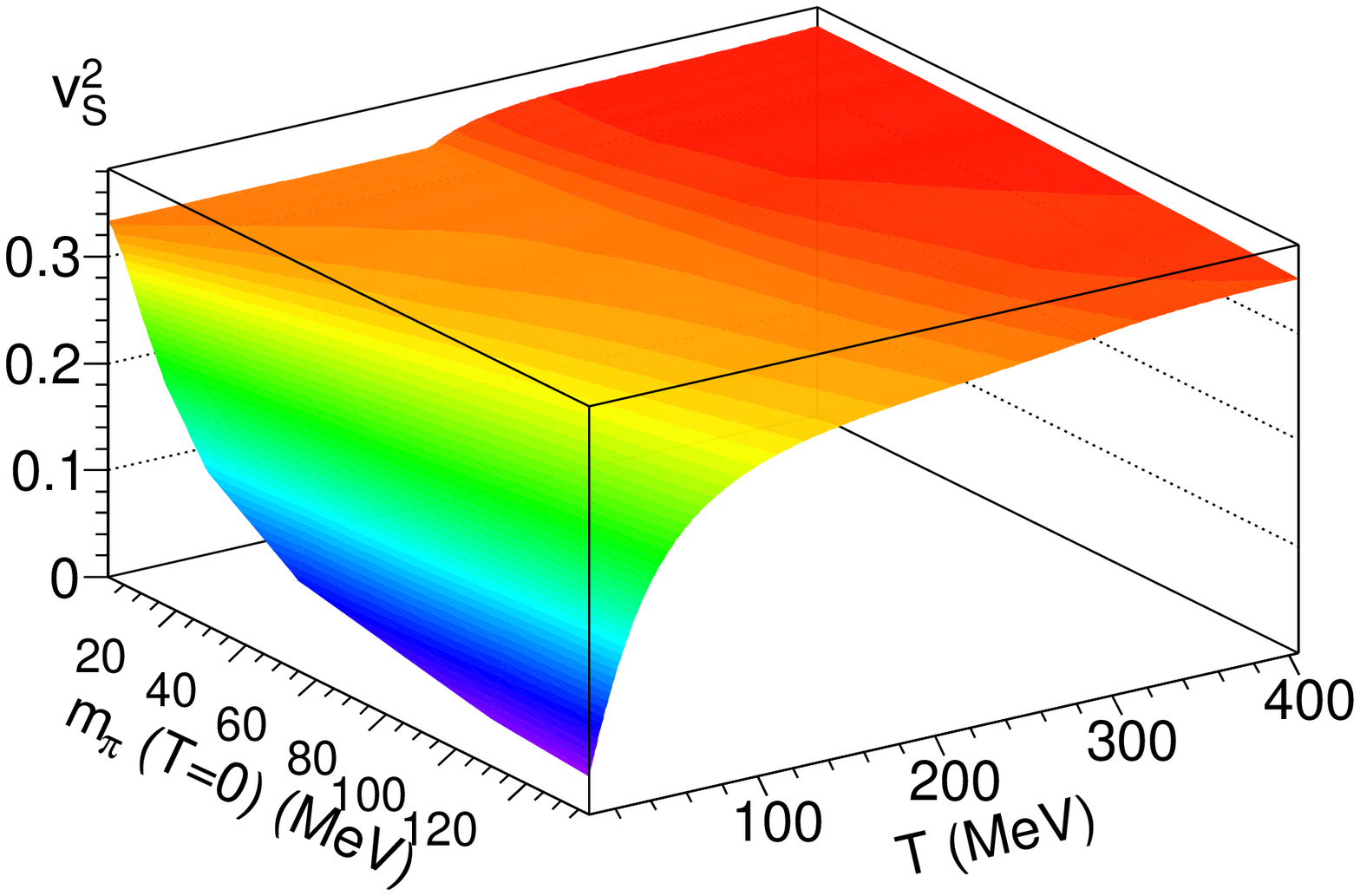}
\includegraphics[scale=0.35]{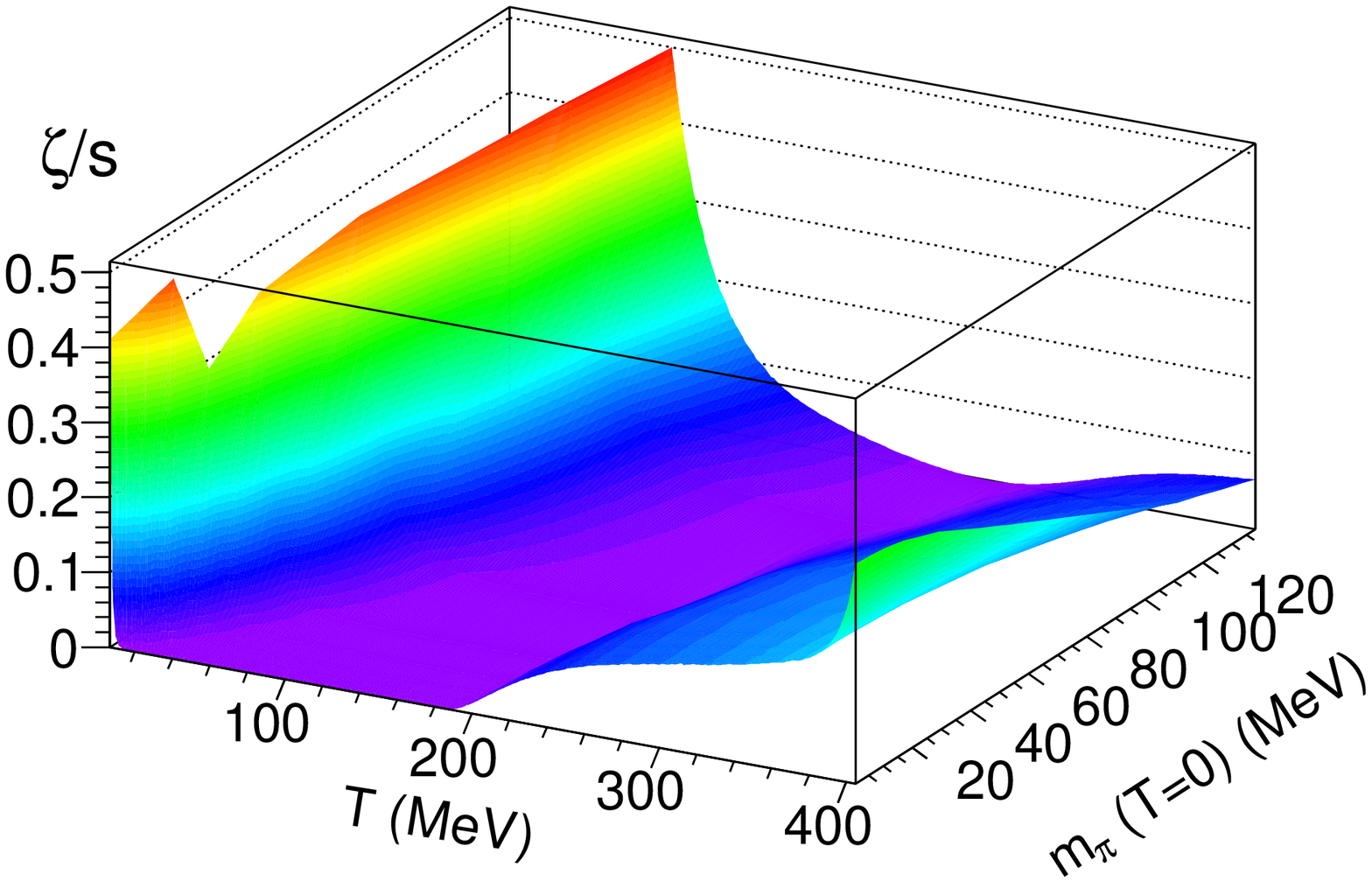}
\caption{\label{fig:3dbulk} Left panel: Speed of sound as a function of temperature and pion mass at zero temperature. Right panel: Bulk viscosity over entropy density. Note the nonstandard direction of the axes for the sake of clarity.}
\end{center}
\end{figure}

Finally, we show how the second-order case connects with the crossover by slowly increasing the pion mass from 0.5 MeV (our chiral value) up to 138 MeV. In Fig.~\ref{fig:3dmasses} we plot the evolution of the pion and Higgs masses
and the order parameter $v(T)$ as a function of the temperature and the pion mass at $T=0$. In Fig.~\ref{fig:3dbulk} we show the squared speed of sound and the bulk viscosity.

\subsection{Discussion of the results}

 In  both cases, the second-order phase transition and the crossover, we have
   not obtained a maximum for the bulk viscosity at the phase transition. We
    have extensively studied the reasons for this fact. One of the keys to
understand this result is the behavior of the speed of sound (or
more generally, the equation of state). Conformality, or the lack of
it,  should be reflected in this factor either by taking the conformal value
$v_S^2=1/3$ or going far from it. One can conclude that, in order to
see a maximum of the bulk viscosity at the critical temperature,
one should have a nonmonotonic behavior of $v_S^2$ near $T_c$. For
instance, the bulk viscosity could approach the conformal value as
$T \rightarrow T_{c}-$,  showing a sudden dip at $T_c$ and increasing again
to the conformal value $1/3$ for  $T > T_c$. This behavior would produce a maximum in the
bulk viscosity. Actually, this is the scenario when the bulk
viscosity is phenomenogically included in the QCD phase transition
(see for instance~\cite{Denicol:2009am,Chakraborty:2010fr,Li:2009by}) and from the
lattice QCD equation of state~\cite{Bazavov:2009zn,Borsanyi:2010cj,Laine:2006cp}.

\begin{figure}[t]
\begin{center}
\includegraphics[scale=0.35]{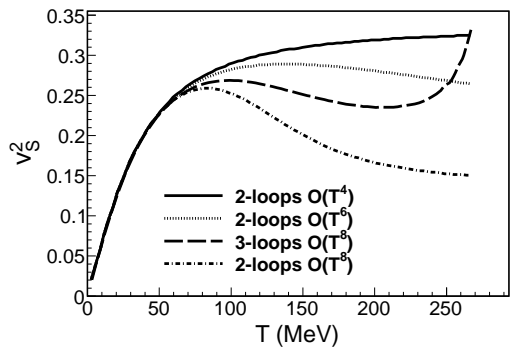}
\caption{\label{fig:cspion} Squared speed of sound of pions with physical mass of 138 MeV. We use thermal chiral perturbation theory~\cite{Gerber:1988tt} at different approximations.}
\end{center}
\end{figure}

   This behavior is also consistent with the second peak in the bulk viscosity showed in~\cite{FernandezFraile:2008vu}. The nature of this peak can easily be understood from the speed-of-sound curve. In that work a dilute gas of pions
with a physical mass of 138 MeV is considered. The speed of sound is obtained
from finite temperature chiral perturbation theory at two loops and
order $T^8$.  In
Fig.~\ref{fig:cspion}   we plot   (dashed line)  the results obtained by direct use of the formulas in~\cite{Gerber:1988tt} for  the pressure of a pion gas. The local
minimum of $v_S^2$ is responsible for the breaking of conformality
around $T\sim 220$ MeV and therefore a maximum in the bulk
viscosity is found in~\cite{FernandezFraile:2008vu} at that temperature.

Focusing again in the L$\sigma$M in the large-$N$ limit, the monotonic
behavior of the speed of sound is inherited by the behavior of the thermal
mass of the pion. The mild increase of the pion mass, without
any indication of phase transition, determines the dependence on temperature of the speed of
sound and eventually the bulk viscosity. This behavior is a
consequence of the large-$N$ approximation, in which some of the
interesting details of the model are washed out. In other words, the simplification of the large-$N$ limit in
some aspects of the model is also accompanied by an
oversimplification of the dynamics that eventually gives rise to the
absence of a maximum in the bulk viscosity. In  fact, as we have
already mentioned, from the dynamical renormalization group is
not expected to have a divergence of the bulk viscosity in the large-$N$
 limit of this model~\cite{Nakano:2011re}. This limit was also
considered for the Gross-Neveu model in~\cite{FernandezFraile:2010gu} and the results found there are
perfectly compatible with ours.

The use of a more complicated dispersion relation or a different
thermal effective mass behavior can notably change this result. A
hint for this fact can be found in~\cite{Chakraborty:2010fr}. In
that reference the authors consider also the L$\sigma$M model but in
a different way. First, they do not take the large-$N$ limit  and
therefore the interactions among pions and Higgs must be included.
However, in order to avoid some divergences in the cross section,
they only consider constant scattering
amplitudes (in fact they do not take into account any correction
to the amplitudes due to the finite pion mass). Second, they do not
solve the BUU equation by the Chapman-Enskog expansion. Instead, they use the
relaxation time approximation to obtain the bulk viscosity. Finally, the effective potential is
calculated in the CJT approximation
which gives qualitative different results on the behavior of the
effective masses and eventually on the speed of sound and bulk
viscosity. In that reference, a peak in the bulk viscosity is found
in the crossover case for some specific values of the Higgs masses.

Therefore, in order to conclude  our study of the bulk viscosity in
the L$\sigma$M, it seems to be interesting to consider a different effective mass
to show that a maximum in the bulk viscosity is obtained without
changing any other aspect  of our calculation. Thus we will use the Hartree
approximation of the CJT theory as presented in~\cite{Petropoulos:2004bt}
for the calculation of the order parameter and thermal masses.
Doing so is not completely consistent  since we are using the
Hartree approximation
 for the effective potential and keeping the large-$N$ limit in the
  scattering amplitudes. However, in the next and last  section we will not try to develop a perfect consistent
computation but just to introduce a different pion effective mass in our previous
calculation to check  if a maximum is obtained for  the bulk viscosity.
Additionally, we have also checked that the use of the scattering
amplitudes in the large-$N$ limit or the use of constant scattering
amplitudes without any further correction give rise to similar
results.

\section{The Cornwall-Jackiw-Tomboulis formalism in the Hartree approximation \label{sec:bulkcjt}}

In this section we introduce a thermal pion mass from the effective potential obtained within the CJT formalism for the L$\sigma$M in the Hartree approximation. Here we will
refer to~\cite{Petropoulos:2004bt} where this approach is nicely presented.
In a nutshell, the CJT method is a tool for the computation of an effective action, not only for the VEV of the field $\phi(x)=\langle \Phi(x) \rangle$ but also for  the two-point function $G(x,y) = \langle T \Phi(x) \Phi(y)\rangle$.
This generalized effective action can be understood as the generating functional of the two-particle irreducible graphs (2PI). Some of the two-particle irreducible graphs that contribute to the CJT effective potential are shown in Fig.~\ref{fig:2pi}.

In~\cite{Petropoulos:2004bt} two different approaches were used to sum up certain sets of diagrams, namely the large-$N$ limit and the Hartree approximation. Of course the large-$N$ approximation gives very close results to ours and it gives no new information about the bulk viscosity.
Therefore, we will consider the Hartree approximation in which only the ``double bubble'' diagram is taken into account, and which is equivalent to summing up all the ``daisy'' and ``superdaisy'' diagrams in the 1PI effective potential. The essential features of the method are sketched in the Appendix~\ref{app:cjt}.

The effective potential is a function of the order parameter $\phi$ and two dressed propagators $G_{\sigma}$ and $G_{\pi}$, one for the Higgs and one for the pion, respectively. From  the effective potential $V$, one can obtain the two gap equations
\be \frac{dV}{dG_{\pi}} =0 ; \quad \frac{dV}{dG_{\sigma}}=0 \ , \ee
and also minimize the effective potential with respect to the order parameter. If the dressed propagators are written as functions of some effective masses $G^{-1}_i = k^2 + M_i^2$, then the three equations give the following nonlinear system for $\phi$, $M_{\sigma}$ and $M_{\pi}$:
\be
\begin{array}{lcr}
 M^2_{\sigma} & = & -2 \overline{\mu}^2 + \frac{12 \lambda}{N} \phi^2 +  4 \lambda F(M_{\pi}) +\frac{12\lambda}{N} F(M_{\sigma})  \ , \\
M^2_{\pi} & =& -2 \overline{\mu}^2 + \frac{4\lambda}{N} \phi^2 + \frac{4 (N+2)\lambda}{N} F(M_{\pi})  + \frac{4 \lambda}{N} F(M_{\sigma}) \ ,  \\
0 & = & \left[ -2 \overline{\mu}^2 + \frac{4\lambda}{N} \phi^2 + \frac{12\lambda}{N}  F(M_{\sigma}) + 4 \lambda F(M_{\pi})  \right] \phi - \epsilon \ , \label{eq:system}
\end{array}
\ee
where $F(M)$ is defined in Eq.~(\ref{eq:integralf}). This system can be  solved numerically in order to obtain the effective masses and the order parameter that enter into our bulk viscosity calculation.

Of course, a renormalization program should be properly performed
here. The technical aspects of the renormalization of the model can
be found in~\cite{Petropoulos:2004bt} and in the references therein.
However, in this work we will deal only with the temperature-dependent part of the effective potential.

\subsection{Crossover case}

\begin{figure}[t]
\begin{center}
\includegraphics[scale=0.35]{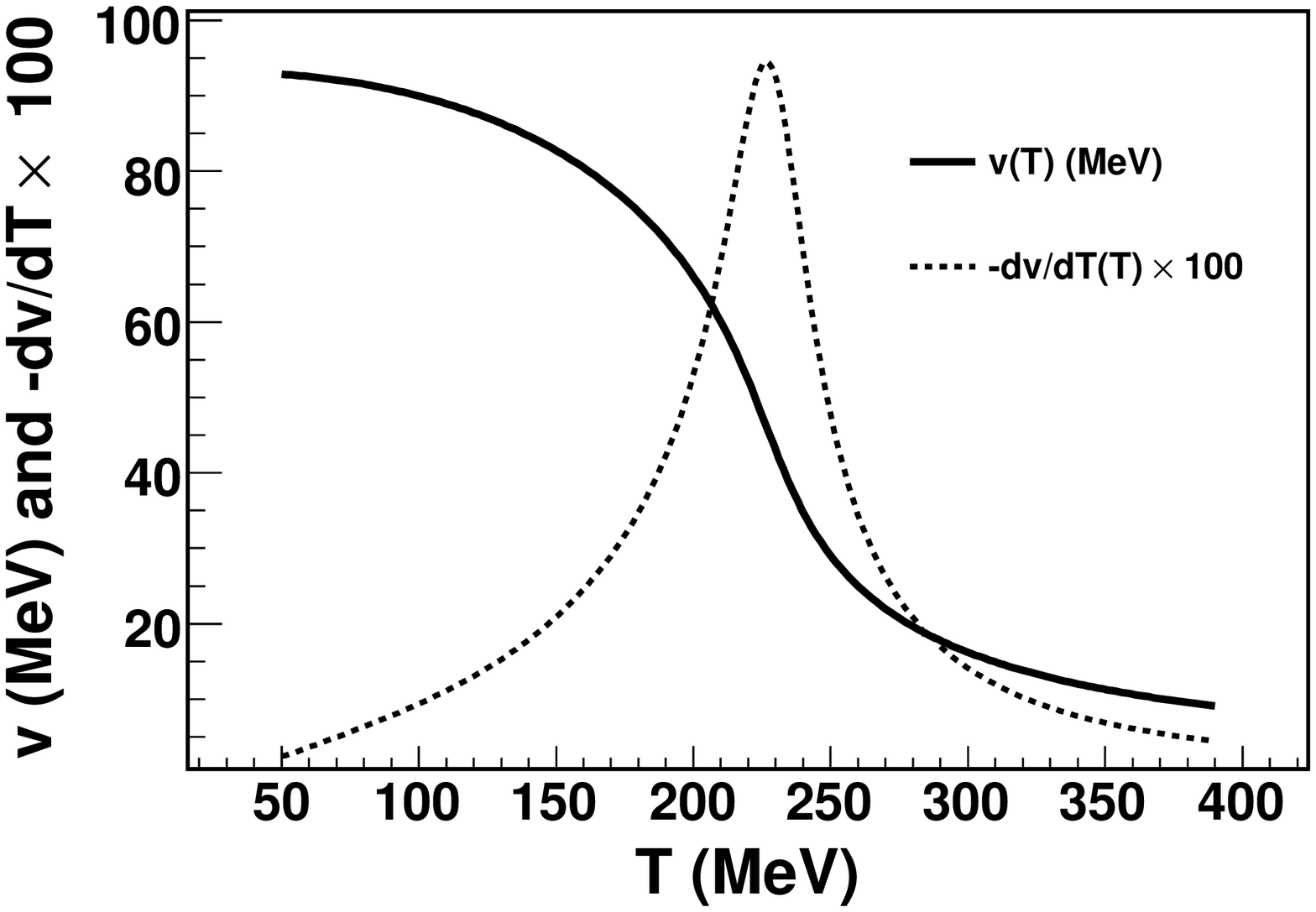}
\includegraphics[scale=0.35]{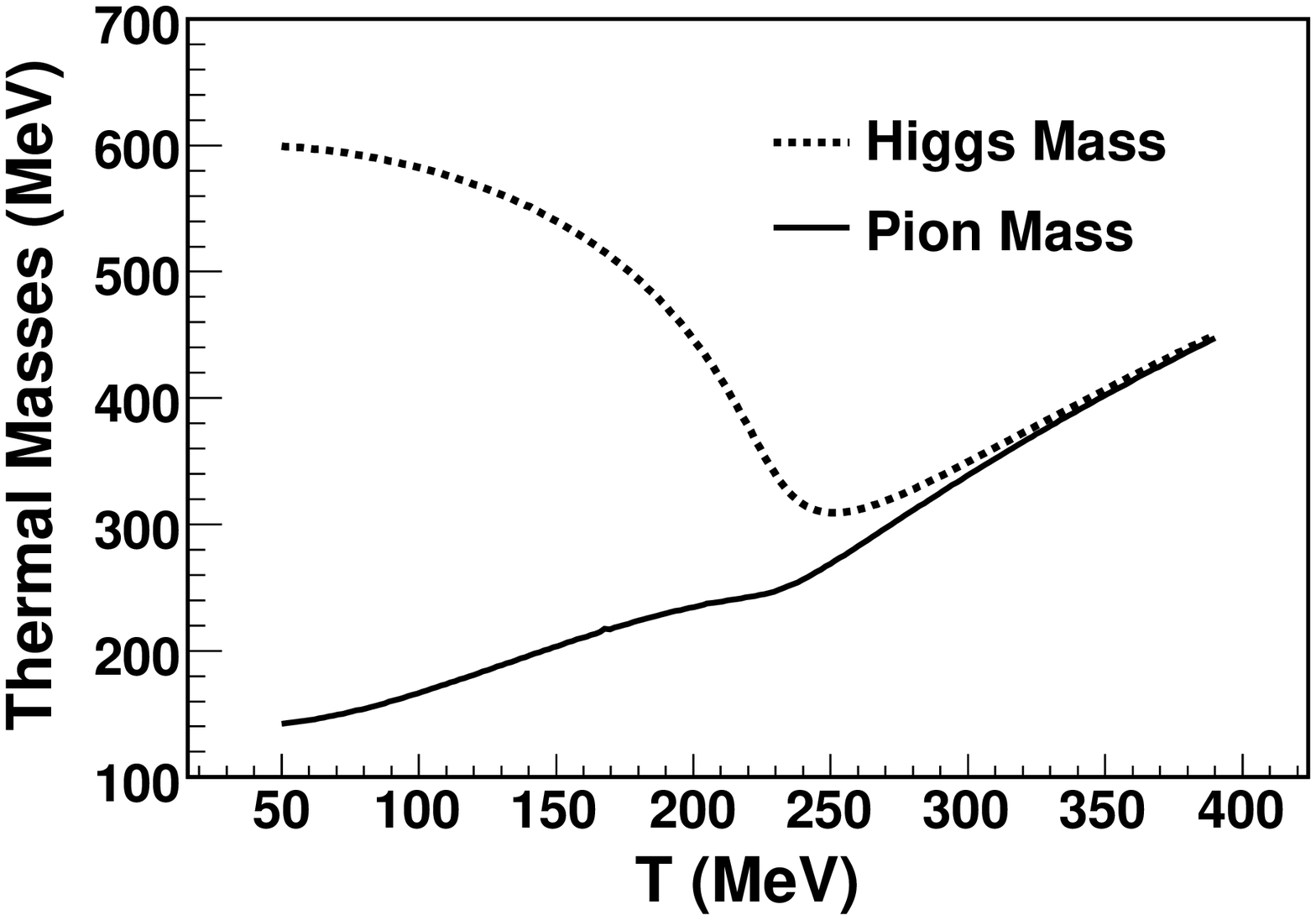}
\caption{\label{fig:thermalc6} Left panel: Order parameter and susceptibility in the CJT formalism in the Hartree approximation with masses $M_{\pi}=138$ MeV and $M_{\sigma}=600$ MeV at $T=0$. Right panel: Thermal masses in the Hartree aproximation
 for the same set of parameters.}
\end{center}
\end{figure}

\begin{figure}[t]
\begin{center}
\includegraphics[scale=0.35]{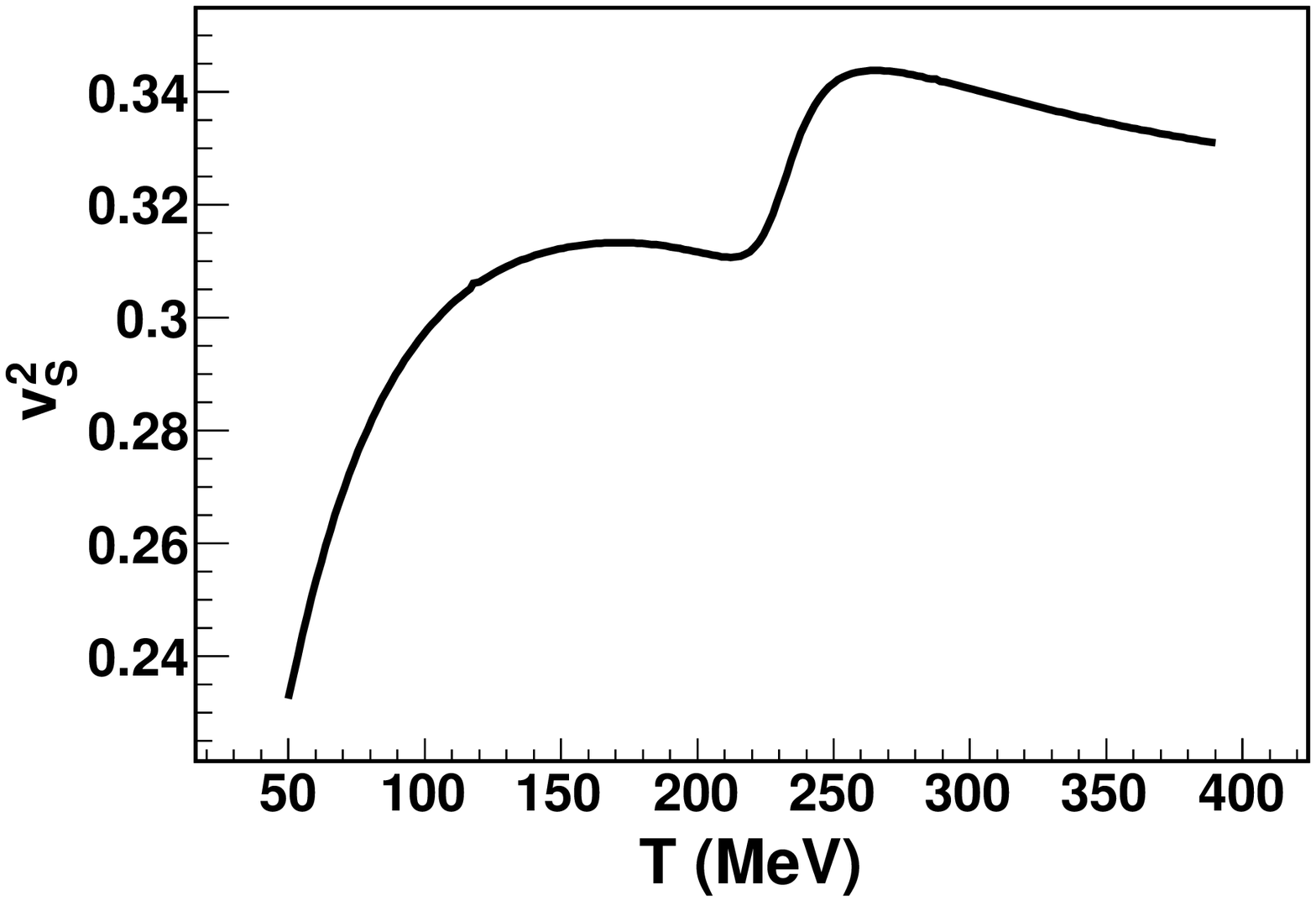}
\includegraphics[scale=0.35]{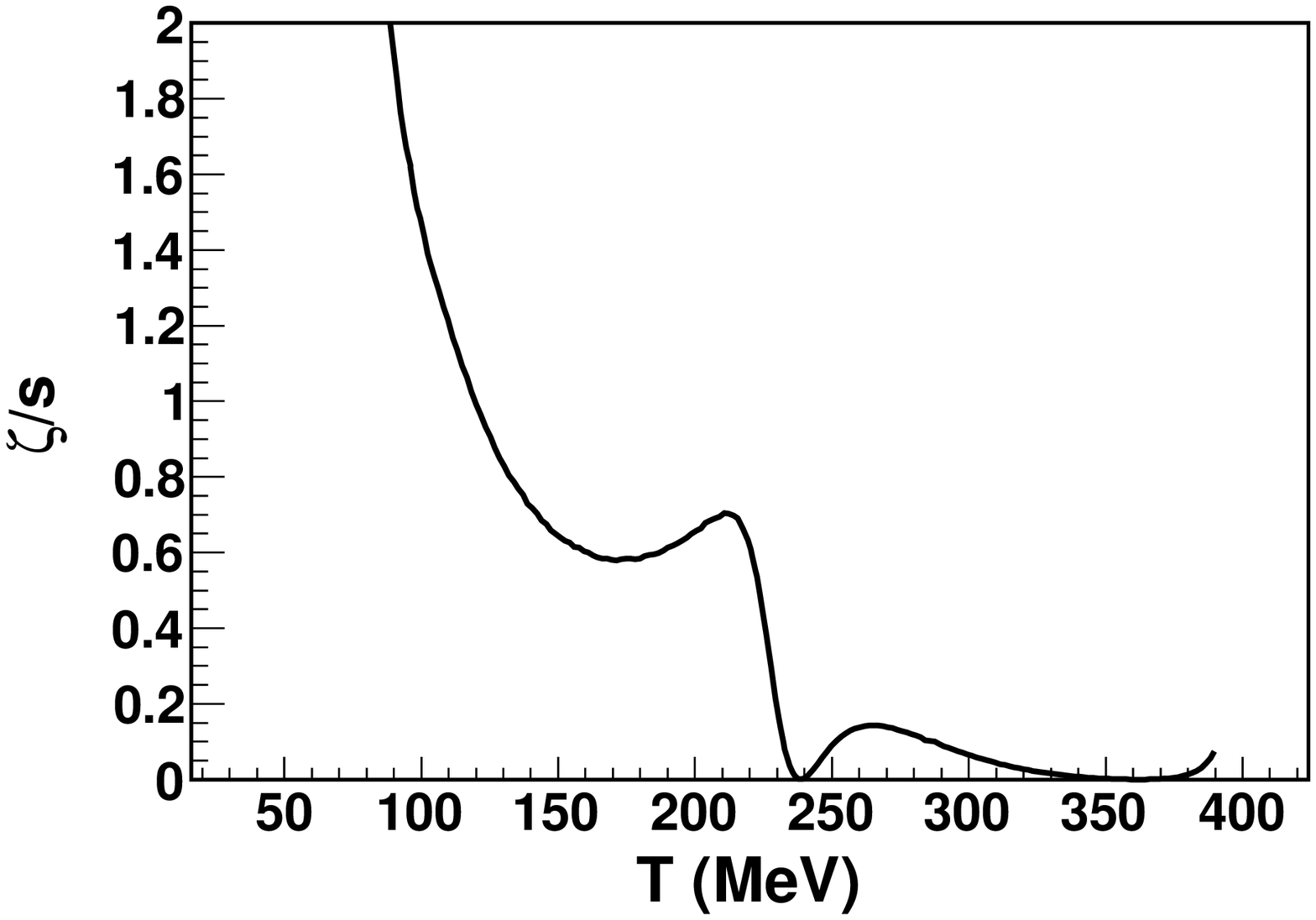}
\caption{ \label{fig:thermal2c6} Left panel: Square speed of sound for the case $M_{\pi} (T=0)=138$ MeV and $M_{\sigma}(T=0)=600$ MeV. A local minimum is seen near the crossover temperature. Right panel: Bulk viscosity over entropy density. The first maximum appears
at the local minimum of the speed of sound and the second maximum at the maximum of $v_S^2$, these points correspond to the zones where the loss of conformality is larger.}
\end{center}
\end{figure}

 Within this new formalism  we start again by considering the crossover case with a Higgs mass of 600 MeV and a pion mass of $138$ MeV at $T=0$. In Fig.~\ref{fig:thermalc6} we present the results obtained by solving  the nonlinear system in Eq.~(\ref{eq:system}). In the left panel we show the order parameter and
minus its derivative with respect to the temperature. They look qualitatively very similar to the results found by using the large-$N$ limit with a crossover temperature of $T_{cr}=230$ MeV. In the right panel we show  the thermal masses, very similar also to those found from the large-$N$ limit. However,
one can see that the pion thermal mass shows a plateau at $T \simeq 220$ MeV with a derivative approaching zero in that region. This behavior produces an  effect which is clearly seen in the speed of sound (where only the pion effects are included
\footnote{In the Hartree approximation the $\sigma$ is not $N$-suppressed anymore and it should be included in the thermodynamical functions. However, given the similar results in~\cite{Chakraborty:2010fr} we will assume that the pertinent changes are minimal and we 
do not include this degree of freedom in order to maintain the simplicity in the kinetic theory formalism.}
) in the left
panel of Fig.~\ref{fig:thermal2c6}. Its shape is drastically different from the one coming from the large-$N$ limit computation. Now a  new minimum is obtained at $T=220$ MeV and it corresponds to a small loss of conformality around this region. The subsequent maximum is also
a signature of the loss of conformality. Among them, the conformal value $1/3$ is reached. Therefore, one expects a zero bulk viscosity surrounded by two maxima.
This is exactly the case  in which the zero of the bulk viscosity resembles our zero in the large-$N$ limit, as we show in the right panel of the same figure.

\begin{figure}[t]
\begin{center}
\includegraphics[scale=0.35]{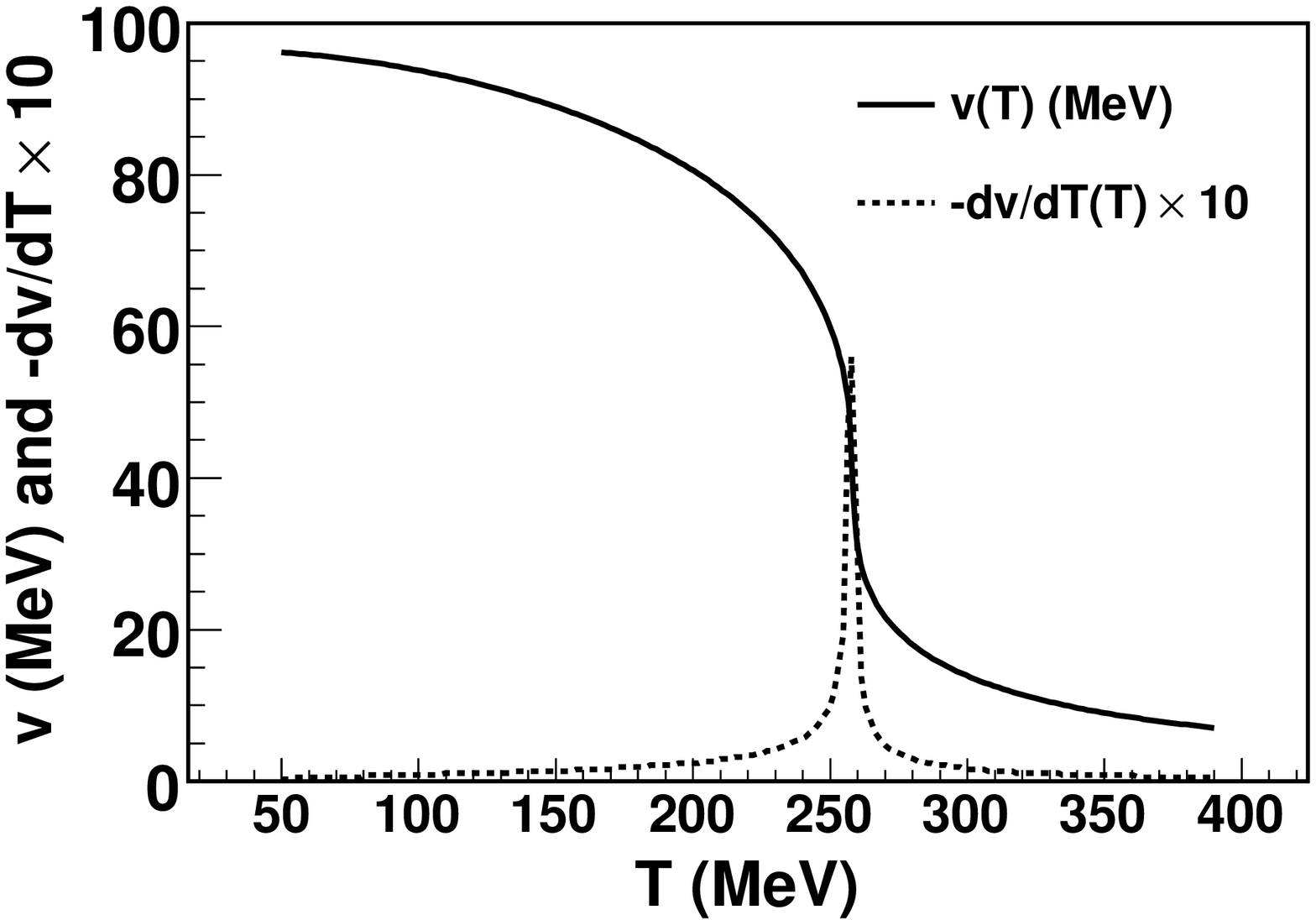}
\includegraphics[scale=0.35]{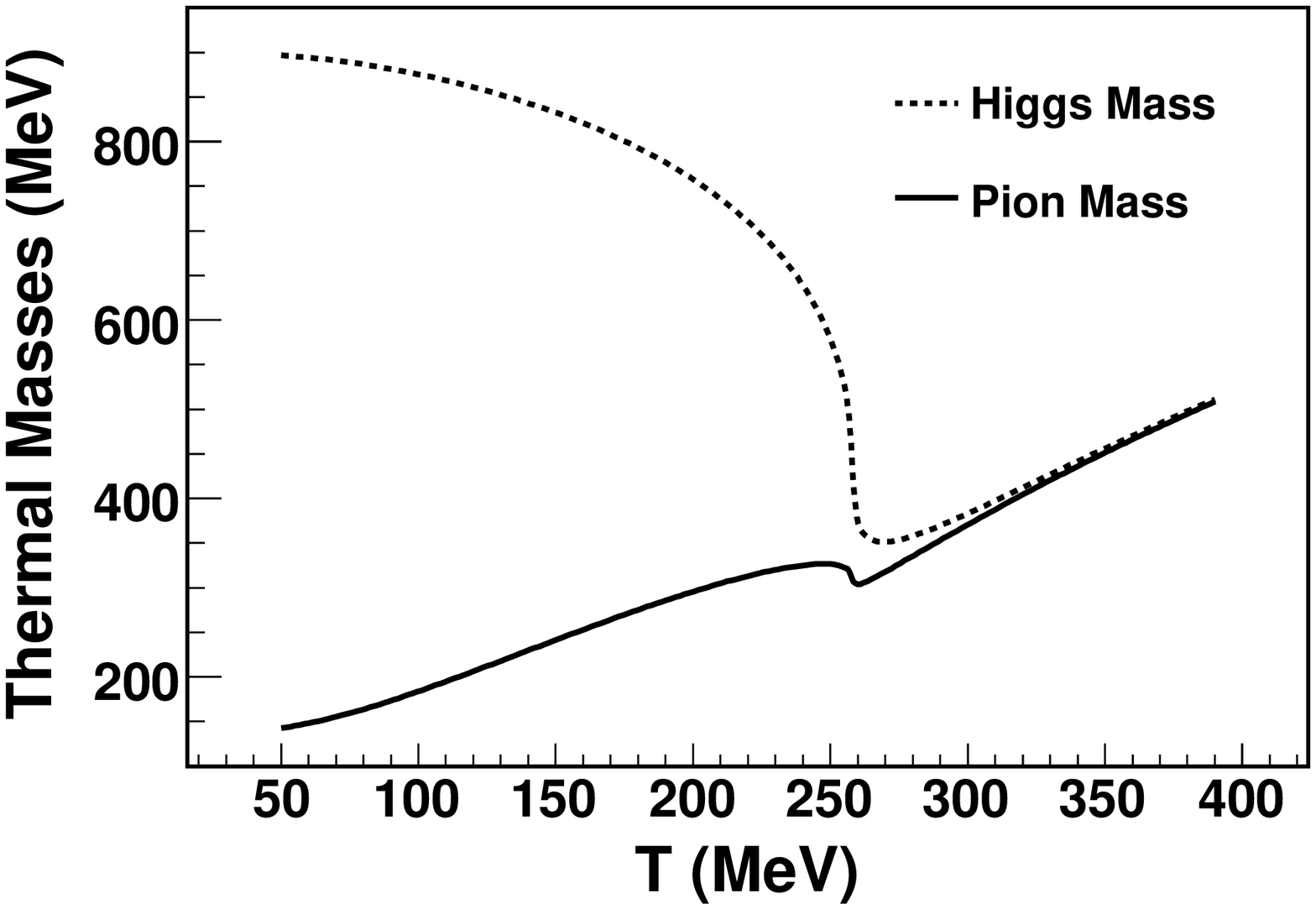}
\caption{\label{fig:thermalc9} Same as Fig.~\ref{fig:thermalc6} but with $M_{\sigma}(T=0)=900$ MeV. The order parameter goes faster toward zero and the susceptibility clearly indicates the position of the crossover temperature.
The pion mass has a zone of negative derivative near the crossover temperature. }
\end{center}
\end{figure}

\begin{figure}[t]
\begin{center}
\includegraphics[scale=0.35]{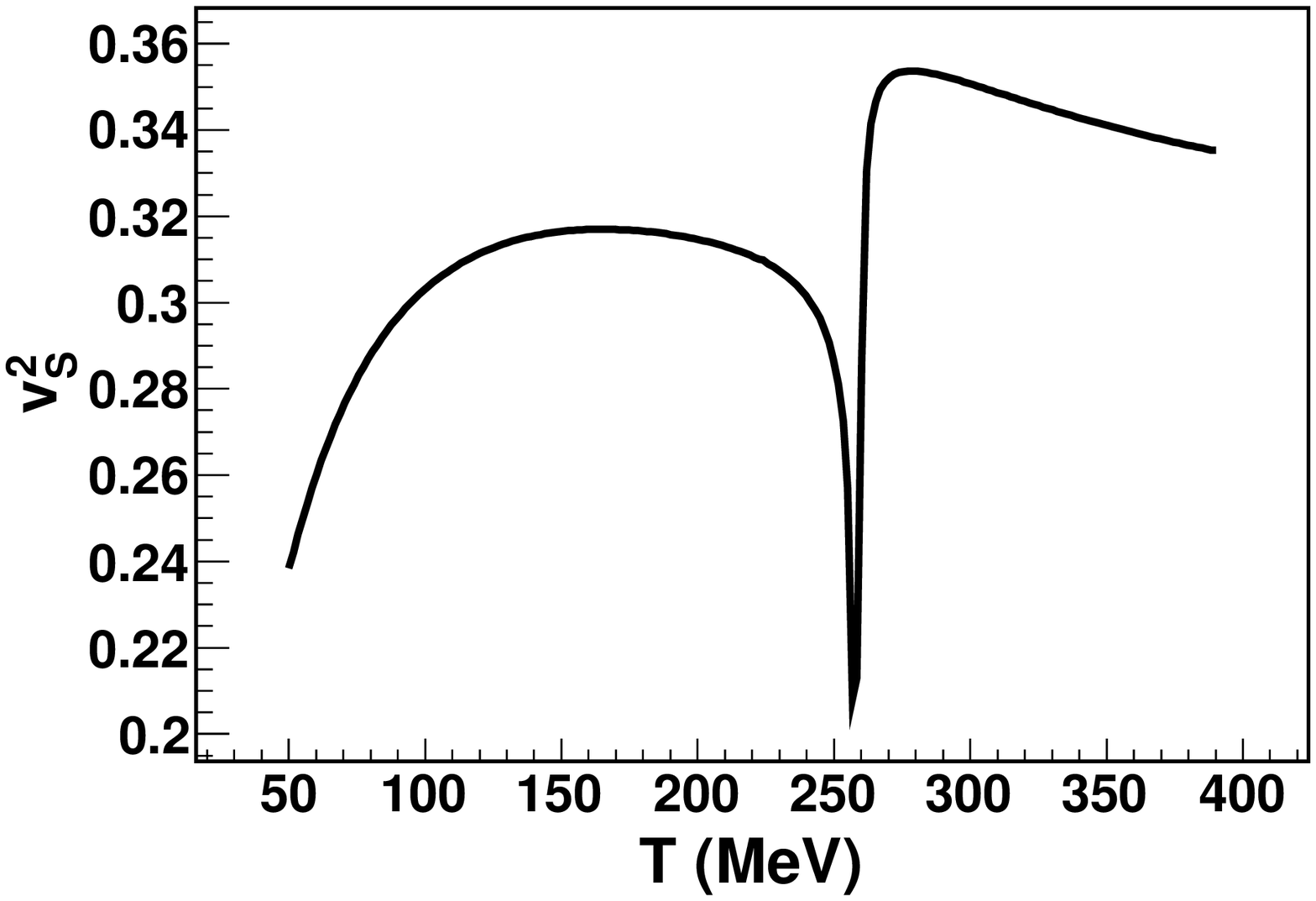}
\includegraphics[scale=0.35]{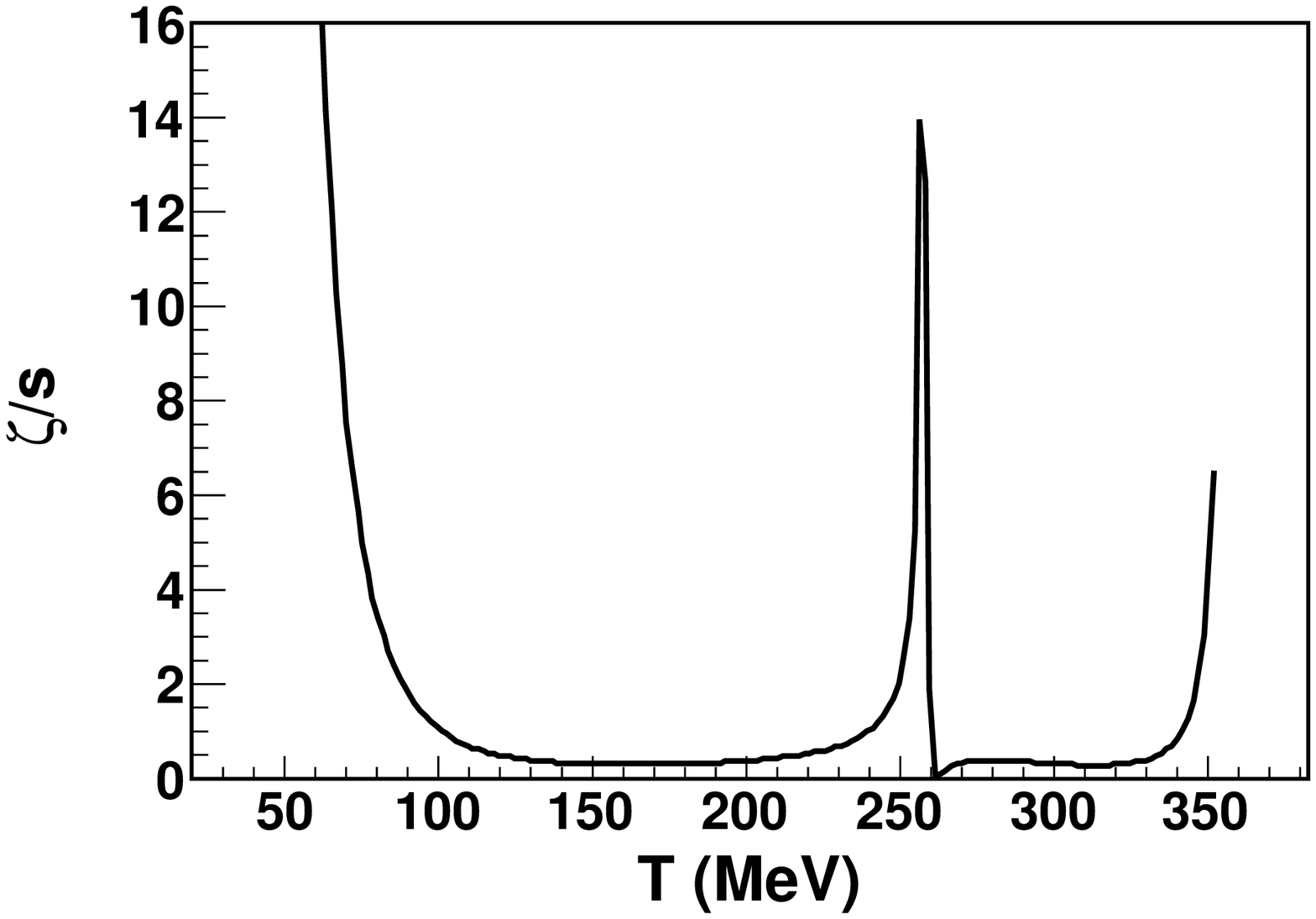}
\caption{ \label{fig:thermal2c9} Same as Fig.~\ref{fig:thermal2c6} but with $M_{\sigma}(T=0)=900$ MeV. At $T_{cr}$ the square speed of sound possesses an abrupt minimum showing a large violation of conformality. The bulk viscosity shows a clear maximum at $T_{cr}$. }
\end{center}
\end{figure}

 Now we can proceed to increase the Higgs mass. The solution of the nonlinear system in Eq.~(\ref{eq:system}) is shown in Fig.~\ref{fig:thermalc9}, where a similar result to the previous case is obtained. The main difference is that the transition is more abrupt
with a nice peak in the susceptibility. Now the curve  shows a
crossover temperature of $T_{cr}=260$ MeV. The effect on the pion
mass is seen in the right panel of that figure showing a clear zone
around the $T_{cr}$ where the pion mass possesses a negative
derivative. The implications of this behavior become clearer in the
speed-of-sound curve that we plot in the left panel of
Fig.~\ref{fig:thermal2c9}. The previous minimum at $T_{cr}$ is much
deeper showing a larger breaking of conformality. This minimum
produces a peak in the bulk viscosity at the crossover temperature.
This is the maximum of $\zeta/s$ that one expects at the phase
transition and the one that is lost in the large-$N$ approximation.

\subsection{First-order phase transition in the chiral limit}

\begin{figure}[t]
\begin{center}
\includegraphics[scale=0.35]{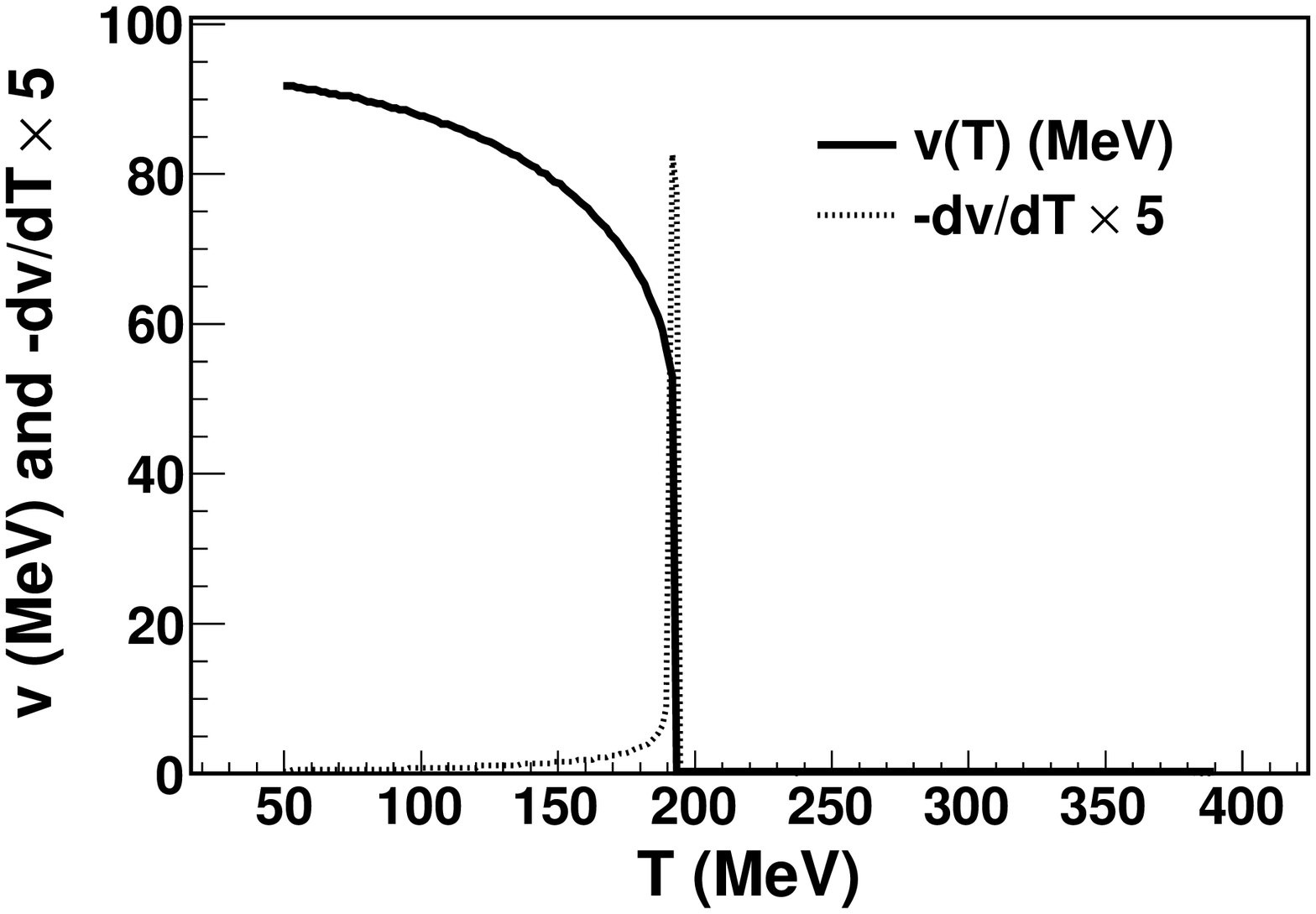}
\includegraphics[scale=0.35]{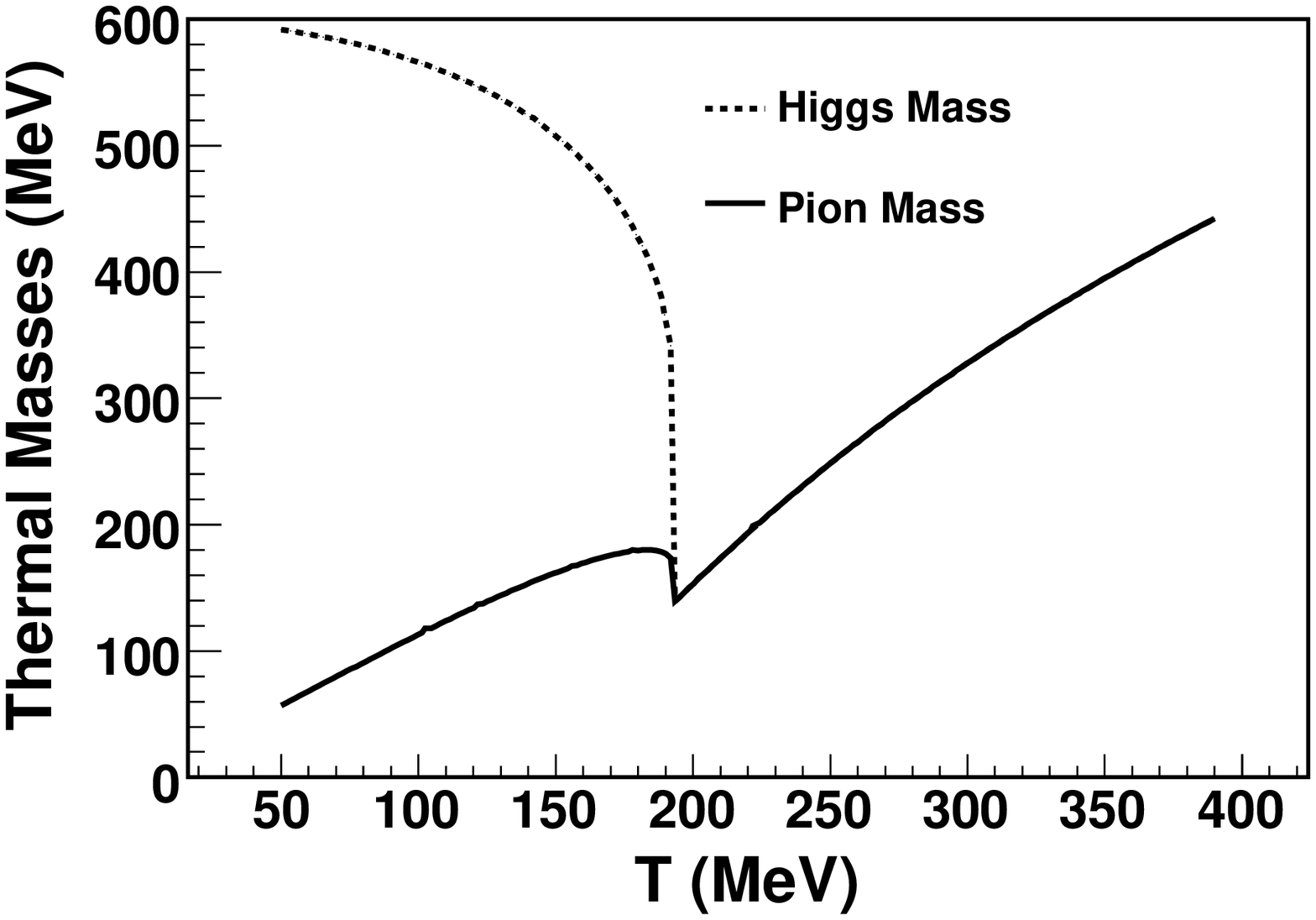}
\caption{\label{fig:thermalf} Same as Fig.~\ref{fig:thermalc6} but with $M_{\pi}(T=0)=0$ MeV. The discontinuity of the order parameter reveals a first-order phase transition. In the right panel one can observe that the Goldstone theorem is not
satisfied for the pion mass within the Hartree approximation.}
\end{center}
\end{figure}

\begin{figure}[t]
\begin{center}
\includegraphics[scale=0.35]{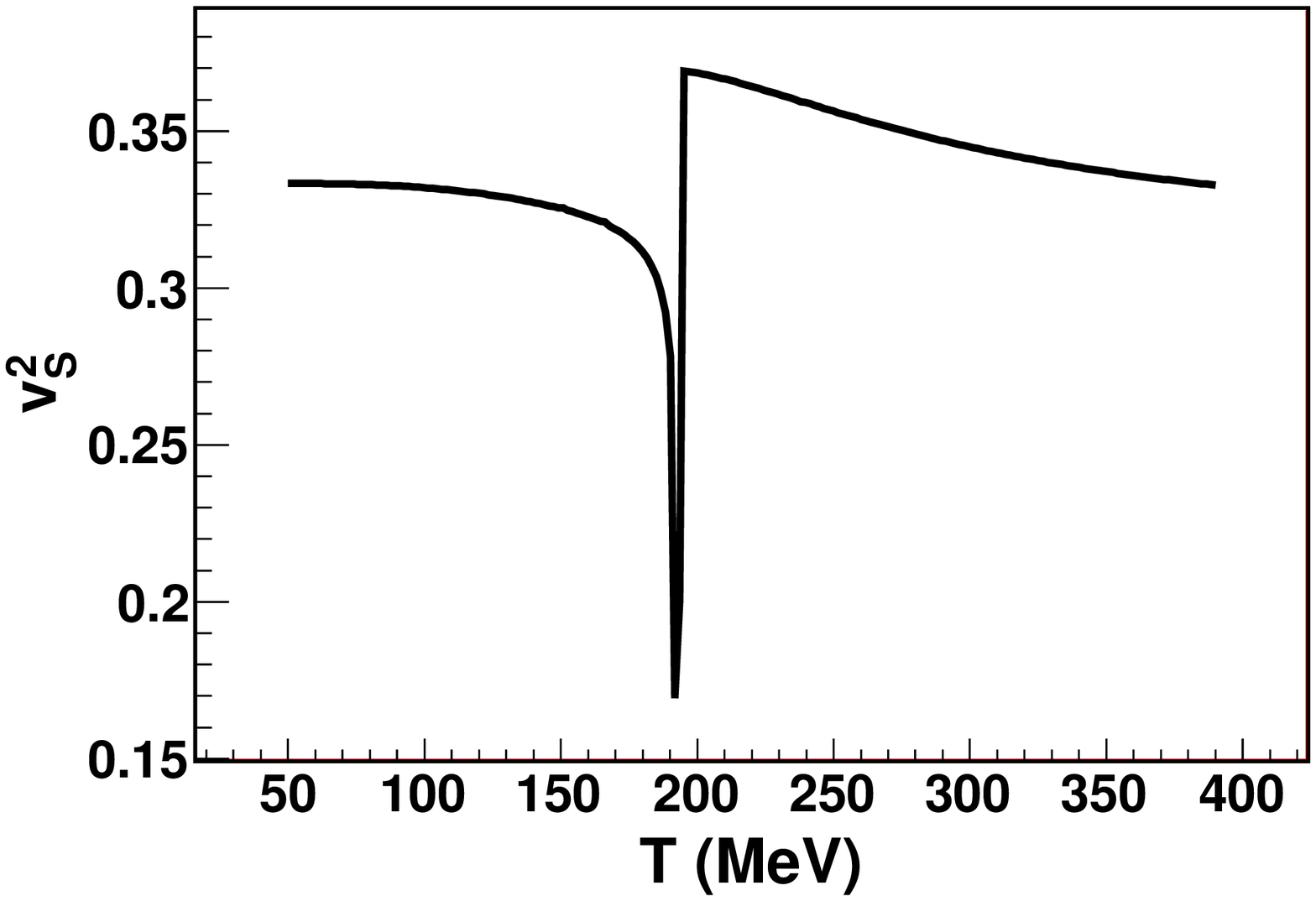}
\includegraphics[scale=0.35]{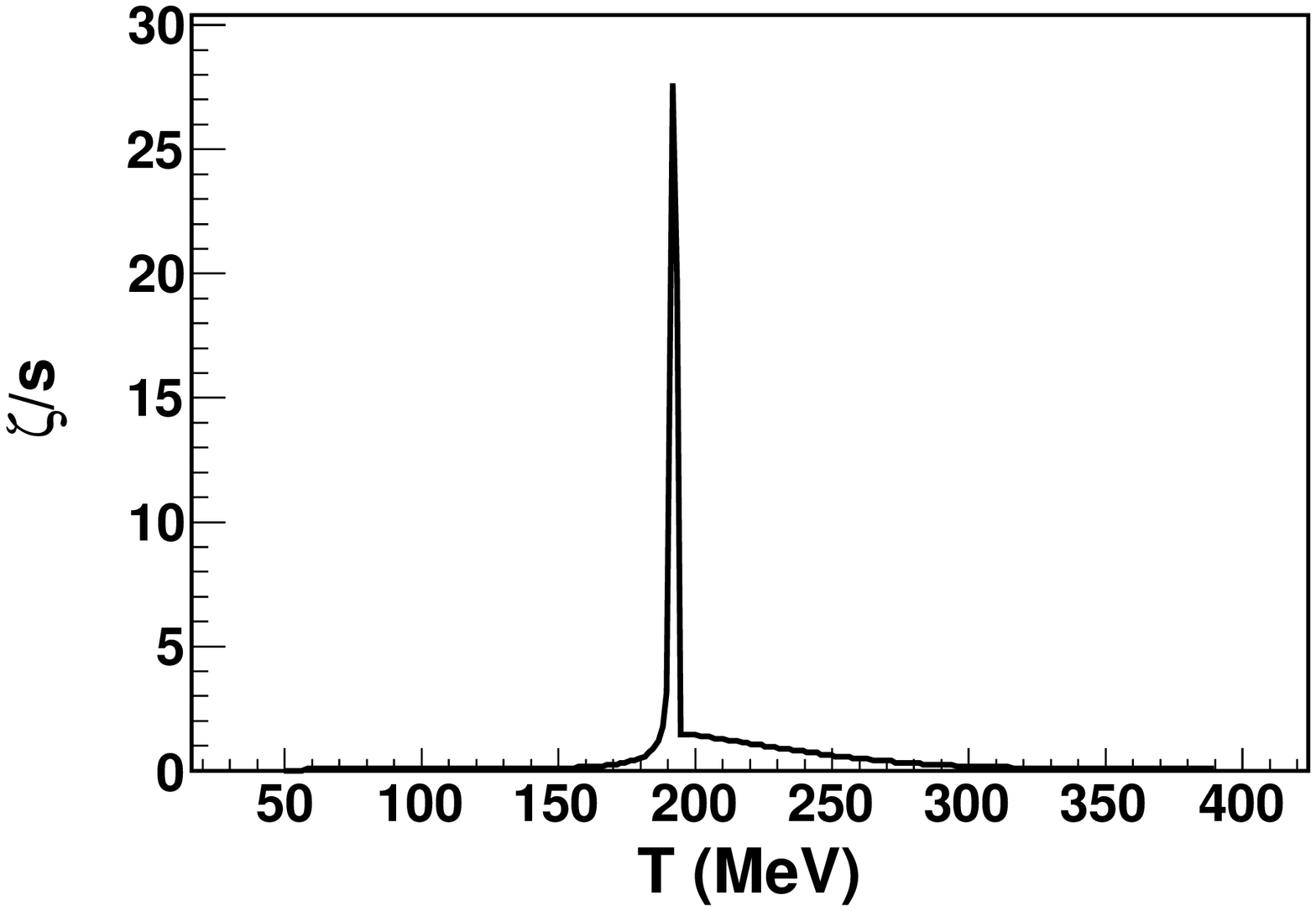}
\caption{ \label{fig:thermal2f} Same as Fig.~\ref{fig:thermal2c6} but with $M_{\pi} (T=0)=0$ MeV. The speed of sound inherits the discontinuity of the first-order transition. A maximum in the bulk viscosity over entropy density is still
found at the critical temperature.}
\end{center}
\end{figure}

Finally, we take the chiral limit for pions at zero temperature. In the left panel
of Fig.~\ref{fig:thermalf} we show how in the chiral limit the phase transition is of first order, with a clear discontinuity in the
order parameter at $T_{c}=190$ MeV. One feature of the Hartree
approximation, as opposed to the large-$N$ limit, is that it  does
not respect the Goldstone theorem. This can be seen in the right
panel of Fig.~\ref{fig:thermalf}, where the pion mass in the low-temperature phase is different from zero. In addition, both masses
show discontinuities at $T_c$. This jump is also seen in the speed
of sound in the left panel of Fig.~\ref{fig:thermal2f}. The bulk
viscosity presents a clear peak in the phase transition temperature
with a finite discontinuity inherited by the first-order
 nature  of the transition in the Hartree approximation.

\newpage

\section{Summary \label{sec:conclusions}}

In this work we have extensively investigated the loss of
conformality and the bulk viscosity in the L$\sigma$M. In the large-$N$ limit, where the dynamics of the Higgs field is washed out,
the phase diagram has been calculated in order to pin down the
location of the critical temperature. The bulk viscosity has been
calculated in this limit and we find that  it vanishes at the
conformal points but it is different from zero otherwise. However, a maximum of the bulk viscosity is
not found in the large-$N$ limit, consistent with other systems in the
same limit as in~\cite{FernandezFraile:2010gu}. Moreover, we find
this result quite natural as recent dynamical renormalization group calculation shows
that in this limit the bulk viscosity remains finite (it does not
diverge) in the critical point~\cite{Nakano:2011re}.

However we have shown that a maximum of the bulk viscosity can be
found if one considers the Hartree approximation in the CJT effective
potential. Just by obtaining the effective pion mass  from this
approach in our kinetic computation of the bulk viscosity, we find
that a maximum is obtained at the crossover temperature for some
appropriate region of the Higgs mass. Moreover, a clear maximum of
$\zeta/s$ is obtained at the critical point when the pion mass is
set to zero at $T = 0$. However, this approximation fails
to reproduce the Goldstone theorem of the L$\sigma$M and gives rise
to a first-order phase transition instead of  a second-order one.
Consequently, the large-$N$ and the Hartree approximations considered
here seem to be relevant and informative to fully understand the behavior of
the bulk viscosity of the L$\sigma$M at the critical point.

\subsection{Acknowledgments}

We thank Felipe J. Llanes-Estrada and Anna W. Bielska for reading the manuscript and for useful suggestions.
We also thank the referee for helping us to improve the whole content of this manuscript.
This work was supported by Grants Consolider-CSD2007-00042, FPA2011-27853-C02-01 and UCM-BSCH GR58/08 910309.
Juan M. Torres-Rincon is a recipient of a FPU Grant from the Spanish Ministry of Education.

\appendix

\section{$g_0(T,M^2)$, $g_1(T,M^2)$ and their relation with the moments of $n_p$ \label{app:moments1}}

The function $g_0(T,M^2)$ and its derivative $g_1(T,M^2)$ are needed for the temperature-dependent part of the effective potential
in Eq.~(\ref{eq:finaleffpot}). The function $g_0(T,M^2)$ is defined as the following integral:
\be \label{eq:g0fun} g_0(T,M^2)=\frac{T^4}{3\pi^2} \int_y^{\infty} dx \ (x^2-y^2)^{3/2} \frac{1}{e^{x}-1} \ , \ee
where $y=M/T$. Note that in Eq.~(\ref{eq:finaleffpot}) $M^2$ corresponds to the function $G^{-1}_{\pi} [0,\chi]$. The notation $M^2$ is used to suggest
that this factor will turn out to be the pion effective mass squared when the gap equation is solved.

The derivative of this function with respect to $M^2$, appearing in Eq.~(\ref{eq:gapeq}), defines the function $g_1(T,M)$:
\be g_1 \equiv - \frac{dg_0}{dM^2} \ , \ee
and can also be written as:
\be \label{eq:g1fun}  g_1(T,M^2)=\frac{T^2}{2\pi^2} \int_y^{\infty} dx \ \frac{\sqrt{x^2-y^2}}{e^{x}-1} \ . \ee
In the limit $y\rightarrow 0$, i.e. the conformal limit as the pion mass goes to zero, the two functions are given by:
\begin{eqnarray}
\label{eq:g0limit} g_0 (T,0) & = & \frac{T^4}{3 \pi^2} \Gamma(4) \zeta(4) = \frac{\pi^2 T^4}{45} \ , \\
\label{eq:g1limit} g_1 (T,0) & = & \frac{T^2}{2 \pi^2} \Gamma(2) \zeta(2) = \frac{T^2}{12} \ .
\end{eqnarray}

These functions can be related to the moments of the Bose-Einstein distribution function, $n_p$~\cite{Muronga:2006zx}. These moments read:
\be \mathcal{I}^{\alpha_1 \alpha_2 \cdots \alpha_n} (T,m) = N \int \frac{d^3p}{(2\pi)^3 E_p} n_p \ p^{\alpha_1} p^{\alpha_2} \cdots p^{\alpha_n} \ , \ee
with $E_p=\sqrt{p^2 + m^2}$. They can be expanded in a tensor basis in terms of  some coefficients $\mathcal{I}_{n,k}$  depending on the temperature. These coefficients read:
\be \mathcal{I}_{n,k} (T,m)= \frac{Nm^{n+2}}{(2k+1)!! 2 \pi^2} \int_1^{\infty} dx \ x^{n-2k} (x^2-1)^{k+1/2}  \frac{1}{e^{yx}-1} \ , \ee
where $x=E_p/m$ and $y=m/T$.
They satisfy the recursion relation
\be \mathcal{I}_{n+2,k+1} (T,m)= \frac{1}{2k+3} \left( \mathcal{I}_{n+2,k} - m^2 \mathcal{I}_{n,k} \right) \ ,\ee
and  can be related to some thermodynamical quantities in equilibrium, e.g. $P=\mathcal{I}_{2,1}, \epsilon=\mathcal{I}_{2,0}$, etc.

Performing a change of variables in (\ref{eq:g0fun}) and (\ref{eq:g1fun}), one can express these two functions in terms of the $\mathcal{I}_{n,k}$ integrals.
In particular:
\begin{eqnarray}
g_0 (T,M^2) & = & \frac{2}{N} \ \mathcal{I}_{2,1} (T,M) \ , \\
g_1 (T,M^2) & = & \frac{1}{N} \ \mathcal{I}_{0,0} (T,M) \ .
\end{eqnarray}

\section{$K^i$, $I^i$ and their relation with the auxiliary moments of $n_p$ \label{app:moments2}}

In this work we have used two different integration measures for the
shear and bulk viscosities, respectively. These measures are
naturally given by the form of the viscosities as an integration
over the distribution function. In terms of the adimensional
variables defined in Eq.~(\ref{eq:adivar}), they explicitly read:
\be d\mu_{\eta}(x;y) = dx \ (x^2-1)^{5/2} \frac{e^{yx}}{\left( e^{yx}-1\right)^2} \ , \ee
\be d\mu_{\zeta}(x;y) = dx \ (x^2-1)^{1/2} \frac{ e^{yx}}{\left( e^{yx}-1 \right)^2}  \ , \ee
where clearly $x
\in \Omega =[1,\infty)$. The moments of these measures are related to the
respective source functions of the BUU equation. They have been
defined as
\be K^i = \int_{\Omega}  \ d\mu_{\eta} \ x^i \ , \ee
\be I^i = \int_{\Omega}  \ d\mu_{\zeta} \ x^i \ . \ee
From these measures one can
define the scalar products:
\be \langle f \mid g \rangle_{\eta} =\int_{\Omega}  \ d\mu_{\eta}(x;y)f(x)g(x)       \ , \ee
\be   \langle f \mid g \rangle_{\zeta} =\int_{\Omega}  \ d\mu_{\zeta}(x;y)f(x)g(x)   \ , \ee and
the corresponding norms in the usual way. In addition, one can
define the corresponding orthogonal polynomial bases that expand the
space generated by the $x^i$. For the shear viscosity we use a
monic orthogonal polynomial basis defined as
\begin{eqnarray}
P_0 (x) &= & 1 \\
P_1(x) & =& x - \frac{K_1}{K_0} \\
P_2(x) & = & x^2 + \frac{K_0 K_3 - K_1 K_2}{K_1^2 - K_0 K_2} x + \frac{K_2^2 - K_1 K_3}{K_1^2 - K_0 K_2} \\
 \nonumber & \cdots &
\end{eqnarray}

However, for the bulk viscosity we follow the same convention except for the polynomial $P_2(x)$, which is
fixed to be the source function in the left-hand side of Eq.~(\ref{eq:buuforlsm}),

\begin{eqnarray}
 P_0(x) & = & 1 \\
 P_1 (x) &= & x \\
 P_2 (x) & = & \left( \frac{1}{3} - v_S^2 \right) x^2 - \frac{1}{3} + \frac{T}{m} \frac{dm}{dT} v^2_S \\
\nonumber &  \cdots &
\end{eqnarray}
Notice that now $ \langle P_1 \mid P_2 \rangle =0$, but  $ \langle P_0 \mid P_2 \rangle$ is different from zero, so this basis is not completely orthogonal.
Further information about the integration measures, scalar products and polynomial basis can be found with more detail (also for the thermal conductivity coefficient, not 
studied here) in~\cite{torresphd}.

The two functions $K^i$ and $I^i$ are particular cases of  more
general coefficients expanding the ``auxiliary moments'' of the
Bose-Einstein distribution function $n_p$~\cite{Muronga:2006zx}. The
auxiliary moments read \be \mathcal{J}^{\alpha_1 \alpha_2 \cdots
\alpha_n} (T,m) = N \int \frac{d^3p}{(2\pi)^3 E_p} n_p (1+n_p) \
p^{\alpha_1} p^{\alpha_2} \cdots p^{\alpha_n} \ . \ee In an
appropriate tensor basis these moments can be expanded in terms of
the coefficients $\mathcal{J}_{n,k}$  depending on the
temperature. By using our adimensional variables, these coefficients
can be written as:

\be \mathcal{J}_{n,k} (T,m)= \frac{Nm^{n+2}}{(2k+1)!! 2 \pi^2} \int_1^{\infty} dx \ x^{n-2k} (x^2-1)^{k+1/2}  \frac{e^{yx}}{(e^{yx}-1)^2} \ . \ee

Some of them can be related to thermodynamical quantities. For example:
\be \label{eq:thermoJint} \frac{\mathcal{J}_{2,1}}{T} = n ; \quad \frac{\mathcal{J}_{3,1}}{T} = \epsilon + P ; \quad \frac{\mathcal{J}_{3,1}}{T^2} =s \ . \ee
The coefficients satisfy the recursion relation
\be \mathcal{J}_{n+2,k+1} = \frac{1}{2k+3} \left( \mathcal{J}_{n+2,k} - m^2 \mathcal{J}_{n,k} \right) \ . \ee

It is not difficult to see that $K^i$ and $I^i$ are related with the $k=2$ and $k=0$ coefficients, respectively
\begin{eqnarray}
 \mathcal{J}_{4+i,2} & = &  \frac{N m^{6+i}}{30 \pi^2} K_i \ , \\
 \mathcal{J}_{i,0} & = &  \frac{N m^{2+i}}{2 \pi^2}  I_i \ .
\end{eqnarray}

Using the formula (\ref{eq:thermoJint}) we can express the entropy
density and the heat capacity in terms of the integrals $I^i$: \be s
= \frac{N m^5}{6 \pi^2 T^2} \left(  I_3 - I_1 \right), \quad C_V = T
\frac{\pa s}{\pa T} = \frac{Nm^5}{2 \pi^2 T^2} \left( I_3 -
\frac{dm}{dT} \frac{T}{m} I_1 \right) \  . \ee
When there is no chemical potential, the temperature is the only independent variable,
and the adiabatic speed of sound and the isochoric one coincide.
Then, they can be written as \be \label{eq:speedsound} v_S^2 =
\frac{s}{C_V} = \frac{1}{3} \frac{I_3-I_1}{I_3 - \frac{dm}{dT}
\frac{T}{m} I_1} \ , \ee which provides an universal relation for
the speed of sound for any system that can be described in terms of
free quasiparticles with masses depending only on the temperature
and, in particular, for a free particle system.

\section{The Cornwall-Jackiw-Tomboulis effective potential \label{app:cjt}}

In this appendix we briefly review the basic ingredients of the effective potential calculation in the context of the CJT formalism and the Hartree approximation. For more details we refer the reader to
~\cite{Cornwall:1974vz,AmelinoCamelia:1992nc,AmelinoCamelia:1997dd,Lenaghan:1999si,Petropoulos:2004bt}. To introduce this technique we start with the derivation of the effective potential in the simple case of standard
$\lambda \Phi^4$ theory at zero temperature and later we will extend the calculation at finite temperature for the L$\sigma$M in the Hartree approximation.

\subsection{$\lambda \Phi^4$ theory at zero temperature}
The classical Euclidean action for the $\lambda \Phi^4$ theory reads
\be \label{eq:action} S[\Phi] = \int_x \left[ \frac{1}{2} \pa^{\mu} \Phi(x) \pa_{\mu} \Phi(x) + \frac{m^2}{2} \Phi(x)^2+\frac{\lambda}{4!} \Phi(x)^4\right] \ ,\ee
where we have defined
\be \int_x \equiv \int d^4x.\ee

In the context of the CJT method, the starting point is a generating functional of one and two point Green functions depending on a local source $J(x)$ and a bilocal one $K(x,y)$, defined as:

\be \mathcal{Z} [J,K] = \int [d \Phi] \ \exp \left[ S[\Phi] + \int_x J(x) \Phi(x) + \frac{1}{2} \int_{x,y} \Phi(x) K(x,y) \Phi(y) \right] \ . \ee

The corresponding generating functional for the connected Green functions, $\mathcal{W}[J,K]$ is defined as
\[ \mathcal{Z}[J,K] = e^{\mathcal{W}[J,K]}.\]

The expectation value of the field $\phi(x)=\langle \Phi(x) \rangle_{J,K}$ and  the connected two-point function $G(x,y)$ in presence of the sources can be obtained through the functional
derivatives of $\mathcal{W}[J,K]$.
\be \label{deriW} \frac{\delta \mathcal{W} [J,K] }{\delta J(x)} = \phi(x) ; \qquad \frac{\delta \mathcal{W}[J,K]}{\delta K(x,y)} = \frac{1}{2} \left[ \phi(x) \phi(y) + G(x,y) \right] \ . \ee

This two-point function should not be confused with the tree-level propagator $D(x,y)$, whose inverse reads
\be D^{-1} (x,y) = \frac{\delta^2 S[\Phi]}{\delta \Phi(x) \delta \Phi(y) } \ . \ee

For the action given in (\ref{eq:action}) this becomes
\be D^{-1} (x,y) = -\left( \square_x - m^2 -\frac{\lambda}{2} \Phi(x)^2 \right) \delta(x-y) \ . \ee

The two-particle irreducible (2PI) effective action can be obtained as the double Legendre transformation of $\mathcal{W} [J,K]$:
\be \Gamma[\phi,G] = \mathcal{W} [J,K] - \int_x \frac{\delta \mathcal{W} [J,K]}{\delta J(x)} J(x) - \int_{x,y} \frac{\delta \mathcal{W} [J,K]}{\delta K(x,y)} K(x,y) \ , \ee

which, by using,  (\ref{deriW}) can be written as:
\be \Gamma[\phi,G] = \mathcal{W} [J,K] - \int_x \phi(x) J(x) - \frac{1}{2} \int_{x,y} K(x,y) \phi(x) \phi(y) - \frac{1}{2} \int_{x,y} G(x,y) K(y,x) \ . \ee

The stationary conditions for the 2PI effective action read \be
\label{estcond} \frac{\delta \Gamma [\phi,G]}{\delta \phi(x)} = -
J(x) - \int_y K(x,y) \phi(y) ; \qquad \frac{\delta
\Gamma[\phi,G]}{\delta G(x,y) } = - \frac{1}{2} K(x,y) \ , \ee which
leads to the VEV and the dressed propagator when the sources are set
to zero.

\begin{figure}[t]
\begin{center}
\includegraphics[scale=0.45]{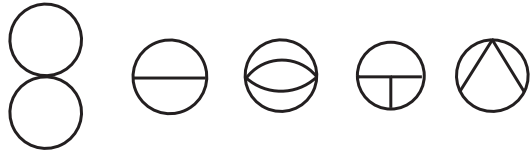}
\caption{ \label{fig:2pi} Two-particle irreducible diagrams which enter in $\Gamma_2$. The solid lines correspond to dressed propagators $G$.}
\end{center}
\end{figure}

Following Cornwall-Jackiw-Tomboulis, the effective action can be written as~\cite{Cornwall:1974vz}

\be \Gamma[\phi,G]  =S[\phi] - \frac{1}{2} \ \textrm{Tr } \log G^{-1} - \frac{1}{2} \ \textrm{Tr } (D^{-1} G-1) + \Gamma_2 [\phi,G] \ ,\ee

where $\Gamma_2 [\phi,G]$ is the sum of 2PI diagrams with $G$ as internal propagators. Some of these diagrams can be seen in Fig.~\ref{fig:2pi} and all of them contribute to the CJT effective action. In the
following we will refer to the first diagram as the double bubble diagram.

As usual, the effective potential (density) is obtained by assuming $\phi$ to be constant, so that one gets:
\be \label{eq:eff_pot} V(\phi,G)  =U(\phi) + \frac{1}{2} \int_k \log G^{-1}(k) + \frac{1}{2} \int_k  (D^{-1} (k) G(k) -1) + V_2 [\phi,G] \ ,\ee
with $U(\phi)=m^2 \phi^2/2 + \lambda \phi^4/ 4!$ being the classical potential.

Now the equations of motion of the effective potential are obtained by minimizing $V$ with respect to $\phi$ and $G$:

\begin{eqnarray}
 \left. \frac{\delta V(\phi,G)}{\delta \phi}\right|_{\phi_0,G_0} & =&  0  \ , \\
 \label{eq:massgap} \left. \frac{\delta V(\phi,G)}{\delta G}\right|_{\phi_0,G_0} & =&  0  \ .
\end{eqnarray}

Here the last equation provides the ``mass gap equation'' for $G_0$ in terms of $\phi$. Substituting this $G_0$ into Eq.~(\ref{eq:eff_pot}), one gets an effective potential in terms of $\phi$ only. Then, by minimizing
this effective potential with respect to $\phi$, one obtains the order parameter (VEV)  $\phi_0$.

In fact, Eq.~(\ref{eq:massgap}) is nothing but the Dyson-Schwinger equation for the propagator:

\be  G^{-1} = D^{-1} + \Sigma(\phi,G) \ ,\ee
where we have defined the self-energy as the functional derivative of the 2PI diagrams:
\be \Sigma(\phi,G) \equiv 2 \  \frac{\delta V_2 (\phi,G)}{\delta G} \ . \ee

On the other hand it is possible to show that each 2PI diagram with $G$ as internal lines, corresponds to an infinite number of 1PI diagrams with the bare propagator $D$ as internal lines.

\begin{figure}[t]
\begin{center}
\includegraphics[scale=0.45]{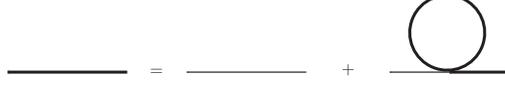}
\caption{ \label{fig:dyson} Dyson-Schwinger equation for the dressed propagator $G$ (represented by wider solid line) in terms of the bare propagator $D$ (narrower solid line) and a self-energy insertion.}
\end{center}
\end{figure}

As the most important example for our work here we  will consider
the Hartree approximation for the calculation of $V_2$. Basically, it
takes into account only the ``double bubble'' diagram
of Fig.~\ref{fig:2pi} (or equivalently diagrams
$\mathcal{O}(\lambda)$). Then for the $\lambda \Phi^4$ theory we
have: \be V_2 = \frac{\lambda}{8} \left[ \int_k G(k, \phi) \right]^2
\  \ee and the gap equation (\ref{eq:massgap}) for the effective
potential reads: \be G^{-1} (k,\phi)= D^{-1} (k,\phi) +
\frac{\lambda}{2} \int_k G(k, \phi) \ , \ee which is depicted in
Fig.~\ref{fig:dyson}.

\begin{figure}[t]
\begin{center}
\includegraphics[scale=0.40]{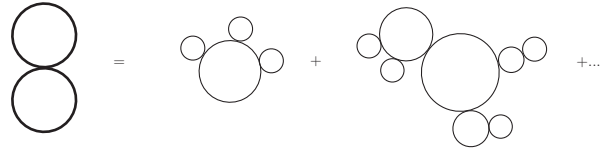}
\caption{ \label{fig:daisy} In the Hartree approximation only the 2PI diagram in the left-hand side is considered. It is equivalent to  all daisy and superdaisy diagrams with the propagator $D$ as internal lines.
We show some examples of those diagrams in the right-hand side.}
\end{center}
\end{figure}

From this equation one realizes that the double bubble diagram with $G$  as internal lines is equivalent to the full resummation of daisies and superdaisies diagrams in Fig.~\ref{fig:daisy} with bare propagator as internal lines.

\subsection{L$\sigma$M at finite temperature}

\begin{figure}[t]
\begin{center}
\includegraphics[scale=0.52]{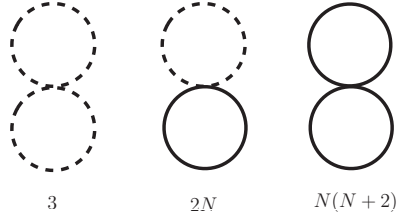}
\caption{ \label{fig:2pilsm} ``Double bubble'' diagrams contributing  to the $V_2$ term of the effective potential in the Hartree approximation. Solid lines represent pions and dashed lines the $\sigma$. Their numerical prefactors in $V_2$ are also shown,}
\end{center}
\end{figure}

The extension of Eq.~(\ref{eq:massgap}) to the L$\sigma$M with $N$ pions and a Higgs is straightforward. In the Hartree approximation the graphs contributing
to $V_2$ (and their numerical prefactors) are those of Fig.~\ref{fig:2pilsm}. Moreover, at finite temperature the $k-$integration should be replaced by Matsubara summations.
Defining
\be \int_{\beta} f (i \omega_n,\mathbf{k}) = T \sum_n \int \frac{d^3k}{(2\pi)^3} f(i \omega_n,\mathbf{k}) \ , \ee
where $\omega_n=2\pi Tn$, the effective potential reads:

\begin{eqnarray}
\nonumber \label{eq:effpotlin} V(\phi,G)  & = & U(\phi) + \frac{1}{2} \int_\beta \log G_{\sigma} ^{-1}(k) + \frac{N}{2} \int_{\beta} \log G_{\pi}^{-1} (\phi,k) \\
  & + &  \frac{1}{2} \int_\beta  [D^{-1}_{\sigma} (k,\phi) G_{\sigma}(\phi,k) -1) + \frac{N}{2} \int_\beta  [D^{-1}_{\pi} (k,\phi) G_{\pi}(\phi,k) -1) + V_2 (\phi,G_{\pi},G_{\sigma}) \ ,
\end{eqnarray}
where $V_2$ is given by
\begin{eqnarray}
\nonumber V_2 (\phi,G_{\sigma},G_{\pi}) &= & 3\frac{\lambda}{N} \left[ \int_{\beta} G_{\sigma} (\phi,k) \right]^2 + N (N+2) \frac{\lambda}{N} \left[ \int_{\beta} G_{\pi} (\phi,k) \right]^2 \\
&  + & 2N \frac{\lambda}{N} \int_{\beta} G_{\sigma} (\phi,k) \int_{\beta} G_{\pi} (\phi,k) \ .
\end{eqnarray}

The two gap equations are obtained by minimizing the effective potential with respect to $G_{\pi}$ and $G_{\sigma}$:
\begin{eqnarray}
G^{-1}_{\sigma} (\phi) & =&  D^{-1}_{\sigma}  (\phi) + 4 \lambda \int_{\beta} G_{\pi} (\phi) + \frac{12 \lambda}{N} \int_{\beta} G_{\sigma} (\phi) \ , \\
G^{-1}_{\pi} (\phi) & =&  D^{-1}_{\pi} (\phi) + 4(N+2)\lambda \int_{\beta} G_{\pi} (\phi) + \frac{4\lambda}{N} \int_{\beta} G_{\sigma} (\phi) \ .
\end{eqnarray}

To solve the system one can use the {\it ansatz} for the dressed propagator:  $G_i^{-1} = k^2 + M^2_i$, with $M_i$ being the effective
 masses. Minimizing the effective potential (\ref{eq:effpotlin}) with respect to $\phi$,  one
obtains the third equation that closes the nonlinear system leading
to the three parameters $\phi$ (eventually called $v$), $M_{\sigma}$ and $M_{\pi}$. Introducing
the explicit form of the bare and dressed propagators, the effective
potential finally reads:

\begin{eqnarray}
 \nonumber V (\phi,M) & =& - \overline{\mu}^2 \phi^2 + \frac{\lambda}{N} \phi^4 - \epsilon \phi + Q(M_{\sigma}) + N Q(M_{\pi}) \\
\nonumber & + & \frac{1}{2} \left( - 2 \overline{\mu}^2 + \frac{12\lambda}{N} \phi^2 -M_{\sigma}^2\right) F(M_{\sigma}) - \frac{N}{2} \left( -2 \overline{\mu}^2 + \frac{4\lambda}{N} \phi^2 -M_{\pi}^2\right) F (M_{\pi}) \\
&  + & 3 \frac{\lambda}{N} [F(M_{\sigma})]^2 + (N+2)\lambda [F(M_{\pi})]^2 +  2 \lambda F(M_{\sigma}) F (M_{\pi}) \ ,
\end{eqnarray}
where
\be \label{eq:integralf} F(M)=\int_{\beta} \frac{1}{k^2+M^2} = \int \frac{d^3k}{(2\pi)^3} \frac{1}{2 E_k} + \int \frac{d^3k}{(2\pi)^3} \frac{1}{E_k} \frac{1}{e^{\beta E_k}-1} \ , \ee
and
\be \label{eq:itnegralq} Q(M)= \frac{1}{2} \int_{\beta} \log (k^2+M^2) = \int \frac{d^3k}{(2\pi)^3} \frac{E_k}{2} + T \int \frac{d^3k}{(2\pi)^3} \log[1-e^{-\beta E_k}] \ , \ee
with $E_k=\sqrt{k^2 + M^2}$. As it is well known, both integrals have
a $T=0$ (divergent) and a finite temperature-dependent part. The renormalization of the effective potential is discussed in~\cite{Petropoulos:2004bt} and references therein. 
In particular, there exist some difficulties in the renormalization under the Hartree approximation. See references~\cite{Berges:2005hc} and~\cite{Fejos:2007ec} for more details~\footnote{We thank the referee for bringing these references to our
attention.}. To our purposes we only take into account the finite thermal contributions $F(M) \rightarrow F_{\beta} (M)$ and $Q(M) \rightarrow Q_{\beta}(M)$.


\begin{thebibliography}{}

\bibitem{Adare:2006ti}
  A.~Adare {\it et al.}  (PHENIX Collaboration),
  Phys.\ Rev.\ Lett.\  {\bf 98}, 162301 (2007)
  [nucl-ex/0608033].

\bibitem{Aamodt:2010pa}
  K.~Aamodt {\it et al.}  (The ALICE Collaboration),
  Phys.\ Rev.\ Lett.\  {\bf 105}, 252302 (2010)
  [arXiv:1011.3914 [nucl-ex]].

\bibitem{Luzum:2008cw}
  M.~Luzum and P.~Romatschke,
  Phys.\ Rev.\ C {\bf 78}, 034915 (2008)
  [Erratum-ibid.\ C {\bf 79}, 039903 (2009)]
  [arXiv:0804.4015 [nucl-th]].

\bibitem{Song:2008hj}
  H.~Song and U.~W.~Heinz,
  J.\ Phys.\ G {\bf 36}, 064033 (2009)
  [arXiv:0812.4274 [nucl-th]].

\bibitem{Bozek:2011ph}
  P.~Bozek,
  J.\ Phys.\ G {\bf 38}, 124043 (2011)
  [arXiv:1106.5953 [nucl-th]].

\bibitem{Kovtun:2004de}
  P.~K.~Kovtun, D.~T.~Son and A.~O.~Starinets,
  Phys.\ Rev.\ Lett.\  {\bf 94}, 111601 (2005)
  [hep-th/0405231].


\bibitem{Csernai:2006zz}
  L.~P.~Csernai, J.~I.~Kapusta and L.~D.~McLerran,
  Phys.\ Rev.\ Lett.\  {\bf 97}, 152303 (2006)
  [nucl-th/0604032].

\bibitem{Dobado:2008vt}
  A.~Dobado, F.~J.~Llanes-Estrada and J.~M.~Torres-Rincon,
  Phys.\ Rev.\ D {\bf 79}, 014002 (2009)
  [arXiv:0803.3275 [hep-ph]].


\bibitem{Dobado:2009ek}
  A.~Dobado, F.~J.~Llanes-Estrada and J.~M.~Torres-Rincon,
  Phys.\ Rev.\ D {\bf 80}, 114015 (2009)
  [arXiv:0907.5483 [hep-ph]].

\bibitem{Dobado:2008ri}
  A.~Dobado, F.~J.~Llanes-Estrada and J.~M.~T.~Rincon,
  AIP Conf.\ Proc.\  {\bf 1031}, 221 (2008)
  [arXiv:0804.2601 [hep-ph]].


\bibitem{Weinberg:1971mx}
  S.~Weinberg,
  Astrophys.\ J.\  {\bf 168}, 175 (1971).

\bibitem{Canuto:1978jj}
  V.~Canuto and S.~H.~Hsieh,
  Nuovo Cim.\ B {\bf 48}, 189 (1978).

\bibitem{Arnold:2006fz}
  P.~B.~Arnold, C.~Dogan and G.~D.~Moore,
  Phys.\ Rev.\ D {\bf 74}, 085021 (2006)
  [hep-ph/0608012].

\bibitem{FernandezFraile:2008vu}
  D.~Fernandez-Fraile and A.~Gomez Nicola,
  Phys.\ Rev.\ Lett.\  {\bf 102}, 121601 (2009)
  [arXiv:0809.4663 [hep-ph]].


\bibitem{Dobado:2011qu}
  A.~Dobado, F.~J.~Llanes-Estrada and J.~M.~Torres-Rincon,
  Phys.\ Lett.\ B {\bf 702}, 43 (2011)
  [arXiv:1103.0735 [hep-ph]].


\bibitem{TorresRincon:2011sa}
  J.~M.~Torres-Rincon,
  Prog.\ Part.\ Nucl.\ Phys.\  {\bf 67}, 461 (2012)
  [arXiv:1111.3770 [hep-ph]].

\bibitem{Onuki}
  A.~Onuki
  Phys.\ Rev.\ E {\bf 55}, 403 (1997)

\bibitem{Meyer:2007dy}
  H.~B.~Meyer,
  Phys.\ Rev.\ Lett.\  {\bf 100}, 162001 (2008)
  [arXiv:0710.3717 [hep-lat]].

\bibitem{Karsch:2007jc}
  F.~Karsch, D.~Kharzeev and K.~Tuchin,
  Phys.\ Lett.\ B {\bf 663}, 217 (2008)
  [arXiv:0711.0914 [hep-ph]].


\bibitem{Paech:2006st}
  K.~Paech and S.~Pratt,
  Phys.\ Rev.\ C {\bf 74}, 014901 (2006)
  [nucl-th/0604008].


\bibitem{Kharzeev:2007wb}
  D.~Kharzeev and K.~Tuchin,
  J. High Energy Phys. {\bf 0809}, 093 (2008)
  [arXiv:0705.4280 [hep-ph]].


\bibitem{Li:2009by}
  B.~-C.~Li and M.~Huang,
  Phys.\ Rev.\ D {\bf 80}, 034023 (2009)
  [arXiv:0903.3650 [hep-ph]].


\bibitem{Chakraborty:2010fr}
  P.~Chakraborty and J.~I.~Kapusta,
  Phys.\ Rev.\ C {\bf 83}, 014906 (2011)
  [arXiv:1006.0257 [nucl-th]].


\bibitem{FernandezFraile:2010gu}
  D.~Fernandez-Fraile,
  Phys.\ Rev.\ D {\bf 83}, 065001 (2011)
  [arXiv:1009.2741 [hep-ph]].


\bibitem{Nakano:2011re} 
  E.~Nakano, V.~Skokov and B.~Friman,
  Phys.\ Rev.\ D {\bf 85}, 096007 (2012)
  [arXiv:1109.6822 [hep-ph]].


\bibitem{Dusling:2011fd}
  K.~Dusling and T.~Schafer,
  Phys.\ Rev.\ C {\bf 85}, 044909 (2012)
  [arXiv:1109.5181 [hep-ph]].


\bibitem{Weinberg:1987vp}
  E.~J.~Weinberg and A.~-q.~Wu,
  Phys.\ Rev.\ D {\bf 36}, 2474 (1987).


\bibitem{Iliopoulos:1974ur}
  J.~Iliopoulos, C.~Itzykson and A.~Martin,
  Rev.\ Mod.\ Phys.\  {\bf 47}, 165 (1975).

\bibitem{vanKessel:2008ht}
  M.~T.~M.~van Kessel,
  arXiv:0810.1412 [hep-ph].

\bibitem{Alexandre:1998ts}
  J.~Alexandre, V.~Branchina and J.~Polonyi,
  Phys.\ Lett.\ B {\bf 445}, 351 (1999)
  [cond-mat/9803007].



\bibitem{Dobado:1994fd}
  A.~Dobado and J.~Morales,
  Phys.\ Rev.\ D {\bf 52}, 2878 (1995)
  [hep-ph/9407321].

\bibitem{Aarts:2004sd}
  G.~Aarts and J.~M.~Martinez Resco,
  J. High Energy Phys. {\bf 0402}, 061 (2004)
  [hep-ph/0402192].

\bibitem{Jeon:1994if}
  S.~Jeon,
  Phys.\ Rev.\ D {\bf 52}, 3591 (1995)
  [hep-ph/9409250].

\bibitem{lifshitz1981physical}
  E.~M.~Lifshitz and L.~P.~Pitaevskii,
  {\it Physical Kinetics (Landau and Lifshitz Course of Theoretical Physics vol 10)}.
  Pergamon, Oxford, 1981.


\bibitem{Denicol:2009am}
  G.~S.~Denicol, T.~Kodama, T.~Koide and P.~Mota,
  Phys.\ Rev.\ C {\bf 80}, 064901 (2009)
  [arXiv:0903.3595 [hep-ph]].

\bibitem{Bazavov:2009zn}
  A.~Bazavov, T.~Bhattacharya, M.~Cheng, N.~H.~Christ, C.~DeTar, S.~Ejiri, S.~Gottlieb and R.~Gupta {\it et al.},
  Phys.\ Rev.\ D {\bf 80}, 014504 (2009)
  [arXiv:0903.4379 [hep-lat]].

\bibitem{Borsanyi:2010cj}
  S.~Borsanyi, G.~Endrodi, Z.~Fodor, A.~Jakovac, S.~D.~Katz, S.~Krieg, C.~Ratti and K.~K.~Szabo,
  J. High Energy Phys. {\bf 1011}, 077 (2010)
  [arXiv:1007.2580 [hep-lat]].

\bibitem{Laine:2006cp}
  M.~Laine and Y.~Schroder,
  Phys.\ Rev.\  D {\bf 73}, 085009 (2006)
  [arXiv:hep-ph/0603048].

\bibitem{Gerber:1988tt}
  P.~Gerber and H.~Leutwyler,
  Nucl.\ Phys.\ B {\bf 321}, 387 (1989).

\bibitem{Petropoulos:2004bt}
  N.~Petropoulos,
  hep-ph/0402136.


\bibitem{torresphd} 
  J.~M.~Torres-Rincon, 
  Ph.D. thesis, Universidad Complutense de Madrid, Spain, 2012,
  arXiv:1205.0782 [hep-ph].


\bibitem{Muronga:2006zx}
  A.~Muronga,
  Phys.\ Rev.\ C {\bf 76}, 014910 (2007)
  [nucl-th/0611091].

\bibitem{Cornwall:1974vz}
  J.~M.~Cornwall, R.~Jackiw and E.~Tomboulis,
  Phys.\ Rev.\ D {\bf 10}, 2428 (1974).


\bibitem{AmelinoCamelia:1992nc}
  G.~Amelino-Camelia and S.~-Y.~Pi,
  Phys.\ Rev.\ D {\bf 47}, 2356 (1993)
  [hep-ph/9211211].

\bibitem{AmelinoCamelia:1997dd}
  G.~Amelino-Camelia,
  Phys.\ Lett.\ B {\bf 407}, 268 (1997)
reedit
  [hep-ph/9702403].



\bibitem{Lenaghan:1999si}
  J.~T.~Lenaghan and D.~H.~Rischke,
  J.\ Phys.\ G {\bf 26}, 431 (2000)
  [nucl-th/9901049].


\bibitem{Berges:2005hc} 
  J.~Berges, S.~Borsanyi, U.~Reinosa and J.~Serreau,
  Ann. Phys.\  {\bf 320}, 344 (2005)
  [hep-ph/0503240].

\bibitem{Fejos:2007ec} 
  G.~Fejos, A.~Patkos and Z.~Szep,
  Nucl.\ Phys.\ A {\bf 803}, 115 (2008)
  [arXiv:0711.2933 [hep-ph]].


\end{thebibliography}
\end{document}